\documentclass[11pt]{article}
\usepackage[matrix,arrow]{xy}
\usepackage{epsfig}
\usepackage[mathscr]{euscript}
\usepackage{amssymb,amsthm}
\usepackage{amsmath}
\usepackage{amscd}
\usepackage{latexsym}
\usepackage{graphics}
\usepackage{color}

\usepackage{graphicx}

\input{xy}
\xyoption{all}

\def\beq{\begin{equation}}
\def\eeq{\end{equation}}
\def\bea{\begin{eqnarray}}
\def\eea{\end{eqnarray}}

\catcode`@=11
\renewcommand{\section}{\@startsection{section}{1}{0pt}{\medskipamount}
{\medskipamount}{\large\bf}}
\numberwithin{equation}{section}
\catcode`@=12

\def\b{\beta}
\def\g{\gamma}

\def\de{\delta}
\def\ve{\varepsilon}

\def\m{\mu}

\def\n{\nu}

\def\vr{\varrho}

\def\p{\phi}
\def\vp{\varphi}

\def\ome{\omega}

\def\1{{\bar 1}}
\def\2{{\bar 2}}
\def\3{{\bar 3}}
\def\4{{\bar 4}}

\def\hra{\,\hookrightarrow\,}
\def\mh{\hat\mu}
\def\nh{\hat\nu}

\newcommand{\zb}{\bar{z}}

\newcommand{\su}{{{\rm SU}(2)}}
\newcommand{\suL}{{{\mathfrak{su}}(2)}}

\newcommand{\sut}{{{\rm SU}(3)}}
\newcommand{\uo}{{{\rm U}(1)}}

\newcommand{\uk}{{{\rm U}(N)}}

\newcommand{\suk}{{{\rm SU}(N)}}

\newcommand{\urm}{{{\rm U}}}
\newcommand{\urmL}{{{\mathfrak u}}}
\newcommand{\surm}{{{\rm SU}}}

\newcommand{\sorm}{{{\rm SO}}}

\newcommand{\slrm}{{{\rm SL}}}
\newcommand{\slrmL}{{{\mathfrak{sl}}}}

\newcommand{\Sp}{{\rm S}}

\newcommand{\gfrak}{{\mathfrak{g}}}

\newcommand{\tfrak}{{\mathfrak{t}}}
\newcommand{\zfrak}{{\mathfrak{z}}}

\newcommand{\Hom}{{\rm Hom}}
\newcommand{\End}{{\rm End}}

\newcommand{\one}{{\bf 1}}
\newcommand{\diag}{{\rm diag}}
\newcommand{\vol}{{\rm vol}}
\newcommand{\C}{\mathbb C}
\newcommand{\R}{\mathbb R}

\newcommand{\Z}{\mathbb Z}
\newcommand{\Q}{\mathbb Q}
\newcommand{\Di}{\mathbb D}
\newcommand{\Oc}{\mathbb O}
\newcommand{\I}{\mathbb I}
\newcommand{\Te}{\mathbb T}
\newcommand{\Hbb}{{\mathbb H}}
\newcommand{\Zcal}{{\cal Z}}
\newcommand{\Acal}{{\cal A}}
\newcommand{\Ncal}{{\cal N}}
\newcommand{\Mcal}{{\cal M}}

\newcommand{\Fcal}{{\cal F}}
\newcommand{\Ecal}{{\cal E}}
\newcommand{\Lcal}{{\cal L}}

\newcommand{\T}{{\cal T}}

\newcommand{\Vcal}{{\cal V}}

\newcommand{\Scal}{{\cal S}}
\newcommand{\Gcal}{{\cal G}}
\newcommand{\Ocal}{{\cal O}}
\newcommand{\Xcal}{{\cal X}}

\newcommand{\Rbun}{{\underline{V_R}}}

\newcommand{\Abun}{{\underline{A}}}
\newcommand{\Fbun}{{\underline{F}}}

\newcommand{\Qsf}{{\sf Q}}
\newcommand{\sfs}{{\sf s}}
\newcommand{\sft}{{\sf t}}
\newcommand{\Rsf}{{\sf R}}
\newcommand{\Rep}{{\mathscr{R}}}
\newcommand{\Bun}{{\mathscr{B}}}

\def\im{{\rm i}}
\def\e{{\,\rm e}\,}
\def\N2{$N{=}2$}
\def\pa{\mbox{$\partial$}}
\def\diff{{\rm d}}
\def\tr{{\rm tr}}
\def\sfrac#1#2{{\mbox{$\frac{#1}{#2}$}}}
\def\>{\rangle}
\def\<{\langle}
\def\+{\dagger}
\def\={\ =\ }
\def\und{\qquad\textrm{and}\qquad}
\def\and{\quad\textrm{and}\quad}
\def\for{\qquad\textrm{for}\quad}
\def\with{\qquad\textrm{with}\quad}

\newcommand{\mbf}[1]{{\boldsymbol {#1} }}

\begin{document}
\begin{titlepage}
\setcounter{page}{0}
\begin{flushright}
ITP--UH--22/14\\
EMPG--14--20
\end{flushright}

\vskip 2cm

\begin{center}

{\Large\bf Sasakian quiver gauge theories \\[10pt]
and instantons on Calabi-Yau cones
}

\vspace{15mm}

{\large Olaf Lechtenfeld}${}^1$ , \ {\large Alexander D. Popov${}^{1}$}
\ \ and \ \ {\large Richard J. Szabo${}^2$}
\\[5mm]
\noindent ${}^1${\em Institut f\"ur Theoretische Physik} and {\em Riemann Center for Geometry and Physics}\\
{\em Leibniz Universit\"at Hannover}\\
{\em Appelstra\ss e 2, 30167 Hannover, Germany}\\
Email: {\tt lechtenf@itp.uni-hannover.de , alexander.popov@itp.uni-hannover.de}
\\[5mm]
\noindent ${}^2${\em Department of Mathematics, Heriot-Watt University\\
Colin Maclaurin Building, Riccarton, Edinburgh EH14 4AS, U.K.}\\
and
{\em Maxwell Institute
  for Mathematical Sciences, Edinburgh, U.K.}\\
and
{\em The Higgs Centre for Theoretical Physics, Edinburgh, U.K.}\\
{Email: {\tt R.J.Szabo@hw.ac.uk}}

\vspace{15mm}

\begin{abstract}
\noindent
We consider $\su$-equivariant dimensional reduction of Yang-Mills theory
on manifolds of the form $M\times S^3/\Gamma$, where $M$ is a smooth
manifold and $S^3/\Gamma$
is a three-dimensional Sasaki-Einstein orbifold. We obtain new quiver
gauge theories on $M$ whose quiver bundles are based on the affine
ADE Dynkin diagram associated to $\Gamma$. We relate them to those
arising through translationally-invariant dimensional reduction over the
associated Calabi-Yau cones $C(S^3/\Gamma)$ which are based
on McKay
quivers and ADHM matrix models, and to those arising through $\su$-equivariant dimensional
reduction over the leaf spaces of the
characteristic foliations of $S^3/\Gamma$ which are K\"ahler orbifolds of $\C P^1$
whose quiver bundles are based on the unextended Dynkin diagram corresponding
to $\Gamma$. We use Nahm equations to describe the vacua of $\su$-equivariant quiver gauge
theories on the cones as moduli spaces of spherically symmetric
instantons. We relate them to the Nakajima quiver varieties which can be realized as Higgs branches of the worldvolume
quiver gauge
theories on D$p$-branes probing D$(p+4)$-branes which wrap an ALE space,
and to the moduli spaces of spherically symmetric solutions in putative non-abelian generalizations
of two-dimensional affine Toda field theories.
\end{abstract}

\end{center}

\end{titlepage}

{\baselineskip=12pt
\tableofcontents
}

\bigskip

\section{Introduction and summary\label{sec:Intro}}

\noindent
Quiver gauge theories naturally arise through equivariant
dimensional reduction over K\"ahler coset
spaces~\cite{AlGar12,Lechtenfeld:2007st,Dolan:2010ur} where they find
applications to Yang-Mills-Higgs theories and the construction of (non-abelian) vortices, and also as the low-energy effective
field theories of D-branes at orbifold
singularities~\cite{DM,JM,DGM,Gubser,Douglas:2000qw} which have important ramifications for the AdS/CFT correspondence and whose Higgs branches provide examples of singularity resolution in string geometry. In this paper we clarify the relations
between these two seemingly disparate appearences of quiver gauge
theories in a particular class of examples.

We consider the simplest coset $\C P^1\cong \su/\uo$ (and certain
orbifolds thereof),
whose associated quiver gauge theories are described in detail
in~\cite{PS1} (see also~\cite{Dolan:2009ie,Szabo:2014zua}); the
underlying graph of the quiver bundles in this case is a
Dynkin diagram of type $A_{k+1}$. To any K\"ahler manifold $Z$ one can
associate a $\uo$-bundle over $Z$ whose total space is a
Sasaki-Einstein manifold with Calabi-Yau metric cone. In particular, when $Z=\C P^1$ any such
Sasaki-Einstein manifold is isomorphic to an orbifold $S^3/\Gamma$ of
the three-sphere for
a finite subgroup $\Gamma$ of $\su$; when $\Gamma=\Z_{k+1}$ is a
cyclic group then $S^3/\Z_{k+1}$ is a lens space. In this paper we
describe new quiver gauge theories which can be associated with the
Sasaki-Einstein manifolds $S^3/\Gamma$ via $\su$-equivariant dimensional
reduction, which in certain limits reduce to those of~\cite{PS1}; the
new quiver bundles are associated to quivers with vertex loop edges. We shall also describe quiver gauge theories
associated to Calabi-Yau cones $C(S^3/\Gamma)$ over these spaces,
which correspond to ADE orbifolds of $\R^4$. In
translationally-invariant dimensional reduction such field theories are
central in the description of D-branes at an orbifold singularity
$\R^4/\Gamma$, and to the McKay correspondence for Calabi-Yau
twofolds. The Sasaki-Einstein manifold $S^3/\Gamma$ interpolates
between the two distinct K\"ahler manifolds $C(S^3/\Gamma)$ and $\C P^1$, and in this paper we
use this bridge to clarify the dynamical relations between the McKay
quiver bundles underlying these worldvolume field theories
and the quiver bundles associated with the K\"ahler coset space $\C P^1$.

For this, we shall study the relations between vacua of these quiver
gauge theories and instantons on the Calabi-Yau cones $C(S^3/\Gamma)$, 
which in translationally-invariant dimensional reduction in string theory correspond to D-branes located at points of the
Calabi-Yau spaces $M_\xi$ which are minimal resolutions of the orbifold
singularities $M_0\cong\R^4/\Gamma$ (where $\xi$ are Fayet-Iliopoulos terms serving as stability
parameters of the resolution). In particular, the moduli spaces of
translationally-invariant solutions of the Hermitian Yang-Mills
equations on $\R^4/\Gamma$ coincide with the resolutions
$M_\xi$~\cite{10,K2}, which is described by a particular matrix
model that can be promoted to an ADHM matrix model~\cite{KN}. This correspondence forms the basis for the
four-dimensional $\Ncal=2$ superconformal quiver gauge theories on
D-branes in which $M_\xi$ parameterize the supersymmetric vacuum
states~\cite{DM,JM,Gubser}; under the AdS/CFT correspondence, this
gauge theory is dual to type~IIB supergravity on an $AdS_5\times
S^5/\Gamma$ background. More generally, in type~II string theory the moduli space of
instantons on the ALE space $M_\xi$ can be identified with the Higgs
branch of the quiver gauge theory with eight real supercharges living
on the worldvolume of D$p$-branes probing a set of D$(p + 4)$-branes
which wrap $M_\xi$.

On the other hand, the quiver gauge theories associated with the
Calabi-Yau spaces $C(S^3/\Gamma)$ are technically much more involved,
because instead of translational-invariance along $C(S^3/\Gamma)$ one
imposes $\su$-equivariance along codimension one subspaces
$S^3/\Gamma$ of the cones; the cones are $\su$-manifolds with
one-dimensional orbit space parameterized by $r\in\R_{\geq0}$ such that
there is one singular orbit $S^3/\Gamma$ over $r=0$ and non-singular
orbits $S^3/\Gamma$ for all $r>0$. The condition of $\su$-equivariance
pulls the model back to the quiver gauge theories associated with
$S^3/\Gamma$, but allows for a residual dependence on the radial coordinate $r$
which leads to Nahm-type equations for spherically symmetric
instantons; it is important to note that these Nahm equations describe
$\su$-invariant instantons on $C(S^3/\Gamma)$ and are not related with
monopoles, a feature which somewhat obscures a direct realization in D-brane field theory. Thus instead of the ADHM-type matrix equations describing
translationally-invariant vacua, imposing $\su$-equivariance on gauge
fields reduces the anti-self-duality equations on the four-dimensional
cone to Nahm-type equations. This difference translates into
significant differences between the well-known McKay quivers and our
new ``Sasakian'' quivers which characterize equivariant dimensional
reduction over the Sasaki-Einstein orbifolds.

Furthermore, the same moduli space $M_\xi$ appears as the
moduli space of $\su$-invariant instantons on the Calabi-Yau cone
$\R^4\setminus\{0\}=C(S^3)$, in which case the anti-self-dual Yang-Mills equations reduce
to the Nahm equations with suitable boundary
conditions~\cite{13,14,15}. This is the moduli space of the
spherically symmetric instanton which has the minimal fractional
topological charge; these moduli spaces are further studied and
extended to the cones $C(S^3/\Gamma)$
in~\cite{Ivanova:2013mea}. Moduli spaces of solutions to Nahm equations with these
boundary conditions also appear as Higgs moduli spaces of supersymmetric
vacua in $\Ncal=4$ supersymmetric Yang-Mills theory on the half-space
$\R^{1,2}\times\R_{\geq0}$ with generalized Dirichlet boundary
conditions~\cite{Gaiotto}; these boundary conditions are realized by brane
configurations in which D$p$-branes end on D$(p+2)$-branes at the
boundary of $\R_{\geq0}$. Thus
the solutions of both the ADHM-type and Nahm-type equations give rise
to the same moduli space $C(S^3/\Gamma)$, which after minimal
resolution of singularities is the Calabi-Yau twofold $M_\xi$. 

We shall show that by considering $\Gamma$-equivariant solutions to the Nahm
equations, the moduli space can be
described as a quiver variety via an infinite-dimensional
hyper-K\"ahler quotient construction based on a flat hyper-K\"ahler Banach space factored by the
action of an infinite-dimensional group of gauge transformations; the resulting finite-dimensional
quiver varieties are based on our Sasakian quivers and have orbifold singularities. On the other hand the ADHM
construction describes a minimal resolution of the same moduli space, via a finite-dimensional
hyper-K\"ahler quotient construction using constant matrices, as a quiver variety based
on McKay quivers instead of our Sasakian quivers; the different occurences of quivers corresponds to the
different constraints imposed on the matrices in the respective
cases. In particular, we argue that the minimal charge instanton on
$\R^4/\Gamma$ (or on its Calabi-Yau resolution) can be constructed
both via the ADHM construction (reduced to the Kronheimer matrix
equations) and by reduction to Nahm equations with respect to a radial
coordinate on $\R^4/\Gamma$ (consistently with the construction on $\R^4$),
wherein one obtains the same four-dimensional hyper-K\"ahler moduli
space of vacua.

The outline of the remainder of this paper is as follows. In
Sect.~\ref{sec:SEgeometry} we give a fairly self-contained account of
the geometry of the Sasaki-Einstein orbifolds $S^3/\Gamma$ in a form
that we will use in this paper. In Sect.~\ref{se:Hvb} we derive the
correspondence between homogeneous vector bundles on $S^3/\Gamma$ and
representations of certain new quivers, which we call ``Sasakian
quivers'' and which play a prominent role throughout the paper. In
Sect.~\ref{sec:dimred} we consider $\su$-equivariant dimensional
reduction over the orbifold $S^3/\Gamma$, and derive the
correspondence between $\su$-equivariant vector bundles on product
manifolds $M\times S^3/\Gamma$ and quiver bundles
($\Gamma$-equivariant vector bundles) on $M$ associated to the
Sasakian quivers; the moduli spaces of vacua of the quiver
gauge theories arising from reduction of Yang-Mills theory are
described and it is shown how they reduce to the quiver gauge theories
obtained via $\su$-equivariant dimensional reduction over $\C P^1$
(and certain orbifolds thereof). In Sect.~\ref{sec:instantons} we
describe the moduli spaces of translationally-invariant instantons on
$\R^4/\Gamma$ in terms of Nakajima quiver varieties, and their
connection to moduli spaces of instantons on an ALE space $M_\xi$
which are based on McKay quivers. Finally, in Sect.~\ref{sec:Nahm} we
consider $\su$-equivariant dimensional reduction over the cones
$C(S^3/\Gamma)$ and study the Higgs branch of vacua of the induced quiver gauge
theories as moduli spaces of solutions to Nahm equations which are
based on extensions of quiver varieties to the setting of Sasakian
quivers; we discuss their relations to the ADHM-type moduli spaces and
to certain Nakajima quiver varieties of $A_n$-type, and also their
realization in a variant of non-abelian affine Toda field theory which
extends the duality between four-dimensional gauge theories and
two-dimensional conformal field theories.

\bigskip

\section{Geometry of Sasaki-Einstein three-manifolds\label{sec:SEgeometry}}

\noindent {\bf Sphere $\mbf{S^3}$. \ } Let $S^3$ be the standard round three-sphere of radius
$r$. It can be described via the embedding $S^3\subset \R^4$ by the equation
\beq\label{2.1} \de_{\m\n}\, y^{\m}\, y^{\n}=r^2  \ , \eeq where
$y^\mu\in\R$ and $\m
,\n ,\ldots=1, 2,3 ,4$. On
$S^3$ one can introduce a basis of left $\su$-invariant one-forms $\{e^a\}$ as 
\beq\label{2.2} e^a:= -
\sfrac{1}{r^2}\ \bar\eta^a_{\m\n}\, y^{\m}\, \diff y^{\n} \ , \eeq where
$\bar\eta^a_{\m\n}$ are
the anti-self-dual 't~Hooft tensors 
\beq
\bar\eta^a_{bc}=\ve^a_{bc} \und \bar\eta^a_{b4}=-\bar\eta^a_{4b}=-\de^a_b
\ ,
\eeq
with $\ve^1_{23}=1$ and $a,b,\ldots=1,2,3$. These one-forms satisfy
the Maurer-Cartan equations \beq\label{2.3} \diff e^a
+\ve^a_{bc}\,e^b\wedge e^c = 0\ , 
\eeq 
and the round metric on $S^3$ is given by
\beq
\diff s^2_{S^3}=r^2\, \delta_{ab}\, e^a\otimes e^b \ .
\eeq

\bigskip

\noindent
{\bf Sasaki-Einstein orbifolds $\mbf{S^3/\Gamma}$. \ }
Sasakian manifolds are the natural odd-dimensional counterparts of K\"ahler
manifolds; a Riemannian manifold is Sasakian if its associated metric
cone is K\"ahler. A Sasakian manifold is Sasaki-Einstein if its
Riemannian metric is an Einstein metric, or equivalently if its metric
cone is a Calabi-Yau space. Thus Sasaki-Einstein spaces are the
natural odd-dimensional versions of K\"ahler-Einstein spaces. For background on Sasaki-Einstein manifolds, see e.g.~\cite{Boyer}.

Three-dimensional Sasaki-Einstein spaces are completely
classified. They correspond to the homogeneous spherical space forms
in three dimensions, which are the smooth orbifolds $S^3/\Gamma$ where
$\Gamma$ is a finite subgroup of $\su$ commuting with $\uo\subset\su$
which acts freely and isometrically
by left translations on the
group manifold $\su\cong S^3$; by the McKay correspondence they have an ADE classification and we run through the complete list below. The orbifold
$S^3/\Gamma$ has a description as a Seifert fibration, i.e. as a $\uo$
V-bundle
\beq
\pi_\Gamma \,:\, S^3/\Gamma \ \longrightarrow \ \C P^1/\Gamma_0 \ ,
\label{Seifertfib}\eeq
which is called the characteristic foliation of the Sasaki-Einstein
space. The space of leaves $\C P^1/\Gamma_0$ is a one-dimensional
complex orbifold with a K\"ahler-Einstein metric; the ADE group
$\Gamma$ is a central extension (by rotations of the regular fibres) of the group
$\Gamma_0\subset\sorm(3)$ which acts isometrically on a local coordinate $z\in\C$ of
the Riemann sphere $\C P^1\cong S^2$ by $\su$ M\"obius
transformations. Concretely, the orbifold base $\C P^1/\Gamma_0$ can be
regarded as a Riemann sphere together with $m$ arbitrary marked points whose
coordinate charts are modelled on $\C/\Z_{k_j}$ for some ramification
indices $k_j\geq0$ with $j=1,\ldots,m$. Let
$\Lcal_\Gamma\to\C P^1/\Gamma_0$ be the line V-bundle
associated to the Seifert fibration (\ref{Seifertfib}); it is defined by
the identifications
\beq
(z,w)\sim \big(\zeta_{k_j} \,z,\zeta_{k_j}^{l_j}\, w \big) \qquad \mbox{with} \quad
\zeta_{k_j} =\exp\big(2\pi\, \im/k_j\big)
\label{orbid}\eeq
of the local coordinates $(z,w)\in\C^2$ of the base and fibre around the
$j$-th orbifold point, for some integer isotropy weights $0\leq
l_j<k_j$. Note that ${\rm deg}(\Lcal_\Gamma)=c_1(|\Lcal_\Gamma|)=d\in\Z$, where
$|\Lcal_\Gamma|\to\C P^1$ is the associated smooth line bundle obtained by smoothing the orbifold
points and eliminating the monodromies $l_j=0$; on the other hand, the
orbifold Chern class of $\Lcal_\Gamma$ is given by
\beq
c_1(\Lcal_\Gamma) = d+\sum_{j=1}^m\, \frac{l_j}{k_j} \ \in \ \Q_{>0} \ .
\eeq
The locally free $\uo$-action on $S^3/\Gamma$ arises
from rotations in the fibres over $\C P^1/\Gamma_0$, with the fibre over
the $j$-th ramification point of $\C P^1/\Gamma_0$ given by
$S^1/\Z_{k_j}$ due to the orbifold identification (\ref{orbid}). Since
$S^3$ is simply connected and $\Gamma$ acts freely, the fundamental group of $S^3/\Gamma$ is
the finite group $\Gamma$ itself. Using the standard
presentation of $\pi_1(S^3/\Gamma)$ in terms of generators and
relations associated to the Seifert invariants, this gives a geometric
presentation of the orbifold group $\Gamma$ in terms of the central
generator $h$ along the generic $S^1$
fibre over $\C P^1/\Gamma_0$ together with one-cycles $\xi_j$ for $j=1,\ldots,m$
encircling the orbifold points on $\C P^1/\Gamma_0$, with the relations
\beq
\xi_j^{k_j}= h^{-l_j} \und \xi_1\cdots \xi_m=h^d \ .
\eeq
In particular, by setting $h=1$ this yields a
presentation of the orbifold fundamental group $\Gamma_0= \pi_1(\C P^1/\Gamma_0)$ in terms of generators $\xi_j$
with relations.

By working on the uniformizing system of local charts of the orbifold $\C
P^1/\Gamma_0$, we can define a local basis of one-forms for $S^3/\Gamma$
as follows. For this, we use the Seifert description of $S^3$ as the
total space of the circle bundle of the line bundle $\Lcal\to\C P^1$
of degree one corresponding
to the Hopf fibration $S^3\to S^2$. Consider the $(k+1)$-tensor power $\Lcal^{k+1}:=(\Lcal )^{\otimes (k+1)}$ of $\Lcal$, which is the Hermitian line bundle
\beq\label{2.4}
\Lcal^{k+1} \ \longrightarrow \ \C P^1
\eeq
with the unique $\su$-invariant connection $a_{k+1}$ having in local coordinates the form
\beq\label{2.5}
a_{k+1}=(k+1)\, a_1 = \frac{k+1}{2(1+z\,\bar z)}\ (\bar z\, \diff z -
z\, \diff \bar z)\ .
\eeq 
Let
\beq
\beta=\frac{ \diff z}{1+z\, \bar z} \qquad \mbox{and} \qquad
\bar\beta=\frac{\diff \bar z}{1+z\, \bar z}
\eeq
be the unique $\su$-invariant forms of types $(1,0)$ and $(0,1)$ on
$\C P^1$; they form a basis of covariantly constant sections of the
canonical line bundles $K=\Lcal^2$ and $K^{-1}=\Lcal^{-2}$ obeying
\beq
\diff \beta-a_2\wedge\beta=0 \und
\diff\bar\beta-a_{-2}\wedge\bar\beta=0 \ .
\label{betacovconst}\eeq
The $\su$-invariant K\"ahler $(1,1)$-form on $\C P^1$ is $\omega
=\frac\im2\, r^2\, \beta\wedge\bar \beta$. Then
basis one-forms on $S^3/\Gamma$ can be introduced via
\beq
e^1+\im\, e^2 = \pi_\Gamma^*\, \beta \und \, e^3= \mbox{$\frac1d$}\, \big(
\diff
\varphi-\im\, c_1(\Lcal_\Gamma) \, \pi_\Gamma^* \, a_1 \big) \
,
\label{eaS3Gamma}\eeq
where $0\leq\varphi<2\pi$ is a local coordinate on the $S^1$ fibre of
the Seifert bundle (\ref{Seifertfib}). The frame
element $e^3$ is a connection one-form on the Seifert fibration
with curvature
\beq
\diff e^3=\sfrac2{d\, r^2} \, c_1(\Lcal_\Gamma) \,
\pi_\Gamma^*\, \omega \ .
\label{diffe3S3Gamma}\eeq
From (\ref{betacovconst}) and (\ref{diffe3S3Gamma}) we obtain similar Maurer-Cartan
equations (\ref{2.3}) for $e^a$ as on $S^3$. In particular, the
Sasaki-Einstein metric
on $S^3/\Gamma$ is given by
\beq
\diff s_{S^3/\Gamma}^2 = \pi_\Gamma^*\,\diff s_{\C P^1}^2+r^2\, e^3\otimes
e^3 \ ,
\label{SEmetric}\eeq
where $\diff s_{\C P^1}^2= r^2\,
\beta\otimes\bar\beta$ is the
K\"ahler-Einstein metric on $\C P^1$.

\bigskip

\noindent
{\bf Calabi-Yau cones $\mbf{C(S^3/\Gamma)}$. \ }
Our Sasaki-Einstein orbifolds can be defined as the bases of affine
Calabi-Yau cones described by polynomial equations in $\C^3$. The equation (\ref{2.1}) defines the embedding of
$S^3$ into the space $\R^4\cong\C^2$ with the complex coordinates
\beq\label{2.8}
z^1=y^1+\im\,y^2 \und z^2=y^3+\im\,y^4\ .
\eeq
In fact, $\R^4$ can be considered as a cone
$C(S^3)$ over $S^3$,
\beq\label{2.9}
\R^4\setminus\{0\} = C(S^3)\ ,
\eeq with the metric
\beq\label{2.10}
\diff s_{C(S^3)}^2 =\diff r^2 + r^2\, \de_{ab}\, e^a\otimes e^b
\eeq
where $r^2=\de_{\m\n}\, y^{\m}\, y^{\n}$ and $e^a$ are the one-forms
(\ref{2.2}) on $S^3$. We can identify $\R^4\cong \C^2$ with
the affine subvariety cut out by the linear relation
$f(x,y,z)=x+y+z=0$ in $\C^3$; by rescaling the affine coordinates
$(x,y,z)\in\C^3$ by $\lambda\in\C^*$, the polynomial
$f (x,y,z)$ may also be regarded as cutting out a copy of the Riemann sphere
$S^2\cong \C
P^1\subset\C P^2$, and in this way the Calabi-Yau cone describes the
standard Hopf fibration $S^3\to S^2$. In
this case the spheres $S^3$ of varying radii are the natural invariant
submanifolds for the action of the Lorentz group
$\sorm(4)\cong\su\times\su$ on $\R^4$, in the sense that the orbits of
the free left action of $\su$ on $(z^1,z^2)$ parameterize families of
three-spheres via the cone relation (\ref{2.9}).

Consider the action of $\Gamma$ on $\C^2$ given by
\beq\label{2.12}
(z^1,z^2) \ \longmapsto \ (g^1{}_\alpha\, z^\alpha, g^2{}_\alpha\,
z^\alpha)
\eeq
for $\alpha,\beta,\ldots=1,2$, where $g^\alpha{}_\beta$ are the matrix
elements of $g\in\Gamma$ in the fundamental two-dimensional
representation of $\Gamma\subset\su$ on $\C^2$. It has a single
isolated fixed
point at the origin $(z^1,z^2)=(0,0)$. The orbifold $\C^2/\Gamma$ is
defined as the set of equivalence classes with respect to the
equivalence relation
\beq\label{2.13}
(g^1{}_\alpha\, z^\alpha, g^2{}_\alpha\,
z^\alpha) \equiv (z^1,z^2)
\eeq
for all $g\in\Gamma$. It has a singularity at the origin and is a cone over $S^3/\Gamma$,
\beq\label{2.14}
\big(\C^2\setminus\{0\} \big)\, \big/\, \Gamma=C\big(S^3/\Gamma \big) \ ,
\eeq
with the metric (\ref{2.10}) for $e^a$ given in
(\ref{eaS3Gamma}). 

A natural description of the cone (\ref{2.14}) is as the complex surface in 
$\C^3$ which is invariant under the $\C^*$-action given by
$(x,y,z)\mapsto (\lambda^{w_x}\, x,\lambda^{w_y}\, y,
\lambda^{w_z}\, z)$; due to the scaling symmetry with respect
to $\lambda\in\C^*=\R_{>0}\times S^1$ these spaces are cones, and
because of the $\R_{>0}$-action the Calabi-Yau metric can be written
in the form (\ref{2.10}). Explicitly, it can be defined as the subvariety cut out
by a polynomial equation
\beq
f_\Gamma(x,y,z)=0 \ ,
\label{fGammaxyz}\eeq
where
$f_\Gamma(x,y,z)$ is a weighted homogeneous polynomial, i.e. $f_\Gamma(\lambda^{w_x}\, x,\lambda^{w_y}\, y,
\lambda^{w_z}\, z) = \lambda^n\, f_\Gamma(x,y,z)$ for all $\lambda\in\C^*$. The degree $n>0$
coincides with the dual Coxeter number of the corresponding ADE Lie group,
while the pairwise coprime weights $w_x,w_y,w_z\in\Z_{\geq0}$ are
divisors of $n$ with $|\Gamma|=2\, w_x\, w_y$ and $w_z=\frac
n2=w_x+w_y-1$. These integers can be expressed in terms of
representation theory data of the orbifold group $\Gamma$: If $n_\ell$ for 
$\ell=0,1,\dots,r_\Gamma$ is the dimension of the $\ell$-th irreducible
representation $V_\ell$ of $\Gamma$, where $r_\Gamma$ is the rank of the
corresponding ADE Lie group and $\ell=0$ labels the trivial
representation with $n_0=1$, then
\beq
n=\sum_{\ell=0}^{r_\Gamma}\, n_\ell \und |\Gamma|=\sum_{\ell=0}^{r_\Gamma}\, n_\ell^2 \ .
\eeq
The base of the cone
is the intersection of $\C^2/\Gamma$ with the sphere $S^5\subset\C^3$ of
radius $r$,
which is just $S^3/\Gamma$. The characteristic foliation (\ref{Seifertfib})
is then generated by the remaining $\uo$-action on $S^3/\Gamma$ inherited from
the original $\C^*$-action on the cone. In fact, this construction
provides an alternative realization of the leaf space $\C P^1/\Gamma_0$
as a quasi-smooth weighted
projective curve cut out by the \emph{same} polynomial equation
(\ref{fGammaxyz}), but now regarded in the weighted projective plane
$\C P^2(w_x,w_y,w_z)$. This yields a natural description of $\C
P^1/\Gamma_0=(S^3/\Gamma)/S^1$ as the complex orbifold
\beq
\C P^1/\Gamma_0=C\big(S^3/\Gamma\big)\, \big/\,
\C^*=\big((\C^2\setminus\{0\})/\Gamma \big)\, \big/\, \C^* \ .
\eeq

\bigskip

\noindent {\bf $\mbf{\Gamma}$-action on one-forms. \ }
Consider a one-form
\beq\label{2.15} Y=Y_{\m}\, \diff
y^{\m} = Y_{z^1}\, \diff z^1 + Y_{z^2}\, \diff z^2 + Y_{\zb^{\1}}\, \diff \zb^{\1} +
Y_{\zb^{\2}}\, \diff \zb^{\2}
\eeq
on $\R^4\cong\C^2$ which is invariant under the action of
$\Gamma\subset\su\subset\sorm(4)$ defined by (\ref{2.12}). Then for
the components
\beq\label{2.16} Y_{z^1} = \sfrac12 \, (Y_1 - \im\, Y_2)\und Y_{z^2} =
\sfrac12 \, (Y_3 - \im\, Y_4) \eeq we have
\beq\label{2.17} Y_{z^1} \ \longmapsto \ (g^{-1})_1{}^\alpha\,
Y_{z^\alpha} \und
Y_{z^2} \ \longmapsto \ (g^{-1})_2{}^\alpha\, Y_{z^\alpha} \eeq
for all $g\in\Gamma$.
On the other hand, for the components $(X_a, X_r)$ in spherical coordinates, defined as
\bea\nonumber
Y\=Y_{\m}\, \diff y^{\m} &=:& X_a\, e^a + X_r\, \diff r \\[4pt] 
&=& \sfrac{1}{2}\, (X_1 - \im\,X_2) \, (e^1+\im\,e^2) +
\sfrac{1}{2}\,(X_3 - \im\,r\,X_r) \, (e^3+\im\,\sfrac{\diff r}{r}) +
\mbox{h.c.} \ ,
\label{2.18}\eea
the $\Gamma$-action on $Y_\mu$ implies
\beq\label{2.19}
X_1 + \im\,X_2 \ \longmapsto \ \pi(g)(X_1 +
\im\,X_2) \und X_3 + \im\,r\,X_r \ \longmapsto \
\pi(g)(X_3)+ \im\,r\,X_r \ , \eeq
which defines the representation $\pi:\Gamma\to\End_{C^\infty(S^3)} \big(\Omega_{S^3}^1 \big)$ of the
orbifold group on one-forms on $S^3$. Once the transformations (\ref{2.12}) are given
explicitly, the
transformations (\ref{2.19}) can be worked out from the formulas (\ref{2.2}) which
in the coordinates (\ref{2.8}) have the form \beq\label{2.22} e^1+\im\,e^2=
\sfrac{\im}{r^2}\, (z^1\, \diff\zb^{\2}- \zb^{\2}\,\diff z^1)\und e^3=\sfrac{\im}{2r^2}\,
(\zb^{\1}\, \diff z^1+z^2\, \diff \zb^{\2}- z^1\, \diff\zb^{\1}-
\zb^{\2}\, \diff z^2)\ . \eeq
Note that $X_a$ and $X_{\tau}:=r\, X_r$ for $\tau = \log r$ can be considered as components of one-forms on the cylinder
$\R\times S^3/\Gamma$ with the metric 
\beq\label{2.20} \diff
s^2_{\R\times S^3/\Gamma} =\diff \tau^2 + \de_{ab}\, e^a\otimes e^b=\frac{\diff
  r^2}{r^2} + \de_{ab}\, e^a\otimes e^b=
\frac{1}{r^2}\, \diff s_{C(S^3/\Gamma)}^2\eeq where $\diff
r=\diff\big(z^1\,\zb^{\1}+z^2\, \zb^{\2}\,\big)^{1/2}$, which is conformally equivalent to the metric (\ref{2.10})
on the cone (\ref{2.14}) for $e^a$ given in (\ref{eaS3Gamma}). 

\bigskip

\noindent
{\bf $\mbf{A_k}$. \ }
In this case $\Gamma=\Z_{k+1}$ for $k\geq1$ is the cyclic group of
order $k+1$, generated by an element $h$ with $h^{k+1}=1$, and
the quotient space $S^3/\Z_{k+1}$ is the lens space $L(k+1,1)$;  in
this instance $\Gamma_0$ is the trivial group.
It can be identified with the total space of the $\uo$-bundle
associated to the line bundle (\ref{2.4}),
regarded as an $S^1/\Z_{k+1}$-fibration over $\C P^1$. Basis
one-forms on this space can be constructed explicitly by using a local
section of the circle bundle of the bundle (\ref{2.4}) given by the matrix
\beq\label{2.6}
g(z,\bar z) =\frac{1}{(1+z\,\bar z)^{1/2}}
\, \begin{pmatrix}1&-\zb\\z&1\end{pmatrix}\, \begin{pmatrix}\exp(\frac{\im\,\vp}{k+1})&0\\[2pt]
0& \exp(-\frac{\im\,\vp}{k+1})\end{pmatrix} \ \in \ \su \ , 
\eeq 
where $0\le\vp<2\pi$ is a local coordinate on $S^1$ and
$\exp(\frac{\im\, \vp}{k+1}) \in S^1/\Z_{k+1}$.
On the trivial rank two complex vector bundle $S^3/\Z_{k+1}\times\C^2$ over $S^3/\Z_{k+1}$ we
introduce the flat connection
\beq\label{2.7}
B:=g^{-1}\, \diff g = \begin{pmatrix}\frac{\im}{k+1}\,\diff\vp +a_1&
  -\exp(- \frac{2\, \im\,\vp}{k+1})\,\bar\beta \, \\[4pt]
\exp(\frac{2\, \im\,\vp}{k+1})\, \beta &-\frac{\im}{k+1}\,\diff\vp -a_1
\end{pmatrix}=:\begin{pmatrix} \im\, e^3&
  -e^1+\im\,e^2\\ e^1+\im \, e^2& -\im\,e^3\end{pmatrix} \ ,
\eeq 
which is an $\suL$-valued one-form on $S^3/\Z_{k+1}$. From flatness of the
connection (\ref{2.7}), $\diff B+B\wedge B=0$, we obtain the same Maurer-Cartan
equations (\ref{2.3}) for $e^a$ as on $S^3$, i.e. for $k=0$.

The relevant group
theory data is encoded in the extended simply laced Dynkin diagram
${\sf A}_k$ of the affine $\widehat
A_k$ Lie algebra given by
\beq
{\scriptsize\xymatrix{
& & {\times}\ar@{-}[dll] \ar@{-}[drr] & & \\
{ 1}\ar@{-}[r] &{ 1}\ar@{-}[r]& \ \cdots \ \ar@{-}[r]& { 1}\ar@{-}[r] & { 1}
}}
\label{AkDynkin}\eeq
with $r_\Gamma+1=k+1$ nodes designating the irreducible representations of $\Gamma$; here and in the following the integers at the
nodes are the Dynkin indices $n_\ell$ of the affine roots, and the node marked $\times$
corresponds to the trivial representation with $n_0=1$.
Any unitary representation of $\Z_{k+1}$ is given by a sum of
one-dimensional irreducible representations
$V_{\ell}$ with $\ell=0,1,\dots,k \ ({\rm mod}\, (k+1))$ on which the generator $h$ acts as multiplication by
\beq\label{2.11}
\zeta_{k+1}^{\ell}:=\exp(\sfrac{2\pi\,\im\,{\ell}}{k+1})\ .
\eeq
As a subgroup of $\su$, the generating element of $\Z_{k+1}$ is given
by
\beq
\big(h^\alpha{}_\beta\big)=\begin{pmatrix}
  \zeta_{k+1}&0\\0&\zeta_{k+1}^{-1} \end{pmatrix} \ ,
\label{hfundrep}\eeq
and the transformation (\ref{2.19}) in this case reads as
\beq
\pi(h)(X_1+ \im\, X_2)=\zeta_{k+1}^{2} \, (X_1+ \im\, X_2) \und
\pi(h)(X_3)= X_3 \ .
\label{pihAk}\eeq
Alternatively, the transformation (\ref{pihAk})
can be derived immediately from the definition (\ref{2.7}) by considering
the fibre monodromy $\vp\mapsto\vp +
2\pi$ which generates \beq\label{2.21} e^1\pm \im\,e^2 \ \longmapsto \
\zeta_{k+1}^{\pm\,2} \, (e^1\pm\im\,e^2) \und e^3 \ \longmapsto \
e^3\ . \eeq

The pertinent weighted homogeneous polynomial is
\beq
f_{\Z_{k+1}} (x,y,z) = x^{k+1}+y^2+z^2 \ ,
\label{fZk+1}\eeq
with the weights
\beq
w_x=1 \qquad \mbox{and} \qquad w_y=w_z=q+1 \qquad
\mbox{for} \quad k=2q+1
\eeq
and
\beq
w_x=2 \qquad \mbox{and} \qquad w_y=w_z=2q+1 \qquad \mbox{for} \quad
k=2q \ .
\eeq
Note that in the special case $k=1$ where
$S^3/\Z_2=\su/\Z_2= \sorm(3)=\R P^3$, this provides a realization
of the complex projective line $\C P^1$ as the smooth Stenzel curve $x^2+y^2+z^2=0$ in~$\C P^2$.

\bigskip

\noindent{\bf $\mbf{D_k}$. \ } 
In this case $\Gamma=\Di^*_{k-2}$ for
$k\geq4$ is the binary extension of the dihedral group
$\Gamma_0=\Di_{k-2}$ in $\su$ of
order $4(k-2)$, and $S^3/\Di^*_{k-2}$ is a prism manifold; the
orbifold group $\Gamma$ in this case is the pullback of $\Gamma_0$
under the covering homomorphism $S^3\to \sorm(3)=\su/\Z_2$ of degree two. It has a presentation as a non-abelian group generated
by elements $\xi_1,\xi_2,\xi_3$ with the relations
\beq
\xi_1^2=\xi_2^2=\xi_3^{k-2}=\xi_1\, \xi_2\, \xi_3 \ .
\eeq
Note that $\xi_3$ generates a cyclic subgroup $\Z_{2(k-2)}$, and
$\xi_1=\xi_2\, \xi_3$. The center of this group has order two and is
generated by the element $h=\xi_1\, \xi_2\, \xi_3$; whence the generic fibre of the Seifert manifold
$S^3/\Di^*_{k-2}$ is $S^1/\Z_2$, and the geometry is analogous to that
of the lens space $S^3/\Z_2$. The branching
indices $(k_1,k_2,k_3)$ on $\C P^1/\Di_{k-2}$ are given by the elliptic triple $(2,2,k-2)$, and so
there are three exceptional fibres $S^1/\Z_2$, $S^1/\Z_2$, and
$S^1/\Z_{k-2}$, respectively, each with isotropy one; in particular
the base orbifold group is a $(2,2,k-2)$ triangle group. Moreover,
the degree of the line V-bundle $\Lcal_{\Di^*_{k-2}}\to \C
P^1/\Di_{k-2}$ is $d=1$; whence the underlying smooth line bundle is
$|\Lcal_{\Di^*_{k-2}}|= \Lcal$ and the orbifold Chern class is given
by
\beq
c_1(\Lcal_{\Di^*_{k-2}})=2+\sfrac1{k-2} \ .
\eeq

The extended simply laced
Dynkin diagram ${\sf D}_k$ for the affine $\widehat D_k$ Lie algebra is
\beq
{\scriptsize\xymatrix{
{\times} \ar@{-}[dr] & & & & & &
 1
\\
 &  2 \ar@{-}[r] &  2 \ar@{-}[r] & \ \cdots \
 \ar@{-}[r] &  2 \ar@{-}[r] &  2 \ar@{-}[ur]
 \ar@{-}[dr] & \\
 1 \ar@{-}[ur] & & & & & &  1
}}
\label{DynkinDk}\eeq
with $r_\Gamma+1=k+1$ nodes. As elements of $\su$ the generators of
$\Di^*_{k-2}$ are given by
\beq
\big(\xi_2{}^\alpha{}_\beta\big) = \begin{pmatrix} 0& 1\\ -1&
  0 \end{pmatrix} \qquad \mbox{and} \qquad \big(\xi_3{}^\alpha{}_\beta\big)= \begin{pmatrix} \zeta_{2(k-2)} & 0\\ 0&
  \zeta_{2(k-2)}^{-1}\end{pmatrix}
\label{Difundrep}\eeq
together with $\xi_1=\xi_2\,\xi_3$; in this representation the central
element is $h=-\one_{\C^2}$. The transformations
(\ref{2.19}) under the generator $\xi_3$ are analogous to those of
(\ref{pihAk}), while under $\xi_2$ one has 
\bea
\pi(\xi_2)(X_1) = -X_1 \ , \qquad \pi(\xi_2)(X_2)=X_2 \und \pi(\xi_2)(X_3)= -X_3 \ .
\eea
The prism manifold $S^3/\Di^*_{k-2}$ can also be described as the
weighted projective curve
\beq
f_{\Di^*_{k-2}}(x,y,z)= x^{k-1}+x\, y^2+z^2=0 \qquad \mbox{in} \quad
\C P^2(2,k-2,k-1) \ .
\eeq

\bigskip

\noindent{\bf $\mbf{E_6}$. \ }
In this case $\Gamma=\Te^*\cong\slrm(2,\Z_3)$ is the binary tetrahedral group of order
$24$ which is a central extension of the tetrahedral group
$\Gamma_0=\Te\cong A_4$ by a cyclic group $\Z_2$. It has a presentation in
terms of generators $\xi_1,\xi_2,\xi_3$ with relations
\beq
\xi_1^2=\xi_2^3=\xi_3^3=\xi_1\, \xi_2\, \xi_3 \ .
\eeq
Again the order two center of $\Te^*$ is generated by
$h=\xi_1\,\xi_2\,\xi_3$, and the branching indices on $\C P^1/\Te$ are given
by the Platonic triple $(2,3,3)$; whence $\Gamma_0=\Te$ is a triangle
group, and there are three exceptional
fibres $S^1/\Z_2$, $S^1/\Z_3$, and $S^1/\Z_3$ with monodromy
one. The line V-bundle $\Lcal_{\Te^*}\to \C P^1/\Te$ with
$|\Lcal_{\Te^*}|=\Lcal$ has orbifold Chern class
\beq
c_1(\Lcal_{\Te^*}) = \sfrac{13}6 \ .
\eeq

The extended simply laced Dynkin diagram ${\sf E}_6$ for the affine $\widehat E_6$ Lie algebra is
\beq
{\scriptsize\xymatrix{
 & & {\times} \ar@{-}[d] & & \\
 & & 2 \ar@{-}[d] & & \\
1 \ar@{-}[r] & 2 \ar@{-}[r] & 3 \ar@{-}[r] & 2 \ar@{-}[r] & 1
}}
\eeq
It is straightforward to write down elements of $\su$ for the
generators $\xi_j$ of $\Te^*$. However, for most practical calculations
it is more convenient to note that the group $\Te^*\subset\su$ is generated by the order eight dicyclic group $\Di^*_2$ and the additional generator
\beq
\big(g^\alpha{}_\beta\big) = \frac1{\sqrt2}\, \begin{pmatrix}
  \zeta_8^{-1}& \zeta_8^{-1} \\ -\zeta_8& \zeta_8 \end{pmatrix} = \frac1{1-\im}\, \begin{pmatrix} 1&
  1\\-\im& \im \end{pmatrix}
\label{gadditional}\eeq
satisfying $g^3=-\one_{\C^2}$. The additional transformations by $g$ in (\ref{2.19})
read as
\beq
\pi(g)(X_1)=X_2 \ , \qquad \pi(g)(X_2)=-X_3
 \und \pi(g)(X_3)= -X_1 \ .
\eeq
The Sasaki-Einstein manifold $S^3/\Te^*$ also has a presentation as
the weighted projective curve
\beq
f_{\Te^*}(x,y,z) = x^4+y^3+z^2=0 \qquad \mbox{in} \quad \C P^2(3,4,6) \
.
\eeq

\bigskip

\noindent{\bf $\mbf{E_7}$. \ }
In this case $\Gamma=\Oc^*$ is the binary octahedral group of order
$48$ which is a central extension of the octahedral group
$\Gamma_0=\Oc\cong S_4$ by a cyclic group $\Z_2$. It has a presentation in
terms of generators $\xi_1,\xi_2,\xi_3$ with relations
\beq
\xi_1^2=\xi_2^3=\xi_3^4=\xi_1\, \xi_2\, \xi_3 \ .
\eeq
The order two center of $\Oc^*$ is generated by
$h=\xi_1\,\xi_2\,\xi_3$, and the branching indices on $\C P^1/\Oc$ are given
by the Platonic triple $(2,3,4)$; whence there are three exceptional
fibres $S^1/\Z_2$, $S^1/\Z_3$, and $S^1/\Z_4$ each of isotropy
one. The line V-bundle $\Lcal_{\Oc^*}\to \C P^1/\Oc$ with
$|\Lcal_{\Oc^*}|=\Lcal$ has orbifold Chern class
\beq
c_1(\Lcal_{\Oc^*}) = \sfrac{25}{12} \ .
\eeq

The extended simply laced Dynkin diagram ${\sf E}_7$ for the
affine exceptional $\widehat E_7$ Lie algebra is
\beq
{\scriptsize\xymatrix{
 & & & 2 \ar@{-}[d] & & & \\
\times \ar@{-}[r] & 2 \ar@{-}[r] & 3 \ar@{-}[r] & 4 \ar@{-}[r] & 3 \ar@{-}[r] &
2\ar@{-}[r] & 1
}}
\eeq
The representation of $\Oc^*$ in $\su$ can be obtained similarly to
that of $\Te^*$ by extending the order~$16$ dicyclic group $\Di^*_4$
by the same generator (\ref{gadditional}).
The spherical three-manifold $S^3/\Oc^*$ also has a presentation as
the weighted projective curve
\beq
f_{\Oc^*}(x,y,z) = x^3\, y +y^3+z^2=0 \qquad \mbox{in} \quad \C P^2(4,6,9) \
.
\eeq

\bigskip

\noindent{\bf $\mbf{E_8}$. \ }
In this final case $\Gamma=\I^*$ is the binary icosahedral
group which is a double cover of the simple icosahedral group
$\Gamma_0=\I\cong A_5$ of order $60$, and $S^3/\I^*$ is the Poincar\'e homology
sphere $L(5,3,2)$. It has a presentation in
terms of generators $\xi_1,\xi_2,\xi_3$ with relations
\beq
\xi_1^2=\xi_2^3=\xi_3^5=\xi_1\, \xi_2\, \xi_3 \ .
\eeq
Once more the order two center of $\I^*$ is generated by
$h=\xi_1\,\xi_2\,\xi_3$, and the branching indices on $\C P^1/\I$ are given
by the Platonic triple $(2,3,5)$; whence there are three exceptional
fibres $S^1/\Z_2$, $S^1/\Z_3$, and $S^1/\Z_5$ each of isotropy
one. The line V-bundle $\Lcal_{\I^*}\to \C P^1/\I$ with
$|\Lcal_{\I^*}|=\Lcal$ has orbifold Chern class
\beq
c_1(\Lcal_{\Te^*}) = \sfrac{61}{30} \ .
\eeq
Note that the orbifold Chern class is always of the form
$c_1(\Lcal_\Gamma)=1+ \frac2{|\Gamma_0|}$.

The extended simply laced Dynkin diagram ${\sf E}_8$ for the
affine exceptional $\widehat E_8$ Lie algebra is
\beq
{\scriptsize\xymatrix{
 & & & & & 3 \ar@{-}[d] & & \\
\times \ar@{-}[r] & 2 \ar@{-}[r] & 3 \ar@{-}[r] & 4 \ar@{-}[r] & 5
\ar@{-}[r] & 6 \ar@{-}[r] & 4 \ar@{-}[r] & 2
}}
\eeq
Similarly to the $E_6$ and $E_7$ cases, the finite subgroup
$\I^*\subset\su$ is generated by $\Di^*_5$ and an additional element
of $\su$. However, it is more convenient to notice that it is also
generated by
\beq
\big(\sigma^\alpha{}_\beta\big) = -\begin{pmatrix}
  \zeta_5^{-2}&0\\0&\zeta_5^{2}\end{pmatrix} \und
\big(\widetilde\sigma\,^\alpha{}_\beta\big) =
\frac1{\zeta_5^2-\zeta_5^{-2}}\, \begin{pmatrix}\zeta_5+\zeta_5^{-1}&1\\
  1&-\zeta_5-\zeta_5^{-1} \end{pmatrix} \ .
\eeq
The additional transformations by $\widetilde\sigma$ in (\ref{2.19}) are given by
$$
\pi(\,\widetilde\sigma\,) (X_1)= \sfrac15\, \big(\zeta_5^2-\zeta_5^{-2}\big)\,
\big(\zeta_5-\zeta_5^{-1}\big)\, (X_1-2X_3) \ , \qquad
\pi(\,\widetilde\sigma\,) (X_2)=-
\sfrac{\big(\zeta_5^2-\zeta_5^{-2}\big)^2}{3+\zeta_5^2+\zeta_5^{-2}}\, X_2
$$
\beq 
 \und \pi(\,\widetilde\sigma\,)(X_3)= - \sfrac15\, \big(\zeta_5^2-\zeta_5^{-2}\big)\,
\big(\zeta_5-\zeta_5^{-1}\big)\, (2X_1 + X_3) \ .
\eeq
Note that, in contrast to the $A_k$ and $D_k$ cases, for the
$E$-series the
$\Gamma$-action generally mixes horizontal and vertical
components of one-forms on the $S^1$-bundle (\ref{Seifertfib}). The Poincar\'e three-sphere $S^3/\I^*$ also has a presentation as
the weighted projective curve
\beq
f_{\I^*}(x,y,z) = x^5+y^3+z^2=0 \qquad \mbox{in} \quad \C P^2(6,10,15) \
.
\eeq

\bigskip

\section{Homogeneous vector bundles and quiver representations\label{se:Hvb}}

\noindent {\bf $\mbf\Gamma$-modules. \ }
As previously, denote by $V_\ell\cong\C^{n_\ell}$ for
$\ell=0,1,\dots,r_\Gamma$ the irreducible unitary representations of the ADE orbifold
group $\Gamma$, and introduce the vector space
\beq\label{3.1}
\widehat V = \bigoplus^{r_\Gamma}_{{\ell}=0}\, V_{\ell}
\eeq
which is the multiplicity space for the regular representation of $\Gamma$ of dimension $|\Gamma|$, i.e. the finite-dimensional vector space of functions on $\Gamma$.
The action of a group element $g\in\Gamma$ on $\widehat V$ is
represented by $n_\ell\times n_\ell$ unitary matrices $V_\ell(g)$ on $V_\ell$
for each $\ell=0,1,\dots,r_\Gamma$. Let us fix the vector
space
\beq
\label{3.5} V_R= \bigoplus^{r_\Gamma}_{{\ell}=0}\, R_\ell\otimes V_{\ell}\cong
\C^N \with R_\ell \cong \C^{N_{\ell}} \und N:=\sum^{r_\Gamma}_{{\ell}=0}\, n_\ell\, N_{\ell} \ .
\eeq
Every $\Gamma$-module is of this form: The action of $\Gamma$ on (\ref{3.5}) is defined by a group homomorphism $\gamma:\Gamma\to \urm(N)$ with
\beq\label{3.6}
w^\ell\otimes v_\ell \ \longmapsto \ \gamma(g)(w^\ell\otimes v_\ell)=
w^\ell\otimes \big(V_\ell(g)(v_\ell)\big)
\eeq
for all $g\in\Gamma$, $w^\ell\in R_\ell$ and $v_\ell\in V_\ell$.

\bigskip

\noindent {\bf $\mbf\Gamma$-projection. \ } 
For any irreducible
representation $V_{\ell}$ of $\Gamma$, the fibred product
\beq\label{3.7} \Vcal_{\ell} := \su\times_{\Gamma}V_{\ell}
\eeq
is a homogeneous complex vector bundle of rank $n_\ell$ over the Sasaki-Einstein
orbifold $S^3/\Gamma$. Using
(\ref{3.7}), one can introduce an $\su$-equivariant complex vector
bundle of rank $N$ over
$S^3/\Gamma$ as the Whitney sum
\beq\label{3.8} \bigoplus^{r_\Gamma}_{{\ell}=0}\, R_\ell \otimes
{\Vcal}_{\ell}\with R_\ell\cong \C^{N_{\ell}} \und \sum^{r_\Gamma}_{{\ell}=0}\, n_\ell\, N_{\ell}=N \ .
\eeq
Since $\Gamma\subset\su$, the bundle (\ref{3.8}) is
$\Gamma$-equivariant. The action of $\Gamma$ on the components of any anti-Hermitian
connection $X= X_a\, e^a$
on the bundle (\ref{3.8}) is given by a combination of the action (\ref{2.19}) and the
adjoint action generated by (\ref{3.6}) as
\beq\label{3.9} 
X_1+\im\,X_2 \ \longmapsto \ \g(g)\, \pi(g)(X_1+\im\,X_2)\,
\g(g)^{-1} \und X_3\ \longmapsto \ \g(g)\, \pi(g)(X_3)\, \g(g)^{-1}\ .
\eeq
If we consider a connection on a bundle over the Calabi-Yau cone
(\ref{2.14}) then one
additionally has the $\Gamma$-action
\beq\label{3.10} X_r\ \longmapsto \ \g(g)\, X_r\, \g(g)^{-1}\ . \eeq
The $\Gamma$-action on
sections of the bundle (\ref{3.8}) is given by (\ref{3.6}), while on
$\bigoplus^{r_\Gamma}_{{\ell}=0}\, \urmL(N_{\ell})$-valued sections $A$ of the
corresponding adjoint
bundle it is given by
\beq\label{3.11} A\ \longmapsto \ \g(g)\, A\, \g(g)^{-1}\ . \eeq 
Alternatively, one may
start from a complex vector bundle over $S^3$ of rank $N$ with gauge
group $\uk$ broken to the subgroup
\beq\label{3.12} \Gcal(R)= \prod\limits^{r_\Gamma}_{\ell =0}\, \urm (N_{\ell}) \eeq
commuting with the $\Gamma$-action (\ref{3.6}), after imposing $\Gamma$-symmetry which reduces the initial bundle to the
vector bundle (\ref{3.8})
over $S^3/\Gamma$.

\bigskip

\noindent{\bf $\mbf\Gamma$-equivariance. \ } After
imposing $\Gamma$-symmetry, any homogeneous rank $N$ Hermitian vector
bundle over $S^3$ decomposes into isotopical components as the bundle (\ref{3.8}) with the gauge
group (\ref{3.12}) under the action of the orbifold group
$\Gamma$. The requirement of $\Gamma$-equivariance of an $\su$-equivariant connection
$X=X_a\, e^a$ states that it defines a covariant representation of
$\Gamma$, in the sense that its components satisfy the equations
\beq\label{covrepXa}
\g(g)\, X_a \, \g(g)^{-1}=\pi(g)^{-1}(X_a)
\eeq
for all $g\in\Gamma$, where $\g(g)$ is given in (\ref{3.6}). This condition
decomposes the connection components fibrewise as
\beq\label{Xablocks}
X_a=\bigoplus_{(\ell,\ell'\,)\in\Qsf_1^a}\, (X_a)^{\ell,\ell'} \with
(X_a)^{\ell,\ell'}\in\Hom_\C\big(\C^{n_{\ell'}\, N_{\ell'}}\,,\,
\C^{n_\ell\, N_{\ell}} \big) \ ,
\eeq
where $\Qsf_1^a$ for $a=1,2,3$ is the set of non-zero blocks; given $\ell=0,1,\dots,r_\Gamma$, the corresponding pairs
$(\ell,\ell'\,)\in\Qsf_1^a$ are found by comparing (\ref{3.6}) with
the covariance conditions
\beq
\gamma(g)\left(X_a\big(w^{\ell'}\otimes v_{\ell'} \big)\right) =
\pi(g)^{-1}(X_a)\left(w^{\ell'}\otimes \big(V_\ell(g)(v_{\ell'})\big) \right)
\eeq
for $g\in\Gamma$, $w^{\ell'}\in R_{\ell'}$ and $v_{\ell'}\in V_{\ell'}$.

The requirement of $\Gamma$-equivariance in this way naturally defines
a representation of
a finite quiver $\Qsf=\Qsf_\Gamma= (\Qsf_0,\Qsf_1,\sfs,\sft)$ associated with the Sasaki-Einstein orbifold $S^3/\Gamma$, i.e. an
oriented graph
given by a finite set of vertices $\Qsf_0$, a finite set of arrows $\Qsf_1\subset \Qsf_0\times \Qsf_0$, and two projection maps
$\sfs,\sft: \Qsf_1 \rightrightarrows \Qsf_0$ taking each arrow to its source vertex and its
target vertex respectively; in the present case the vertices are just
the nodes of the affine ADE Dynkin diagram
corresponding to $\Gamma$, while the arrows are determined by the
non-zero blocks of the horizontal and vertical connection components
$X_1+\im\,X_2$ and $X_3$. A (linear) representation of the quiver
$\Qsf$ is a $\Qsf_0$-graded vector space
$R=\bigoplus_{\ell=0}^{r_\Gamma} \, R_\ell$, $R_\ell\cong\C^{N_\ell}$
together with a collection of linear transformations $B= (B_e:R_{\sfs(e)}\to R_{\sft(e)})_{e\in\Qsf_1}$. Given a $\Qsf_0$-graded vector space $R$, the representation space
\beq
\Rep {\rm ep}_\Qsf(R):= \bigoplus_{e\in \Qsf_1}\, \Hom_\C\big(R_{\sfs(e)},R_{\sft(e)}\big)
\label{repspace}\eeq
is the affine variety parameterizing representations of the quiver
$\Qsf$ into $R$. Note that $R$ is the multiplicity space of the
$\Gamma$-module (\ref{3.5}), and hence there is a one-to-one correspondence between representations of the
discrete group $\Gamma\subset\su$ and the representation varieties
(\ref{repspace}). The gauge group (\ref{3.12}) acts naturally on (\ref{repspace}) as
\beq
B_e \ \longmapsto \ g_{\sft(e)}\, B_e\, g_{\sfs(e)}^{-1}
\label{Bgaugetransf}\eeq
where $g_\ell\in {\rm U}(N_\ell)$; note that the diagonal $\uo$ subgroup of scalars in (\ref{3.12}) acts trivially, so we will often factor it out and work instead with the projective gauge group ${\rm P}\Gcal(R):=\Gcal(R)/\uo$.
If the quiver is equiped with a set of relations $\Rsf$, i.e. formal
$\C$-linear combinations of arrow compositions of the quiver, then we denote by $\Rep {\rm ep}_{\Qsf,\Rsf}(R)$ the subvariety of (\ref{repspace}) consisting of representations of $\Qsf$ into $R$ which satisfy the relations $\Rsf$. For background on quivers and their
representations in the context of this paper, see
e.g.~\cite[Sect.~5]{PS2} and~\cite{PS1}. 

\bigskip

\noindent{\bf $\mbf{A_k}$. \ }
In this case $n_\ell=1$ for all $\ell=0,1,\ldots,k \ ({\rm mod}\,(k+1))$ and the generator
$h$ of the cyclic group $\Z_{k+1}$ acts on $V_\ell\cong\C$ as
$V_\ell(h)(v_\ell)=\zeta_{k+1}^\ell\, v_\ell$. On the vector space
(\ref{3.1}) the generating element acts as the diagonal matrix
 \beq\label{3.2} \diag\big(1,\zeta_{k+1},\dots,\zeta_{k+1}^k\big) \ . \eeq
In this case we will also often consider the vector spaces
 \beq\label{3.3}
V_1^{\oplus n}\oplus V_{-n}\cong \C^{n+1} \ \ni \ (z^1,\dots,z^n, z^{n+1}) \eeq
for $n\geq1$, on which the generator $h$ acts as the map
\beq\label{3.4}
(z^1,\dots,z^n,z^{n+1}) \ \longmapsto \ (\zeta_{k+1}\, z^1,
\dots,\zeta_{k+1} \,
z^n,\zeta_{k+1}^{-n}\, z^{n+1}) \eeq 
which defines a homomorphism of the cyclic group $\Z_{k+1}$ into the
Lie group
$\surm(n+1)$.

The covariant representations of $\Z_{k+1}$ are characterized by the equations
\beq\label{3.13}
\g(h)\, (X_1+\im\,X_2)\, \g(h)^{-1}=\zeta_{k+1}^{-2}\, (X_1+\im\,X_2) \und 
\g(h)\, X_3\, \g(h)^{-1}=X_3\ ,
\eeq
and for $k\geq2$ it is easy to see that the non-zero blocks are given fibrewise
by the matrix elements
\bea
(X_1+\im\,X_2)^{{\ell},{\ell}+2}&=:& \vp_{\ell} \ \in \ \Hom_\C
\big(\C^{N_{{\ell}+2}}\,,\, \C^{N_{{\ell}}} \big) \ , \nonumber \\[4pt]
(X_1-\im\,X_2)^{{\ell}+2,{\ell}} &=:&-\vp_{\ell}^\+ \ \in \
\Hom_\C \big(\C^{N_{{\ell}}}\,,\, \C^{N_{{\ell}+2}} \big)\ , \label{3.14} \\[4pt]
(X_3)^{{\ell},{\ell}} &=:& \chi_{\ell} \ \in \ \End_\C
\big(\C^{N_{{\ell}}} \big) \nonumber
\eea
for ${\ell}=0,1,\dots,k$,
where we used the relation
\beq\label{3.15}
X_1-\im\,X_2=-(X_1+\im\,X_2)^\+\ .
\eeq
In the case $k=1$, when $\zeta_2^2=1$ and the $\Z_2$-projection is
given by $S^3 \to \R P^3$, one has only non-vanishing blocks $(X_1\pm\im\,
X_2)^{\ell,\ell},(X_3)^{{\ell},{\ell}}
\in\End_\C(\C^{N_\ell})$ for $\ell=0,1$. Analysis of the
explicit form of the matrices (\ref{3.14}) and of the corresponding quivers
shows that the general cases of even and odd rank $k$ should be treated
separately.

\begin{itemize}
\item[{\tt a)}] \ \underline{$k=2q,\ S^3/\Z_{2q+1}$:}
\end{itemize}
Using the property $\zeta_{k+1}^{2q+1}=1$, one can show that the matrix
\beq\label{3.16}
\diag\big(1,\zeta_{k+1}^2, \dots,\zeta_{k+1}^{2k}\big)=\diag\big(1,\zeta_{k+1}^2,
\dots
,\zeta_{k+1}^{2q},
\zeta_{k+1},\zeta_{k+1}^3, \dots, \zeta_{k+1}^{2q-1}\big)
\eeq
is equivalent to the matrix (\ref{3.2}) with permuted diagonal elements. Then by using the matrix
\beq\label{3.17}
\g(h)=\diag\big(\one_{\C^{N_0}}\otimes 1 ,
\one_{\C^{N_1}}\otimes \zeta^2_{k+1} ,\dots,
 \one_{\C^{N_q}}\otimes \zeta_{k+1}^{2q} ,
\one_{\C^{N_{q+1}}}\otimes \zeta_{k+1} , \one_{\C^{N_{q+2}}} \otimes \zeta_{k+1}^3,
\dots, \one_{\C^{N_{2q}}}\otimes \zeta_{k+1}^{2q-1}
\big)
\eeq
in (\ref{3.13}) we obtain the solution
\bea
(X_1+\im\,X_2)^{{\ell},{\ell}+1}&=:& \phi_{\ell +1}\ \in \ \Hom_\C(\C^{N_{\ell +1}}, \C^{N_{\ell}})
 \und (X_1+\im\,X_2)^{k,0} \ =: \ \phi_{k+1} \ , \label{3.18}
\\[4pt]
(X_1-\im\,X_2)^{{\ell}+1,{\ell}} &=:& -\phi_{\ell +1}^{\+} \
\in \ \Hom_\C (\C^{N_{{\ell}}}, \C^{N_{{\ell}+1}})\und
(X_1-\im\,X_2)^{0,k} \ =: \ -\phi_{k+1}^\+\ \nonumber
\eea
with $\ell =0,1,\dots,2q-1$, and
\beq
(X_3)^{{\ell},{\ell}}=:\vr_{\ell} \ \in \ \End_\C(\C^{N_{{\ell}}})
\label{3.18a}\eeq
with $\ell=0,1,\dots,2q$, where $\vr^{\+}_{\ell}=-\vr_{\ell}$.
Note that in these equations we use the same symbol $N_{\ell}$ as in
(\ref{3.14}), but they are in fact related by permutation, as are
$\chi_{\ell}$ and $\vr_{\ell}$, and $\vp_{\ell}$ and $\phi_{\ell}$.
Finally we obtain the irreducible affine $\widehat A_{2q}$-type
quivers $\Qsf_{\widehat{A}_{2q}}$ given by

\medskip

\begin{equation}
 \label{3.19}
 \xymatrix@C=20mm{
& & {\bullet}\ar[dll]\ar@{.>}@(ur,ul)[] & & \\
{\bullet}\ar@{.>}@(ul,dl)[]\ar[r] & {\bullet}\ar@{.>}@(dl,dr)[] \ar[r] & \ \cdots
\ \ar[r] & {\bullet}\ar@{.>}@(dl,dr)[]\ar[r] & {\bullet}\ar@{.>}@(dr,ur)[]\ar[ull]
}
\end{equation}

\bigskip
\noindent
with $2q+1$ vertices, arrows and loop edges. For clarity, throughout we
designate arrows associated to the horizontal components
$X_1+\im\,X_2$ by solid lines and arrows associated to the vertical
components $X_3$ with dashed lines; in particular, here
the loop edges are
associated with the bundle endomorphisms $\vr_{\ell}$. The underlying
graph of this quiver is the extended affine Dynkin diagram
${\sf A}_{2q}$ from (\ref{AkDynkin}).

\begin{itemize}
\item[{\tt b)}] \ \underline{$k=2q+1,\ S^3/\Z_{2q+2}$:}
\end{itemize}
In this case one finds that the $\Z_{2q+2}$-equivariant vector bundle over $S^3/\Z_{2q+2}$ is a direct sum of two irreducible bundles
and the associated quiver splits into two connected quivers
of the type (\ref{3.19}).
Arguing in a similar way as above, we now have
\beq\label{3.20}
\g(h)=\diag\big(\one_{\C^{N_0}}\otimes 1 ,
\one_{\C^{N_1}}\otimes\zeta_{k+1}^2,\dots,
\one_{\C^{N_{q}}}\otimes\zeta_{k+1}^{2q},
\one_{\C^{N_{q+1}}}\otimes 1,\one_{\C^{N_{q+2}}}\otimes\zeta_{k+1}^2,\dots,
\one_{\C^{N_{2q+1}}}\otimes\zeta_{k+1}^{2q} \big)
\eeq
with $\zeta_{k+1}^{2\ell}=\zeta_{q+1}^\ell$, and the corresponding
reduced quivers are given by
\bea
\xymatrix{
{\bullet}\ar@{.>}@(ul,dl)[] \ar@(dr,ur)[]
} 
\qquad \qquad 
\xymatrix@C=20mm{
{\bullet}\ar@{.>}@(ul,dl)[] \ar@(dr,ur)[]
} \qquad &\for& q=0 \ , \\[15pt]
 \label{3.21}
 \xymatrix{
{\bullet}\ar@/^/[rr]{}\ar@{.>}@(ul,dl)[] &
&{\bullet}\ar@/^/[ll]{} \ar@{.>}@(dr,ur)[] 
}
\qquad\qquad
\xymatrix{
{\bullet}\ar@/^/[rr]{}\ar@{.>}@(ul,dl)[] & &{\bullet}\ar@/^/[ll]{}\ar@{.>}@(dr,ur)[]
} \qquad &\for& q=1 \ , \\[15pt]
\label{3.22}
{\sf Q}_{\widehat{A}_q} \ \sqcup \ {\sf Q}_{\widehat{A}_q} \qquad &\for& q\geq 2 \ ,
\eea
with the quiver (\ref{3.22}) having $(q+1)+(q+1)$ vertices, arrows and loop edges.

\bigskip

\noindent{\bf $\mbf{D_k}$. \ }
The dicyclic group $\Di^*_{k-2}$ has $k-1$ two-dimensional
representations $W_j\cong\C^2$ on which the generators $\xi_2$ and
$\xi_3$ act as the matrices
\beq
W_j(\xi_2)=\begin{pmatrix} 0&1\\(-1)^j &0 \end{pmatrix} \und 
W_j(\xi_3)=\begin{pmatrix} \zeta_{2(k-2)}^j & 0\\
  0&\zeta_{2(k-2)}^{-j} \end{pmatrix}
\eeq
for $j=0,1,\dots,k-2$; in particular, $W_1$ is the fundamental
representation (\ref{Difundrep}). For $\ell=1,\dots,k-3$ the representations $V_\ell:=W_\ell$ are
irreducible, while $W_0=V_0\oplus V_{k}$ and
$W_{k-2}= V_{k-2}\oplus V_{k-1}$ simultaneously
diagonalize into two eigenlines, with $V_0$ the trivial representation and
\beq
V_k(\xi_2)=V_{k-2}(\xi_3)= V_{k-1}(\xi_3)= -1 = -V_k(\xi_3) \und
V_{k-2}(\xi_2)=-\im^{k} = -V_{k-1}(\xi_2) \ .
\eeq

The covariant representations of $\Di^*_{k-2}$ are characterized by
the equations
\beq
\gamma(\xi_2)\, (X_1+\im\, X_2)\, \gamma(\xi_2)^{-1}= (X_1+\im\,X_2)^\dag
\und \gamma(\xi_2)\, X_3\, \gamma(\xi_2)^{-1}=-X_3
\label{gammac2}\eeq
together with
\beq
\gamma(\xi_3)\, (X_1+\im\,X_2)\, \gamma(\xi_3)^{-1}= \zeta_{2(k-2)}^{-2}\,
(X_1+\im\, X_2) \und \gamma(\xi_3)\, X_3\, \gamma(\xi_3)^{-1}=X_3 \ ,
\label{gammac3}\eeq
where we have used the relation (\ref{3.15}). By working in the
canonical basis of $W_j\cong\C^2$, from these equations it is
straightforward to see that all representation spaces $W_j$ for
$j=0,1,\dots,k-2$ are $X_3$-invariant, while under the horizontal
connection components they transform as
\beq
(X_1+\im\, X_2)(W_0) \ \subset \ W_2 \und (X_1+\im\, X_2)(W_1)\ \subset\ W_3 \ ,
\eeq
and
\beq
(X_1+\im\, X_2)(W_{k-3})\ \subset \ W_{k-5} \und (X_1+\im\,
X_2)(W_{k-2})\ \subset\ W_{k-4} \ ,
\eeq
together with
\beq
(X_1+\im\, X_2)(W_j)\ \subset \ W_{j-2}\oplus W_{j+2} \for j=2,\dots,k-4 \
.
\eeq
After diagonalising the reducible representations $W_0$ and $W_{k-2}$
into their simultaneous eigenlines, some straightforward linear
algebra shows that $(X_1+\im\, X_2)(V_\ell)\subset V_2$ for $\ell=0,k$
and $(X_1+\im\, X_2)(V_\ell)\subset V_{k-4}$ for $\ell=k-2,k-1$, while
$X_3(V_0)\subset V_k$, $X_3(V_k)\subset V_0$ and $X_3(V_{k-2}) \subset
V_{k-1}$, $X_3(V_{k-1})\subset V_{k-2}$.

Following our treatment of the $A_{2q+1}$ family above, we use the
block diagonal matrices
\beq
\gamma(\xi_j)= \diag\big(\one_{\C^{N_0}}\oplus\one_{\C^{N_k}},\one_{\C^{N_1}}\otimes
W_2(\xi_j),\ldots, \one_{\C^{N_{k-3}}}\otimes W_{2(k-3)}(\xi_j),
\one_{\C^{N_{k-2}}}\oplus \one_{\C^{N_{k-1}}}\big)
\eeq
in (\ref{gammac2}) and (\ref{gammac3}) to obtain the solution
\bea
(X_1+\im\, X_2)^{\ell,\ell-1}&=:& \phi_{\ell-1}^+ \ \in \
\Hom_\C\big(\C^{2N_{\ell-1}}\,,\, \C^{2N_\ell}\big) \ , \nonumber \\[4pt]
(X_1-\im\,X_2)^{\ell-1,\ell}&=:& - \phi_{\ell-1}^+\,^\dag\ \in \ \Hom_\C\big(\C^{2N_\ell}\,,\,
\C^{2N_{\ell-1}}\big) \ , \nonumber \\[4pt]
(X_1+\im\,X_2)^{\ell,\ell+1} &=:& \phi_{\ell+1}^- \ \in \
\Hom_\C\big(\C^{2N_{\ell+1}}\,, \, \C^{2N_\ell}\big) \ , \nonumber
\\[4pt]
(X_1-\im\,X_2)^{\ell+1,\ell} &=:& - \phi_{\ell+1}^-\,^\dag \
\in \ \Hom_\C\big(\C^{2N_\ell}\,,\, \C^{2N_{\ell+1}}\big)
\eea
for $\ell=1,\dots,k-4$, together with
\bea
(X_1+\im\,X_2)^{\ell',\ell}&=:& \varphi_\ell \ \in \
\Hom_\C\big(\C^{N_\ell}\,,\, \C^{2N_{\ell'}}\big) \ , \nonumber \\[4pt]
(X_1-\im\, X_2)^{\ell,\ell'}&=:& - \varphi_\ell^\dag \ \in \
\Hom_\C\big( \C^{2N_{\ell'}}\,,\, \C^{N_\ell}\big) \ , 
\eea
for $\ell'=1$ (resp. $\ell'=k-3$) and $\ell=0,k$ (resp. $\ell=k-2,k-1$), while
for the vertical components we find
\bea
(X_3)^{\ell,\ell}&=:& \varrho_\ell\=-
\varrho_\ell^\dag \ \in \ \End_\C\big(\C^{2N_\ell}\big) \ ,
\nonumber\\[4pt]
(X_3)^{\ell',\ell^{\prime\prime}}&=:& \chi_{\ell'} \ \in \
\Hom_\C\big(\C^{N_{\ell^{\prime\prime}}}\,,\, \C^{N_{\ell'}}\big) \ ,
\\[4pt]
(X_3)^{\ell^{\prime\prime},\ell'} &=:& -\chi_{\ell'}^\dag \ \in
\ \Hom_\C\big(\C^{N_{\ell^{\prime}}}\,, \, \C^{N_{\ell^{\prime\prime}}}\big) \nonumber
\eea
for $\ell=1,\dots,k-3$, and $\ell'=k$
(resp. $\ell^{\prime\prime}=0$) and $\ell'=k-1$
(resp. $\ell^{\prime\prime}=k-2$). 
We thereby arrive at the quiver
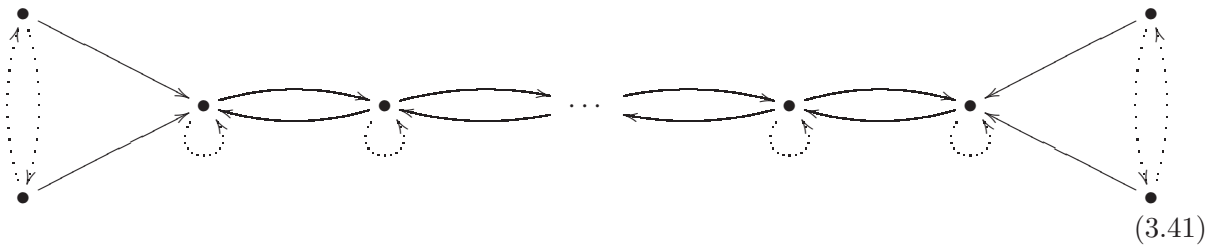
\beq
\xymatrix@C=20mm{
{\bullet}\ar@{.>}@/^/[dd]{} \ar[dr]& & & & & & \bullet
\ar@{.>}@/_/[dd]{} \ar[dl] \\
 & \bullet \ar@{.>}@(dl,dr)[] \ar@/^/[r]{}
 &{\bullet}\ar@/^/[l]{}\ar@{.>}@(dl,dr)[]\ar@/^/[r]{} & \ \cdots \
 \ar@/^/[r]{} \ar@/^/[l]{} & {\bullet}\ar@/^/[l]{} \ar@/^/[r]{}
 \ar@{.>}@(dl,dr)[] & {\bullet}\ar@/^/[l]{}
 \ar@{.>}@(dl,dr)[] & \\
\bullet \ar@{.>}@/^/[uu]{}\ar[ur] & & & & & & \bullet \ar@{.>}@/_/[uu]{}\ar[ul]
}
\eeq
with $k+1$ vertices, $2k$ arrows and $k-3$ loop edges; its
underlying graph is the affine Dynkin diagram ${\sf D}_k$ from
(\ref{DynkinDk}). Note that the horizontal segment of this quiver
consists of a chain of $k-4$ connected $\widehat{A}_1$-type quivers from~(\ref{3.21}).

\bigskip

\noindent{\bf $\mbf{E_k}$. \ }
The constructions above can in principle be extended to the
exceptional series. For example, the binary tetrahedral group $\Te^*$
has seven irreducible representations consisting of three
one-dimensional representations given by the quotient $\Te^*\to\Z_3$,
three two-dimensional representations obtained by taking tensor
products of these one-dimensional representations with the fundamental
representation of $\Te^*\subset\su$, and one three-dimensional
representation given by the quotient $\Te^*\to\Te\subset{\rm SO}(3)$;
we leave it to the interested reader to work out the details of the
corresponding ${\sf E}_6$ quiver diagram. For the extended Dynkin
diagrams ${\sf E}_7$ and ${\sf E}_8$ the representation theory becomes
somewhat more complicated.

\bigskip

\section{Equivariant dimensional reduction and quiver bundles\label{sec:dimred}}

\noindent
{\bf Equivariant vector bundles. \ }
In this section we consider the dimensional reduction of invariant connections on equivariant
vector bundles over product manifolds. Let $\Ecal$ be an $\su$-equivariant Hermitian
vector bundle of rank $N$ over $M\times S^3$,
where $M$ is a smooth, closed and oriented manifold of real dimension $D$; the group $\su$ acts trivially on $M$ and by isometries on
$S^3\cong\su$. The sphere $S^3$ can be regarded as a coset space $G/H$ with $G=\su$
and the trivial stabilizer subgroup $H=\{1\}$; for the relevant background on equivariant dimensional reduction over coset spaces, see e.g.~\cite{AlGar12,Lechtenfeld:2007st,Dolan:2010ur}.

We will use the same symbol $\Ecal$ for the vector bundle
\beq \label{4.2}
\Ecal \ \longrightarrow \ M\times S^3/\Gamma
\eeq
obtained by projection to the
orbifold $M\times S^3/\Gamma$. By standard induction and reduction, there is an equivalence between
$\su$-equivariant vector bundles (\ref{4.2}) and 
$\Gamma$-equivariant vector bundles over $M$ which are
described by the quivers $\Qsf_\Gamma$ from Sect.~\ref{se:Hvb}; the
finite orbifold
group $\Gamma\subset\su$ also acts trivially on $M$. A representation
of the quiver $\Qsf_\Gamma$ in the category $\Bun{\rm un}(M)$ of
complex vector bundles on $M$ is called a quiver bundle on $M$.

Every $\su$-equivariant complex vector bundle (\ref{4.2}) can be
decomposed uniquely up to isomorphism into isotopical components as a Whitney sum 
\beq
\label{4.3} \Ecal=\bigoplus^{r_\Gamma}_{{\ell}=0} \, E_{\ell}\otimes \Vcal_{\ell}\ ,
\eeq 
where $E_{\ell}\to M$ are Hermitian vector bundles of rank $N_{\ell}$ with
$\sum^{r_\Gamma}_{{\ell}=0}\, n_\ell\, N_{\ell} =N$ and trivial
$\Gamma$-action, and the homogeneous bundles
$\Vcal_{\ell}\to S^3/\Gamma$ are defined in
(\ref{3.7}). As we
showed in Sect.~\ref{se:Hvb}, the gauge group $\Gcal(R)$ of the bundle
(\ref{4.3}) is given by (\ref{3.12}).

\bigskip

\noindent
{\bf $\mbf{\Gamma}$-equivariant connections. \ }
Let $\Acal$ be an $\su$-equivariant gauge connection on $\Ecal$ and $\Fcal = \diff\Acal + \Acal\wedge\Acal$ its curvature,
both with values in the Lie algebra $\urmL(N)$. It has the form
\beq \label{4.1}
\Acal = A+X=A_{\mh}\, e^{\mh} + X_a\, e^a\ ,
\eeq
where $e^{\mh}$ and $e^a$ are basis one-forms on $M$ and $S^3$, respectively, and $A_{\mh}$ and $X_a$ are $\urmL(N)$-valued matrices which
depend only on the coordinates of $M$ with $\mh , \nh ,\ldots=1,\dots,
D$. Since $S^3$ is a group manifold, there are no further restrictions on $A_{\mh}$ and $X_a$ coming from
$\su$-invariance. 

We shall also use the same symbol $\Acal$ for the connection obtained by
projecting (\ref{4.1}) to the orbifold $M\times S^3/\Gamma$. The gauge
potential projection from $M\times S^3$ to $M\times S^3/\Gamma$ is
defined by the
equations
\beq \label{4.5} \g(g)\, A_{\mh}\, \g(g)^{-1} = A_{\mh} \eeq 
for all $g\in\Gamma$, together with the equations (\ref{covrepXa})
for $X_a$ which are resolved by the matrices (\ref{Xablocks}).

The calculation of the curvature $\Fcal =\diff\Acal + \Acal\wedge\Acal$
for $\Acal$ of the form (\ref{4.1}) yields
\beq \label{4.9}
\Fcal = F + \big(\diff X_a + [A, X_a]\big)\wedge e^a +\sfrac12\,
\big([X_a,X_b]-2\, \ve^c_{ab}\, X_c\big)\, e^a\wedge e^b\ ,
\eeq
where $F=\diff A + A\wedge A$ is the curvature of the gauge potential
$A$ on $M$ with gauge group $\Gcal(R)$. Given local real coordinates $x^{\mh}$ on $M$ one can choose $\diff x^{\mh}$ as basis one-forms
$e^{\mh}$ on $M$. Then from (\ref{4.9}) we find the non-vanishing
components of the field strength tensor
\bea \label{4.10}
\Fcal_{\mh\nh}&=& \pa_{\mh}A_{\nh} - \pa_{\nh}A_{\mh} +[A_{\mh}, A_{\nh}]\ ,
\\[4pt]
\label{4.11}
\Fcal_{\mh a}&=:& D_{\mh}X_a \= \pa_{\mh}X_a +[A_{\mh}, X_a]\ ,
\\[4pt]
\label{4.12}
\Fcal_{ab}&=&[X_{a}, X_b] - 2\, \ve^c_{ab}\, X_c\ .
\eea

\bigskip

\noindent
{\bf Quiver gauge theory. \ } 
The dimensional reduction of the Yang-Mills equations on $M\times
S^3/\Gamma$ can be seen at the level of the Yang-Mills Lagrangian;
reduction of the Yang-Mills action functional defines a quiver gauge
theory on $M$ associated to the quiver $\Qsf_\Gamma$. Let
$\diff\,\vol_M$ be the Riemannian volume form with
respect to an arbitrarily chosen metric on the manifold $M$, and let
$\diff\,\vol_{S^3/\Gamma}$ denote the Riemannian volume form
associated to the metric (\ref{SEmetric}) on the Sasaki-Einstein
manifold $S^3/\Gamma$; the
corresponding Hodge duality operator for the product metric on
$M\times S^3/\Gamma$ is denoted $\star\,$. With $\tr_N$ denoting the trace
in the fundamental representation of the $\urm(N)$ gauge group, by substituting in (\ref{4.10})--(\ref{4.12}) and integrating over $S^3/\Gamma$ we arrive at the action
\bea\nonumber
S_{\sf{YM}}&:=&-\frac{1}{4}\, \int_{M\times S^3/\Gamma}\,
\tr_N\,\Fcal\wedge\star\, \Fcal\\[4pt]\nonumber 
&=&-\frac{1}{8}\, \int_{M\times S^3/\Gamma}\, \diff\,\vol_M\wedge
\diff\, \vol_{S^3/\Gamma}\ \tr_N \left(\Fcal_{\mh\nh}\, \Fcal^{\mh\nh}+
2\,\Fcal_{\mh a}\, \Fcal^{\mh a} + \Fcal_{ab}\, \Fcal^{ab}\right )\\[4pt]
&=&-\frac{\pi\, r^3}{6\, d}\, c_1(\Lcal_\Gamma)\, \int_{M}\, \diff\,
\vol_M\ \tr_{N} \Big(F_{\mh\nh}\, F^{\mh\nh}+
\frac2{r^2} \, D_{\mh}X_a \, D^{\mh}X_a \label{SYMreduced} \\ && \nonumber \qquad 
\qquad \qquad \qquad \qquad \qquad
\qquad +\,\frac1{r^4}\,\sum_{a,b=1}^3\,
\big([X_a,X_b]-2\, \ve^c_{ab}\, X_c\big)^2 \Big)
\ .
\eea

In the sector of this field theory with $A_{\mh}=0$ and locally translationally-invariant scalar fields $X_a$, the global
minima of the action are described by the matrix equations
\beq
\Fcal_{ab}=[X_{a}, X_b] - 2\, \ve^c_{ab}\, X_c = 0 \ .
\label{flatS3gamma}\eeq
In general these equations will contain both holomorphic F-term
constraints on the scalar fields, which define the set of relations $\Rsf_\Gamma$
among the arrows
for the quiver $\Qsf_\Gamma$, and also non-holomorphic D-term
constraints, which yield stability conditions for the corresponding
moduli variety of quiver representations. Hence solutions of the BPS
equations are determined by stable representations of the quiver with
relations $(\Qsf_\Gamma,\Rsf_\Gamma)$. The corresponding stable quotient
$\Rep {\rm ep}_{\Qsf_\Gamma,\Rsf_\Gamma}(R) \, \big/\!\!\big/ \ {\rm P}\Gcal(R)$ is a finite set
whose points are in one-to-one correspondence with representations
of the Lie algebra $\suL$ in $\Rep {\rm ep}_{\Qsf_\Gamma}(R)\subset\urmL(N)$.

\bigskip

\noindent
{\bf $\mbf{A_k}$. \ }
In this case from (\ref{3.17}), (\ref{3.20})
and (\ref{4.5}) we obtain a block diagonal form similar to $X_3$ for the gauge
potential $A=A_{\mh} \, e^{\mh}$ given by
\beq \label{4.6}
A=\diag\big(
   A^0,A^1,\dots,A^k \, \big) \ .
\eeq
After projection from $M\times S^3$ to $M\times S^3/\Z_{k+1}$ the scalar
field $X=X_a\, e^a$
from (\ref{4.1}) has horizontal components
\beq \label{4.7}
X_1+\im \, X_2=\begin{pmatrix}
   0&\phi_1&0&\cdots&0\\0&0&\phi_2&\ddots&\vdots\\
   \vdots&\vdots&\ddots&\ddots&0\\0&0& \cdots &0&\phi_k \\
   \phi_{k+1}&0& \cdots &0&0 \end{pmatrix} \und X_1-\im \, X_2=-(X_1+\im\, X_2)^{\+}\ ,
\eeq
and vertical component
\beq \label{4.8}
X_3= \diag (\vr_0, \vr_1,\ldots,\vr_k)
\eeq
for $k=2q$. We will not consider here the case $k=2q+1$ given by the
formulas (\ref{3.20})--(\ref{3.22}), as it can be simply reduced to a
doubling of the matrices (\ref{4.7})--(\ref{4.8}) with $k$ substituted by $q$.

From (\ref{4.6})--(\ref{4.8}) one sees that the $\urmL(N)$-valued
gauge potential $\Acal$ splits into $N_\ell \times N_{\ell'}$
blocks~$\Acal^{\ell\ell'}$ as
\begin{equation}\label{4.13}
\Acal=\big(\Acal^{\ell\ell'}\, \big) \qquad\mbox{with}\quad
\Acal^{\ell\ell'}~\in~\mbox{Hom}_\C\bigl(\C^{N_{\ell'}}\,,\,
\C^{N_\ell}\bigr) \ ,
\end{equation}
where the indices $\ell,\ell',\ldots$ run over $0,1,\ldots,k$,
 and
\bea\label{4.14}
\Acal^{\ell\ell}&=&A^\ell\otimes1 + \vr_\ell\otimes e^3 \ , \\[4pt]
\label{4.15}
\Acal^{\ell\,\ell+1}&=:&\Phi^{~}_{\ell+1} \=
\sfrac{1}{2}\,\phi^{~}_{\ell+1}\otimes (e^1-\im\, e^2\, ) \ ,\qquad
\Acal^{k0} \ =: \ \Phi^{~}_{k+1}\=
\sfrac{1}{2}\,\phi^{~}_{k+1}\otimes (e^1-\im\, e^2\, ) \ ,\\[4pt]
\Acal^{\ell+1\,\ell}&=& -\Phi^{\+}_{\ell+1}\=-\sfrac{1}{2}\,\phi_{\ell+1}^\+
\otimes (e^1+\im \, e^2\, ) \ , \qquad \Acal^{0k}\=-\Phi^{\+}_{k+1} \= -
\sfrac{1}{2}\,\phi^{\+}_{k+1}\otimes (e^1+\im\, e^2\, ) \ , \nonumber \\ &&
\label{4.16}\eea
with all other components vanishing.

For the curvature
\begin{equation}\label{4.17}
\Fcal=\big(\Fcal^{\ell\ell'}\, \big) \with
\Fcal^{\ell\ell'} = \diff\Acal^{\ell\ell'} +
\sum_{\ell^{\prime\prime}=0}^{k}\, \Acal^{\ell\ell^{\prime\prime}}\wedge \Acal^{\ell^{\prime\prime}\ell}
\ ,
\end{equation}
we obtain the
non-vanishing field strength components
\bea\label{4.18}
\Fcal^{\ell\ell}&=&F^\ell - \sfrac{1}{4}\,\bigl(2\, \im\,\vr_\ell
+\phi_\ell^\+\,\phi^{~}_\ell - 
\phi^{~}_{\ell+1}\,\phi^\+_{\ell+1}\bigr)\,
\beta\wedge\bar\beta + \diff\vr_\ell\wedge e^3\ , \\[4pt]
\label{4.19}
\Fcal^{\ell\,\ell+1}&=&\sfrac{1}{2}\,D \phi_{\ell+1}\wedge\bar\beta
+\sfrac{1}{2}\,\bigl(\im\, \phi_{\ell+1} +\vr_\ell\, \phi_{\ell+1} -
\phi^{~}_{\ell+1}\,\vr_{\ell+1}\bigr)\, e^3\wedge\bar\b\ , \\[4pt]
\label{4.20}
\Fcal^{\ell+1\,\ell}&=&- \big( \Fcal^{\ell\,\ell+1}\big)^\+\=- \sfrac{1}{2}\,
\bigl(D \phi_{\ell+1}\bigr)^\+ \wedge \beta
-\sfrac{1}{2}\,\bigl(\im\, \phi_{\ell+1}^\+ +\phi_{\ell+1}^\+\, \vr_\ell{-}
\vr^{~}_{\ell+1}\,\phi_{\ell+1}^\+\bigr)\, e^3\wedge\b
\ , \\[4pt]
\label{4.21}
\Fcal^{k\,0}&=&\sfrac{1}{2}\,D \phi_{k+1}\wedge\bar\beta
+\sfrac{1}{2}\,\bigl(\im\, \phi_{k+1} +\vr_k\, \phi_{k+1} -
\phi^{~}_{k+1}\,\vr_{0}\bigr)\, e^3\wedge\bar\b\ , \\[4pt]
\label{4.22}
\Fcal^{0\,k}&=& -\big(\Fcal^{k\,0}\big)^\+\=-\sfrac{1}{2}\,\big(D \phi_{k+1}\big)^\+\wedge\beta
-\sfrac{1}{2}\,\bigl(\im\, \phi_{k+1}^\+ + \phi_{k+1}^\+ \, \vr_k{-}
\vr_{0}\, \phi_{k+1}^\+\bigr)\, e^3\wedge\b
\ .
\eea
Here we defined $F^\ell:=\diff A^\ell + A^\ell \wedge A^\ell =\sfrac12\,
F^\ell_{\mh\nh}(x)\, \diff x^{\mh}\wedge\diff x^{\nh}$
and introduced the bifundamental covariant derivatives
\beq\label{4.23}
D\phi_{\ell+1}:=\diff\phi_{\ell+1}+A^\ell\, \phi_{\ell+1} - \phi_{\ell+1}\, A^{\ell+1}\ ,
\eeq
with $A^{k+1}:=A^0$, $\vr_{k+1}:=\vr_0$ and $\ell=0,1,\ldots,k$.

The quiver gauge theory action functional
(\ref{SYMreduced}) is obtained by substituting in
(\ref{4.17})--(\ref{4.22}) to get
\bea\nonumber
S_{\sf{YM}}&=& \frac{\pi \,r^3}{6\, (k+1)}\, \int_{M}\, \diff\,\vol_M\
\sum\limits_{\ell=0}^k\, \tr_{N_\ell} \Big( F_{\mh\nh}^\ell\,^\+\, F^{\ell\,
  \mh\nh} +
\frac{1}{r^2}\, \big(D_{\mh}\phi_{\ell+1} \big) \,
\big(D^{\mh}\phi_{\ell+1}\big)^\+ \\ && \qquad \qquad \qquad \qquad +\, \frac{1}{r^2}\, \big(D_{\mh}\phi_{\ell}\big)^\+\,
\big(D^{\mh}\phi_{\ell} \big) +
\frac{1}{2r^4}\, \big(2\,\im\, \vr_\ell + \phi_\ell^\+\,
\phi^{~}_\ell-\phi^{~}_{\ell+1}\, \phi_{\ell+1}^\+ \big)^2 \nonumber \\ &&
\qquad \qquad \qquad \qquad +\,
\frac{1}{r^4}\, \big(\im\, \phi_{\ell+1} + \vr_\ell\, \phi_{\ell+1}-\phi_{\ell+1}\,
\vr_{\ell+1} \big) \, \big(\im\, \phi_{\ell+1} + \vr_\ell\, \phi_{\ell+1}-\phi_{\ell+1}\,
\vr_{\ell+1} \big)^\dag  \nonumber \\ && \qquad \qquad \qquad \qquad +\, \frac{1}{r^4}\, \big(\im\, \phi_{\ell} + \vr_{\ell-1}\, \phi_{\ell}-\phi_{\ell}\,
\vr_{\ell} \big)^\dag \, \big(\im\, \phi_{\ell} + \vr_{\ell-1}\, \phi_{\ell}-\phi_{\ell}\,
\vr_{\ell} \big) \Big)
\ .
\label{4.24}\eea
The
corresponding F-term relations are
\beq
\phi_{\ell+1}\,
\vr_{\ell+1}=\vr_\ell\, \phi_{\ell+1} + \im\, \phi_{\ell+1}
\label{Ftermrels}\eeq
for $\ell=0,1,\dots,k$, which give the relations ${\sf R}_k$ for the
quiver ${\sf Q}_{\widehat{A}_k}$. The D-term constraints are given by
\beq
\phi^{~}_{\ell+1}\, \phi_{\ell+1}^\+ - \phi_\ell^\+\,
\phi^{~}_\ell = 2\,\im\, \vr_\ell
\label{Dtermrels}\eeq
for $\ell=0,1,\dots,k$.

\bigskip

\noindent
{\bf Reduction to $\mbf{A_{k+1}}$ quiver gauge theory. \ }
We will now compare the $\widehat{A}_{k}$-type quiver gauge theory for $k=2q$ with that
based on the $A_{k+1}$
quiver which arises from $\su$-equivariant dimensional reduction over $\C
P^1$ \cite{GP,PS1}. For this, we notice that for $\vp =0$ in (\ref{2.7}) the field $B$ becomes an $\suL$-valued
one-form on $\C P^1\hra S^3/\Z_{k+1}$ with $e^3=-\im\, a_1$, i.e. fixing $\vp =0$ reduces our geometry to the base $\C P^1$
of the fibration (\ref{2.4}). The same effect can also be achieved by taking the limit $k\to\infty$ as was discussed
in e.g.~\cite{LM}; however, here we will keep $k$ finite since we want
to compare the $\widehat{A}_{k}$-type
and $A_{k+1}$ quiver gauge theories.

Let us now describe the dynamical transition from the cyclic
$\widehat{A}_{k}$-type quiver (\ref{3.19}) to the linear $A_{k+1}$
quiver
\medskip
\begin{equation}
 \label{Akquiver}
 \xymatrix@C=20mm{
{\bullet} \ar[r] & {\bullet} \ar[r] & \ \cdots
\ \ar[r] & {\bullet} \ar[r] & {\bullet} 
}
\end{equation}
which arises by restricting an irreducible representation of $\su$ on $\C^{k+1}$ to the subgroup $\uo\subset\su$~\cite{PS1}. In terms of the matrices (\ref{3.18}) and (\ref{3.18a}) it can be realized by putting
\beq
\label{5.1}
\p_{k+1}=0 \ ,
\eeq
and fixing
\beq
\label{5.2}
\vr_\ell=-\mbox{$\frac{\im}{2}$}\, (k-2\ell )\ \one_{\C^{N_\ell}} \for \ell=0,1,\dots,k\ .
\eeq
For this choice the relations (\ref{Ftermrels})
are automatically satisfied, 
and the expressions (\ref{4.18})--(\ref{4.24}) reduce
to the expressions for the $A_{k+1}$ quiver gauge theory derived in~\cite{PS1}.

Note that the case $k=2q+1$ reduces in this limit to an $A_{q+1}\sqcup
A_{q+1}$ quiver gauge theory, which can be obtained within the framework
of~\cite{PS1} by restricting instead a
{\it reducible} representation of $\su$ on $\C^{q+1}\oplus\C^{q+1}$.

\bigskip

\noindent
{\bf Reductions to $\mbf{D_{k}}$ and $\mbf{E_k}$ quiver gauge theories. \ }
For a general ADE group quotient $\Gamma\to\Gamma_0$, setting $\varphi=0$ in (\ref{eaS3Gamma}) again reduces the geometry to the base $\C P^1/\Gamma_0$ of the Seifert fibration (\ref{Seifertfib}). Similar reductions as in the $\widehat{A}_k$ case above then \emph{define} a quiver gauge theory on $M$ based on the ordinary (unextended) Dynkin diagram of ADE type, which arises from a putative equivalence between $\su$-equivariant vector bundles on $M\times \C P^1/\Gamma_0$ and $\Gamma_0\times\uo$-equivariant vector bundles on $M$, where $\su$ acts trivially on the manifold $M$; this V-bundle equivalence generalizes the equivalences of~\cite{GP,PS1} to the equivariant dimensional reduction over two-dimensional orbifolds $\C P^1/\Gamma_0$.

\bigskip

\section{Instantons on orbifolds and quiver varieties\label{sec:instantons}}

\noindent {\bf McKay quivers. \ }
We begin by describing a class of quivers of paramount importance to
the study of instantons on the Calabi-Yau cones $C(S^3/\Gamma)$ and
their relevance to the McKay correspondence; see
e.g.~\cite{Reid,Cirafici:2012qc} for further details. 
Given the representation $V_R$ of the orbifold group $\Gamma$ from (\ref{3.5}), consider the decomposition
\beq
V_R\otimes V_\ell=\bigoplus_{\ell'=0}^{r_\Gamma} \, a_{\ell\ell'}\ V_{\ell'}
\label{VRAdecomp}\eeq
with tensor product multiplicities $a_{\ell\ell'}=\dim_\C \Hom_\Gamma(V_\ell,V_R\otimes V_{\ell'}) \in\Z_{\geq0}$. The
McKay quiver $\Qsf_{\Gamma,R}$ has vertices labelled by the
irreducible representations of the orbifold group $\Gamma$, i.e. the
vertices of the associated extended ADE Dynkin diagram, and
$a_{\ell\ell'}$ arrows from vertex $\ell$ to vertex $\ell'$. In
general, the matrix $A=(a_{\ell\ell'})$ is not symmetric unless $V_R$ is
a self-dual representation of $\Gamma$; in that case $\Qsf_{\Gamma,R}$
is the \emph{double} $\overline{\Qsf}$ of some quiver $\Qsf$, i.e. the quiver
with the same set of nodes $\overline{\Qsf}_0=\Qsf_0$ and with arrow set
$\overline{\Qsf}_1=\Qsf_1\sqcup \Qsf_1^{\rm op}$, where $\Qsf^{\rm
  op}$ is the opposite quiver obtained from $\Qsf$ by reversing the
orientation of the edges. The quiver $\Qsf_{\Gamma,R}$
contains no loop edges if and only if the trivial representation $V_0$
does not appear in the decomposition (\ref{3.5}) of $V_R$ into
irreducible $\Gamma$-modules, i.e.~$N_0=0$.

The McKay correspondence is the observation that for the self-dual fundamental
representation $V_R= \C^2$ of $\Gamma\subset\su$, the matrix $A=A_\Gamma$
is the adjacency matrix of the simply-laced extended Dynkin diagram
corresponding to $\Gamma$; hence the McKay quiver $\Qsf_{\Gamma,\C^2}$
associated to the fundamental representation is the double quiver of the affine
ADE Dynkin graph with any choice of orientation. A simple application of Schur's lemma
shows that the representations of $\Qsf_{\Gamma,\C^2}$ into $R$
correspond bijectively to $\Gamma$-equivariant homomorphisms $V_R\to
\C^2 \otimes V_R$, since by (\ref{3.5}) and (\ref{VRAdecomp}) one has
\beq
\Hom_\Gamma(V_R,\C^2\otimes V_R)=\bigoplus_{\ell,\ell'=0}^{r_\Gamma} \,
a_{\ell\ell'}\ \Hom_\C(R_\ell,R_{\ell'}) \ ,
\label{eq:GammaeqMcKay}\eeq
and so given a morphism in $\Hom_\Gamma(V_R,\C^2 \otimes V_R)$ one can
pair it with an arrow from $\ell$ to $\ell'$ to get a map $R_\ell\to
R_{\ell'}$; hence we have
\beq
\Hom_\Gamma(V_R,\C^2 \otimes V_R) = \Rep {\rm ep}_{\Qsf_{\Gamma,\C^2}}(R) \ .
\label{repspaceMcKay}\eeq
The McKay quiver also comes equiped with a set of relations
$\Rsf_{\Gamma,\C^2}$ that are determined by mapping $B\in\Rep {\rm
  ep}_{\Qsf_{\Gamma,\C^2}}(R)\cong \urmL(N)$ to the corresponding
matrices
$B_1,B_2\in \End_\C(V_R)$ under (\ref{repspaceMcKay}) with
respect to the canonical basis of $\C^2$, which obey $\Gamma$-equivariance conditions derived from (\ref{2.12}) and (\ref{3.6}) as
\beq
\gamma(g)\, B_1\, \gamma(g)^{-1}=(g^{-1})_1{}^\alpha\, B_\alpha \qquad
\mbox{and} \qquad \gamma(g)\, B_2\,
\gamma(g)^{-1}=(g^{-1})_2{}^\alpha\, B_\alpha
\label{Bequivariance}\eeq
for all $g\in\Gamma$.
Then the relations $\Rsf_{\Gamma,\C^2}$ for $\Qsf_{\Gamma,\C^2}$ are given by the commutation relations
\beq
[B_1,B_2]=0 \ .
\label{McKayrels}\eeq
Since $\Gamma\subset\su$, the commutator $[B_1,B_2]$ is
$\Gamma$-invariant and hence is valued in the Lie algebra
\beq
\gfrak(R):=\End_\Gamma^a(V_R) = \bigoplus_{\ell=0}^{r_\Gamma} \, \urmL(N_\ell)
\label{gfrakR}\eeq
of the broken gauge group (\ref{3.12}).

\bigskip

\noindent
{\bf Generalized instanton equations. \ }
Introduce closed two-forms on $\R^4$ given by
\beq \label{6.1}
\ome^a:=\sfrac12\;\eta^a_{\m\n}\, \diff y^\m\wedge\diff y^\n\ ,
\eeq
where $\eta^a_{\m\n}$ are components of the self-dual 't~Hooft tensor given by
\beq \label{6.2}
\eta^a_{bc}=\ve^a_{bc}\und \eta^a_{b4}=-\eta^a_{4b}=\de^a_b\ .
\eeq
The two-forms $\ome^a$, $a=1,2,3$, are self-dual,
\beq \label{6.3}
\ast\,\ome^a=\ome^a\ ,
\eeq
where $\ast$ is the Hodge duality operator for the standard flat Euclidean metric on $\R^4$. They define a hyper-K\"ahler structure on $\R^4$ with complex structures
\beq
(J^a)_\nu^\mu= \omega^a_{\nu\lambda}\, \delta^{\lambda\mu} \ ,
\label{Jadef}\eeq
where the complex structure $J^3$ identifies $\R^4=\C^2$ with the
complex coordinates (\ref{2.8}).

Let $\Abun=W_\mu\, \diff y^\mu$ be a connection on the (trivial) V-bundle $\Rbun:= \R^4\times V_R$ of rank $N$ over $\R^4/\Gamma$ with curvature $\Fbun=\diff\Abun+\Abun\wedge\Abun$. In the following we shall study the moduli space of solutions (with finite topological charge) to the \emph{generalized instanton equations}~\cite{Ivanova:2013mea}
\beq
*\, \Fbun+\Fbun=2\, \omega^a\, \Xi_a \ ,
\label{geninsteqs}\eeq
where
\beq
\Xi_a:= \diag\big(\, \im\,\xi^a_0\,
\one_{ N_0}, \im\,\xi^a_1\,
\one_{ N_1},\dots,\im\, \xi^a_{r_\Gamma}\,
\one_{ N_{r_\Gamma}}\big) 
\eeq
for $a=1,2,3$ are elements of the center of the Lie algebra
(\ref{gfrakR}) and the Fayet-Iliopoulos
 parameter $\xi=(\xi^a_\ell)$ is the linearization of the bundle action. For $\xi=0$ the equations (\ref{geninsteqs}) are the anti-self-dual Yang-Mills
equations on the orbifold $\R^4/\Gamma$, while for $\xi^1=\xi^2=0$, $\xi^3\neq0$ they become the Hermitian Yang-Mills equations which imply that the V-bundle $\Rbun$ is a (semi-)stable vector bundle~\cite{D,UY}; for generic
$\xi\neq0$ they correspond to BPS-type
equations for Yang-Mills theory with sources~\cite{Ivanova:2013mea}.

\bigskip

\noindent
{\bf Moduli spaces of translationally-invariant instantons. \ }
We will provide a geometric interpretation of the McKay quiver in terms of moduli spaces of translationally-invariant $\Gamma$-equivariant instantons on the V-bundle $\Rbun$. Dimensional reduction of the equations (\ref{geninsteqs}) leads to the matrix equations
\beq\label{1.1}\begin{array}{l}
[W_2,W_3]+[W_1,W_4]=\Xi_1\ ,\\[4pt]
[W_3,W_1]+[W_2,W_4]=\Xi_2\ ,\\[4pt]
[W_1,W_2]+[W_3,W_4]=\Xi_3\ ,
\end{array}
\eeq
where the constant matrices $W_\m$ with $\m=1,2,3,4$ take
values in the Lie algebra $\urmL(N)$ and can be regarded as components
of the gauge potential along the internal space of the dimensional reduction. The reduced equations (\ref{1.1}) can be interpreted as hyper-K\"ahler moment map equations, and hence the moduli space of translationally-invariant instantons is given by a hyper-K\"ahler quotient~\cite{Ivanova:2013mea}.

When $R=\widehat{V}$ is the multiplicity space (\ref{3.1}) of the self-dual regular
representation of $\Gamma$ (so that $N=|\Gamma|$), this finite-dimensional hyper-K\"ahler quotient
construction was used by Kronheimer in~\cite{10} to construct a family
of four-dimensional hyper-K\"ahler manifolds $M_\xi$. The
representation theory of the orbifold group $\Gamma$ and the McKay
correspondence are encoded in the property that $M_\xi$ for generic
$\xi\neq0$ is an ALE gravitational instanton, i.e. it is diffeomorphic
to the minimal smooth resolution of the Kleinian singularity
$M_0=\C^2/\Gamma$. The birational morphisms $\pi: M_\xi\to M_0$ are
isomorphisms over the cone $C(S^3/\Gamma)$ whose exceptional fibre
$\pi^{-1}(0)$ is a graph of rational curves $\Sigma_\ell\cong\C P^1$,
$\ell=1,\dots,r_\Gamma$ which is dual to the ordinary (unextended)
ADE Dynkin graph associated to $\Gamma$; the parameters
$\xi=(\xi^a_\ell)$ are the periods of the trisymplectic structure over
$\Sigma_\ell$ under the isomorphism ${\rm H}^2(M_\xi;\R)\cong \R^{r_\Gamma+1}$~\cite{10}. The variety $M_\xi$ also inherits a natural hyper-K\"ahler metric over $C(S^3/\Gamma)$ parameterized by $\xi$ which is asymptotically locally Euclidean (ALE), i.e. it approximates the flat Euclidean metric on the orbifold $\R^4/\Gamma$ at the end of $M_\xi$ up to order~$r^{-4}$.

One way to regard the complex deformation $M_\xi$ is by modifying the
polynomial equation (\ref{fGammaxyz}) in $\C^3$ which deforms the cone $C(S^3/\Gamma)$ to
\beq
f_\Gamma(x,y,z; \vec t \  )=0 \ ,
\eeq
where $\vec t=(t_0,t_1,\dots,t_{r_\Gamma} )$ are coordinates on the base of the deformation related to the periods $\xi^a_\ell$; see e.g.~\cite[Sect.~2.1]{Gubser} for details of this construction. Explicitly, for the five classes of Sasaki-Einstein three-manifolds we have
\bea
f_{\Z_{k+1}}(x,y,z;\vec t \ ) &=& P_{k+1}(x;\vec t \ )+y^2+z^2 \ , \nonumber \\[4pt]
f_{\Di_{k-2}^*}(x,y,z;\vec t \ ) &=& x^{k-1}+Q_{k-2}(x;\vec t \ )+t_0\, y+x\, y^2+z^2 \ , \nonumber \\[4pt]
f_{\Te^*}(x,y,z;\vec t \ ) &=& y^3+Q_{2}(x;\vec t \ )+P_{4}(x;\vec t \ )+z^2 \ , \nonumber \\[4pt]
f_{\Oc^*}(x,y,z;\vec t \ ) &=& y^3+P_{3}(x;\vec t \ )+Q_{4}(x;\vec t \ )+z^2 \ , \nonumber \\[4pt]
f_{\I^*}(x,y,z;\vec t \ ) &=& y^3+Q_{3}(x;\vec t \ )+P_{5}(x;\vec t \ )+z^2 \ ,
\eea
where $P_k(x;\vec t \ )=x^k+\sum_{\ell=0}^k\, t_\ell\, x^{k-\ell}$ and
$Q_{k}(x;\vec t \ )=\sum_{\ell=1}^{k+1}\, t_\ell\, x^{k-\ell+1}$. This
realizes $M_\xi$ as a fibration over the $x$-plane with generic fibre
$\C^*$ (for A and D series) or elliptic curves (for E series).

\bigskip

\noindent
{\bf Nakajima quiver varieties. \ }
Kronheimer's construction can be interpreted in terms of moduli spaces
of representations of the McKay quiver with relations
$(\Qsf_{\Gamma,\C^2},\Rsf_{\Gamma,\C^2})$ into the regular
representation space (\ref{3.1}) of $\Gamma$; see~\cite{Cassens} for
further details. This moduli space is a particular example of a
Nakajima quiver variety~\cite{Nakajima}.

For a quiver $\Qsf$ based on
$\Gamma$ and a
$\Qsf_0$-graded vector space $R$, introduce stability parameters
$\xi:\Qsf_0\to\R^3 \cong\C\oplus\R$ and identify the Lie algebra
$\gfrak(R)$ with its dual $\gfrak(R)^*$ using the Cartan-Killing
form. The vector space $\Rep{\rm ep}_{\overline{\Qsf}}(R)$ carries a metric defined by the Hilbert-Schmidt norm
\beq
\|B\|^2:= \sum_{e\in\overline{\Qsf}_1}\, \tr_{N_{\sfs(e)}} \,B_e\, B_e^\dag
\label{quivermetric}\eeq
and a
holomorphic symplectic form
\beq
\omega_\C(B,B'\,):= \sum_{e\in\overline{\Qsf}_1}\, \epsilon(e)\, \tr_{N_{\sfs(e)}} \, B_e\, B_{\bar e}' \ ,
\eeq
where $\bar{e}\in \Qsf^{\mathrm{op}}_1$ is the reverse edge of $e$
with $\epsilon(e)=1$ and $\epsilon(\bar{e})=-1$ for
$e\in\Qsf_1$. It decomposes as a sum of Lagrangian subspaces $\Rep{\rm ep}_{\overline{\Qsf}}(R)=\Rep{\rm ep}_{{\Qsf}}(R)\oplus\Rep{\rm ep}_{{\Qsf^{\rm op}}}(R)$ with $\Rep{\rm ep}_{{\Qsf^{\rm op}}}(R)\cong\Rep{\rm ep}_{{\Qsf}}(R)^*$, which geometrically identifies it as the cotangent bundle $\Rep{\rm ep}_{\overline{\Qsf}}(R)\cong T^*\Rep{\rm ep}_{{\Qsf}}(R)$; hence it is naturally a quaternionic vector space which gives it the structure of a flat hyper-K\"ahler manifold
that is preserved by the bifundamental action (\ref{Bgaugetransf}) of the gauge group (\ref{3.12}). The corresponding $(1,1)$-form is
\beq
\omega_\R(B,B'\,):= \frac12\, \sum_{e\in\overline{\Qsf}_1}\, \epsilon(e)\, \tr_{N_{\sfs(e)}}\big(\, B_e\, B_e'\,^\dag - B_{\bar e}^\dag\,  B_{\bar e}' \, \big) \ .
\eeq

Then the quiver variety associated to $\Qsf$ and $R$ is the hyper-K\"ahler quotient
\beq
\Xcal_\xi(\Qsf,R):= \Rep{\rm ep}_{\overline{\Qsf}}(R) \,
\big/\!\!\big/\!\!\big/\!^{~}_\xi \ {\rm P}\Gcal(R)
\eeq
by the corresponding hyper-K\"ahler moment map $\mu=(\mu_\C,\mu_\R):\Rep{\rm ep}_{\overline{\Qsf}}(R)\to \R^3\otimes\gfrak(R)$ vanishing at the origin, where
$\mu_\C:\Rep{\rm ep}_{\overline{\Qsf}}(R)\to \gfrak(R)\otimes \C$ is defined by its components
\beq
\mu_\C(B)_\ell= \sum_{e\in\sfs^{-1}(\ell)}\, \epsilon(e)\, B_e\, B_{\bar e}
\label{complexmoment}\eeq
while $\mu_\R:\Rep{\rm ep}_{\overline{\Qsf}}(R)\to
\gfrak(R)$ is defined by
\beq
\mu_\R(B)_\ell=\frac\im2\, \sum_{e\in\sfs^{-1}(\ell)}\,\big( B_e\, B_e^\dag- B_{\bar e}^\dag\, B_{\bar e}\big)
\label{realmoment}\eeq
for $\ell=0,1,\dots,r_\Gamma$. There are relations
\beq
\sum_{\ell=0}^{r_\Gamma}\, \tr_{N_\ell}\, \mu_\C(B)_\ell = 0
=\sum_{\ell=0}^{r_\Gamma}\, \tr_{N_\ell}\, \mu_\R (B)_\ell
\eeq
which follow from cyclicity of the traces.

The quiver variety is then constructed via suitable quotients of the level
set $\mu^{-1}(\Xi)$; this necessitates the traceless condition
\beq
\sum_{\ell=0}^{r_\Gamma} \, \xi_\ell^a \, N_\ell =0 \qquad \mbox{for}
\quad a=1,2,3 \ .
\eeq
The canonical map $\Xcal_\xi(\Qsf,R)\to \Xcal_0(\Qsf,R)$ is a smooth
(hyper-K\"ahler) resolution of singularities for generic values of
$\xi$. By choosing $\xi$ such that the gauge group ${\rm P}\Gcal(R)$
acts freely, the dimension of the hyper-K\"ahler quotient (and of
the vacuum moduli space of the corresponding quiver gauge theory) is given by~\cite[Sect.~3.3]{12}
\bea
\dim_\R \Xcal_\xi(\Qsf,R)= \dim_\R \Rep{\rm ep}_{\overline{\Qsf}}(R)- 4\dim_\R {\rm
  P}\Gcal(R) = 4-2 \vec N\cdot C_\Qsf \vec
N 
\label{dimquiver}\eea
where the generalized Cartan matrix $C_\Qsf$ of the quiver $\Qsf$ is defined by
\beq
\vec N\cdot C_\Qsf\vec N' = 2 \, \sum_{\ell=0}^{r_\Gamma}\, N_\ell\, N_\ell' - \sum_{e\in \overline{\Qsf}_1}\, N_{\sfs(e)}\,
N_{\sft(e)}'
\eeq
for dimension vectors $\vec N:=(N_0,N_1,\dots,N_{r_\Gamma})$ and $\vec
N':=(N_0',N_1',\dots,N_{r_\Gamma}')$ of two quiver representations $R$
and $R'$; note that $\vec N\cdot C_\Qsf \vec
N \in 2\, \Z$. It follows that if $\vec N\cdot C_\Qsf \vec
N >2$ then the representation $R$ is decomposable, while if $\vec N\cdot C_\Qsf \vec
N = 2$ the representation is \emph{rigid}, i.e. it has no moduli. Dimension vectors $\vec N$ of indecomposable quiver representations are called {roots} of the quiver (see~\cite[Sect.~2]{Kac}); in particular rigid representations correspond to real roots. Imaginary roots have Cartan form with $\vec N\cdot C_\Qsf \vec
N \leq0$ and parameterize moduli spaces of dimension $\dim_\R
\Xcal_\xi(\Qsf,R)\geq4$.

\bigskip

\noindent
{\bf McKay quiver varieties. \ } 
The relations (\ref{McKayrels}) for the McKay quiver
$\overline{\Qsf} = \Qsf_{\Gamma,\C^2}$
can be written using the
isomorphism (\ref{repspaceMcKay}) as $\mu_\C(B)=0$; in fact, one may set $\Xi_1=\Xi_2=0$
by a non-analytic change of coordinates on the representation space
(\ref{repspace})~\cite{10,BFRM}. On the other hand, the real moment map
equations $\mu_\R(B)=\Xi_3$ in this case can be written as
\beq
\big[B_1, B_1^\+\, \big]+\big[B_2, B_2^\+\, \big]=\sfrac{\im}{2}\, \Xi_3 \ .
\label{reallevelsets}\eeq
In this instance the generalized Cartan matrix $C_{\Qsf}$ coincides
with the extended Cartan matrix $\widehat{C}_\Gamma=2\,
\one_{\C^{r_\Gamma+1}}-A_\Gamma$ of the simply-laced affine Lie
algebra $\widehat{\gfrak}_\Gamma$ associated to~$\Gamma$.

The Kronheimer construction is then obtained by specialising to the quiver variety associated to the McKay quiver with relations $(\Qsf_{\Gamma,\C^2},\Rsf_{\Gamma,\C^2})$ and the multiplicity space $R=\widehat{V}$, so that
\beq
M_\xi = \Rep{\rm
  ep}_{\Qsf_{\Gamma,\C^2},\Rsf_{\Gamma,\C^2}}\big(\,\widehat{V}\,
\big) \, \big/\!\!\big/\!\!\big/\!^{~}_\xi \ {\rm P}\Gcal\big(\,\widehat{V}\, \big) \ .
\label{ALEquiver}\eeq
Explicitly, the equations (\ref{1.1}) and $\mu(B)=\Xi$ are related by identifying the $\Gamma$-equivariant matrices $B_1:=\frac12\,(-W_4+ \im\, W_3)$ and $B_2:=\frac12\,(W_1+\im\, W_2)$ with $B\in\Rep {\rm
  ep}_{\Qsf_{\Gamma,\C^2}}(\,\widehat{V} \,)$ under the isomorphism
(\ref{repspaceMcKay}). Since $\Gamma$ acts freely on the cone
$C(S^3)=\C^2\setminus\{0\}$, the variety (\ref{ALEquiver}) is
coordinatized by a fixed simultaneous eigenvalue pair $(z_1,z_2)$ of
the commuting matrices $B_1,B_2$ modulo its $\Gamma$-orbit, where the orbit is the regular representation on the coordinates. Since the
dimension vector $\vec n= (n_0,n_1,\dots,n_{r_\Gamma})$ spans the
kernel of the extended Cartan matrix $\widehat{C}_\Gamma$, it follows from the general dimension formula (\ref{dimquiver}) that the quiver variety (\ref{ALEquiver}) has dimension $\dim_\R M_\xi=4$.

\bigskip

\noindent
{\bf Superconformal quiver gauge theory. \ } 
The construction of quiver varieties for $R=\widehat{V}$ has a natural interpretation in
the four-dimensional $\Ncal=2$ superconformal quiver gauge theory on the worldvolume
of $n$ D$p$-branes probing a single D($p+4)$-brane placed at the orbifold singularity of
$\C^2/\Gamma$~\cite{DM,JM,Gubser}. The field theory has gauge group
(\ref{3.12}), with $N_\ell=n\, n_\ell$ the number of constituent
fractional D$p$-branes, and is based on the McKay quiver
$\Qsf_{\Gamma,\C^2}$: At each vertex $\ell=0,1,\dots,r_\Gamma$ there is an
$\Ncal=2$ vector multiplet which corresponds to an $\Ncal=1$ chiral
multiplet $\varphi_\ell$ transforming in the adjoint representation of
$\urm(N_\ell)$, while for each edge $e$ there is a bifundamental
hypermultiplet which corresponds to a pair of $\Ncal=1$ chiral
multiplets $(B_e,B_{\bar e})$ with $B_e$ a complex matrix transforming
as in (\ref{Bgaugetransf}). The cubic $\Ncal=1$ superpotential is
then determined from the complex moment map given by
(\ref{complexmoment}) as
\beq
W_\Gamma(B,\varphi) = \sum_{\ell=0}^{r_\Gamma}\, \tr_{N_\ell}\,
\mu_\C(B)_\ell\, \varphi_\ell \ .
\eeq
One can modify the superpotential by shifting it with complex Fayet-Iliopoulos terms to
\beq
W_\Gamma(B,\varphi)_\xi=W_\Gamma(B,\varphi)-\sum_{\ell=0}^{r_\Gamma}\, \im\, \xi^\C_\ell\, \tr_{N_\ell}\, \varphi_\ell
\eeq
whilst still preserving $\Ncal=2$ supersymmetry.
In the supersymmetric vacuum state all fermion fields are set to zero, with the
scalar fields taking constant vacuum expectation values. The F-term equations
\beq
\frac{\partial W_\Gamma(B,\varphi)_\xi}{\partial\varphi_\ell}=0
\eeq
then reproduce the deformed McKay quiver relations
$\mu_\C(B)_\ell=\im\, \xi^\C_\ell$ for $\ell=0,1,\dots,r_\Gamma$. The
D-term equations instead encode qualitative stability information
about the (non-degenerate) kinetic term given by the metric
(\ref{quivermetric}), and they correspond to the level sets (\ref{reallevelsets}) of the real moment map from
(\ref{realmoment}). By factoring solutions to these equations by the
action of the gauge group (\ref{3.12}) we obtain the vacuum moduli
space. It follows that the quiver variety $\Xcal_\xi(\Qsf,R)$ for $\overline{\Qsf}=
\Qsf_{\Gamma,\C^2}$ and $R=\widehat{V}$ can be realized as the Higgs
branch of this quiver gauge theory.

\bigskip

\noindent
{\bf Moduli spaces of framed instantons. \ }
We can work instead with \emph{framed} McKay
quivers~\cite{KN,Nakajima,Cirafici:2012qc} which are obtained by adding a node
with doubled arrows to each vertex of the McKay quiver with relations
$(\Qsf_{\Gamma,\C^2},\Rsf_{\Gamma,\C^2})$, where in general the extra nodes
correspond to $\Gamma$-modules $V_S$ which parameterize the holonomies
of instanton gauge connections at infinity. The extended representation
space is given by
\beq
\Rep {\rm ep}_{\overline{\Qsf}}(R,S):= \Rep {\rm
  ep}_{\overline{\Qsf}}(R) \ \oplus \ 
\bigoplus_{\ell=0}^{r_\Gamma}\, \Hom_\C(R_\ell,S_\ell)\oplus\Hom_\C(
S_\ell,R_\ell) \ .
\eeq
The corresponding quiver
varieties $\Xcal_\xi(\Qsf;R,S)$ parameterize moduli of framed instantons on the ALE spaces
$M_\xi$, or alternatively the Higgs branches of quiver gauge theories of 
$N>1$ D$(p+4)$-branes at the orbifold singularity with framing corresponding
to the addition of probe D$p$-branes to the D$(p+4)$-branes~\cite{DM}; in this case the moduli arise as
ADHM data.
A connection on a framed vector bundle $E\to M_\xi$ of rank
$N$ is specified by the Chern classes of $E$ and a flat connection at
the end of $M_\xi$, which is isomorphic to the
Sasaki-Einstein space $S^3/\Gamma$. The McKay correspondence can then
be stated as a one-to-one correspondence between flat $\urm(N)$
connections on $S^3/\Gamma$, which correspond to representations $V_S$ of
the fundamental group $\pi_1(S^3/\Gamma)=\Gamma$ in $\urm(N)$, and
integrable highest weight representations of the affine Lie algebra
$\widehat{\gfrak}_\Gamma$ of ADE type associated to $\Gamma$ at
level~$N$. In~\cite{Nakajima} Nakajima constructs natural
representations of $\widehat{\gfrak}_\Gamma$ at level $N$ on the
cohomology of the quiver varieties in terms of geometric Hecke
correspondences.

In general, although the ADHM construction of Yang-Mills instantons on
ALE spaces is a finite-dimensional hyper-K\"ahler quotient
construction, it cannot be presented as translationally-invariant
anti-self-duality equations with Fayet-Iliopoulos terms
$\xi$ as in (\ref{1.1}). However, for the instanton configuration of minimal topological charge $c_1(E)=0$ and
$c_2(E)=(|\Gamma|-1)/|\Gamma|$ corresponding to $N_\ell=n_\ell$ and a
single D$(p+4)$-brane above, the ``outer'' framing fields decouple~\cite[Sect.~8.1]{DM} and the
relevant quiver is the unframed McKay quiver with relations, i.e. 
the ADHM equations in this case reduce to
(\ref{1.1})~\cite[Ex.~3]{KN}. This fact is consistent with the
constructions of moduli spaces $\Xcal_\xi(\Qsf;\widehat{V},S)$ of minimal charge Yang-Mills instantons
on the ALE spaces $M_\xi$, described in~\cite{Nak90} and~\cite[Sect.~7]{BFRM}, as the four-dimensional non-compact hyper-K\"ahler manifold $M_\xi$ itself. Note that
the orbifold group $\Gamma$ acts non-trivially on the ADHM
matrices but trivially on the vector bundle $E\to M_\xi$, while the
embedding of $\Gamma$ in $\urm(N)$ is defined by the asymptotic
holonomy of the gauge connection rather than by covariance equations
such as (\ref{covrepXa}).

\bigskip

\noindent
{\bf $\mbf{A_k}$. \ }
For the cyclic group
$\Gamma=\Z_{k+1}$, the fundamental representation (\ref{hfundrep}) is
reducible with splitting $\C^2 =V_1\oplus V_k$, and the decomposition (\ref{VRAdecomp}) reads as
\beq
\C^2 \otimes V_\ell= V_{\ell+1}\oplus V_{\ell-1}
\label{Akfundrep}\eeq
for $\ell=0,1,\dots,k$. Thus only $a_{\ell,\ell\pm1}=1$ are non-zero,
which gives precisely the adjacency matrix of the $\widehat{A}_k$-type
Dynkin diagram (\ref{AkDynkin}). The McKay quiver
$\Qsf_{\Gamma,\C^2}$ is the corresponding double quiver
\medskip
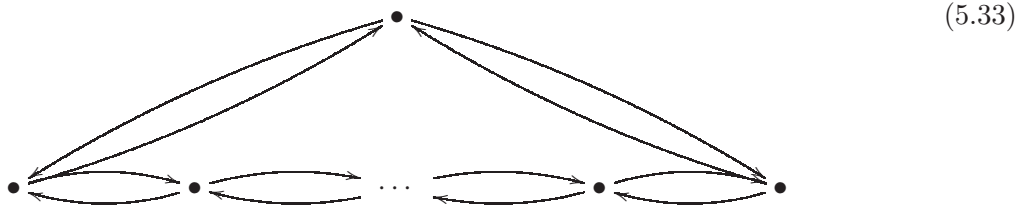
\begin{equation}
 \label{Akdouble}
 \xymatrix@C=20mm{
& & {\bullet}\ar@/_/[ddll]{} \ar@/^/[ddrr]{} & & \\ & & & & \\
{\bullet}\ar@/^/[r]{}\ar@/_/[uurr]{} & {\bullet} \ar@/^/[l]{}\ar@/^/[r]{} & \ \cdots
\ \ar@/^/[l]{}\ar@/^/[r]{} & {\bullet}\ar@/^/[l]{}\ar@/^/[r]{} & {\bullet}\ar@/^/[l]{}\ar@/^/[uull]{}
}
\end{equation}

\medskip
\noindent
The solutions to the $\Gamma$-equivariance conditions are matrices
\beq \label{B1B2cyclic}
B_1=\begin{pmatrix}
   0&0&\cdots&0&\psi_{k+1}\\\psi_1&0&\ddots&0&0\\0&\psi_2&\ddots&\vdots&\vdots\\\vdots&\ddots&\ddots&0&0\\
   0&\cdots&0&\psi_k&0 \end{pmatrix} \und
B_2=\begin{pmatrix}
   0&\varphi_1&0&\cdots&0\\0&0&\varphi_2&\ddots&\vdots\\
   \vdots&\vdots&\ddots&\ddots&0\\0&0& \cdots &0&\varphi_k \\
   \varphi_{k+1}&0& \cdots &0&0 \end{pmatrix}
\eeq
corresponding respectively to the two summands in (\ref{Akfundrep}), where the arrows between vertices of the quiver (\ref{Akdouble}) are represented by linear maps $\psi_{\ell+1}\in\Hom_\C(R_\ell,R_{\ell+1})$ and $\varphi_{\ell+1}\in\Hom_\C(R_{\ell+1},R_{\ell})$ for $\ell=0,1,\dots,k$. The relations (\ref{McKayrels}) read as
\beq
\varphi_{\ell+1}\, \psi_{\ell+1}=\psi_{\ell} \, \varphi_\ell
\label{AkMcKayrels}\eeq
while the stability conditions (\ref{reallevelsets}) are given by
\beq
\varphi_{\ell+1}\,
\varphi_{\ell+1}^\dag-\varphi_\ell^\dag\,\varphi_\ell=
\psi_{\ell+1}^\dag\,\psi_{\ell+1}-\psi_{\ell}\,
\psi_{\ell}^\dag-\mbox{$\frac12$}\, \xi_\ell^3\, \one_{\C^{N_\ell}}
\eeq
on $R_\ell$ for each $\ell=0,1,\dots,k$. By (\ref{dimquiver}) the
dimension of the corresponding Nakajima quiver variety is given by
\beq
\dim_\R \Xcal_\xi(\Qsf_{\Gamma,\C^2},R) = 4+4\, \sum_{\ell=0}^k\, N_\ell\,
N_{\ell+1}-4\, \sum_{\ell=0}^k\, N_\ell^2 \ .
\eeq
This is equal to four when $N_\ell=n$ for all $\ell=0,1,\dots,k$, and
for $n=1$ the quiver variety is a minimal resolution $M_\xi$ of the
quotient singularity $\Xcal_0\big(\Qsf_{\Gamma,\C^2},\widehat{V}\,
\big)= \C^2/\Z_{k+1}$.

\bigskip

\section{Instantons on cones and Nahm equations\label{sec:Nahm}}

\noindent
{\bf Moduli spaces of spherically symmetric instantons. \ }
In this section we describe the moduli spaces of representations of the ``Sasakian" quivers from Sect.~\ref{se:Hvb} in an analogous way to those of the McKay quivers from Sect.~\ref{sec:instantons}. Comparing the two quiver diagrams (\ref{Akdouble}) and
(\ref{3.19}), the two translationally-invariant
$\Gamma$-equivariant complex Higgs fields
(\ref{B1B2cyclic}) yield a quiver based on the Dynkin diagram of type
$\widehat{A}_k$ involving an arrow between vertices for each Higgs field, while
the $\su$-equivariant Higgs fields (\ref{4.7})--(\ref{4.8}) yield a
quiver based on the same Dynkin graph involving an arrow for the
complex horizontal Higgs field together with a loop edge for the vertical Higgs field.

The $\su$-invariant (spherically symmetric) reduction of
$\Gamma$-invariant connections on the trivial V-bundle $\Rbun$ over $\R^4/\Gamma$ is most conveniently done by exploiting conformal invariance of the $\xi$-deformed anti-self-duality
equations (\ref{geninsteqs}) and considering them on the cylinder
$\R\times S^3/\Gamma$. The connection in these coordinates is written
as
\beq
\Abun=W_\tau\, \diff\tau+W_a\, \hat e^a
\eeq
where, in
contrast to (\ref{2.2}), the basis of left
$\su$-invariant one-forms
\beq \label{5.3}
\hat e^a:=-\mbox{$\frac{1}{r^2}$} \,\eta^a_{\m\n}\, y^{\m}\, \diff y^{\n}
\eeq
on $S^3$ are \emph{invariant} with respect to the action of the
orbifold group $\Gamma\subset\su$; this $\su$ group is the
right-acting factor of the Lorentz
group ${\rm SO}(4)=\su\times\su$ on $\R^4$ which preserves the complex
structure $J^3$ from (\ref{Jadef}), while the left-acting $\su$ factor does
not (only a $\uo$ subgroup preserves $J^3$). Now the matrices $W_a$ with $a=1,2,3$ and $W_\tau$ depend on the radial coordinate $r=\e^\tau$, and by defining
\beq
Y_a:=\e^{2\tau}\, W_a \ , \qquad Y_\tau:= \e^{2\tau}\, W_\tau \qquad \mbox{and} \qquad s=\e^{-2\tau}
\eeq
the equations (\ref{geninsteqs}) reduce to the ordinary differential equations~\cite{Ivanova:2013mea}
\beq \label{5.21}
2\, \frac{\diff Y_a}{\diff s}= [Y_\tau , Y_a]- \frac{1}{2}\,\ve_{a}^{bc} \,
[Y_b , Y_c]+\frac{1}{s^2}\, \Xi_a
\eeq
for the $\Gamma$-invariant functions $Y_a,Y_\tau:\R_{>0}\to \gfrak(R)$ with $a=1,2,3$.
For $\Xi_a=0$ these equations coincide with the Nahm equations. In
contrast to the ADHM-type matrix model, no stability parameters $\Xi$ will
be required to define the resolution of singularities of the moduli
space of solutions~\cite{13,14,12} which will instead arise from the
chosen boundary conditions at large $s$; henceforth we set $\Xi=0$
without loss of generality (see also~\cite[Sect.~5]{Ivanova:2013mea}).

Let us momentarily consider the case with trivial orbifold group
$\Gamma=\{1\}$, and let $\Gcal$ be a compact simple Lie group with
maximal torus $\T$; let $\Gcal^\C=\Gcal\otimes\C$ and
$\T^\C=\T\otimes\C$ denote the corresponding complexified groups. Then the equations (\ref{5.21}) are equivalent to the
equations considered by Kronheimer~\cite{13,14} (see also~\cite{15})
in the description of $\su$-invariant
instantons~\cite{Ivanova:2013mea}. Kronheimer shows in~\cite{13} that,
for the spherically symmetric Yang-Mills instantons which have minimal
topological charge on the cone $\R^4\setminus\{0\}=C(S^3)$, under
certain boundary conditions the moduli space is again a Calabi-Yau twofold
$M_\xi'$ which is a minimal resolution of $C(S^3/\Gamma'\,)$. This is the moduli
space of $\su$-invariant framed $\Gcal$-instantons on $C(S^3)$ with a ``pole" type singularity at the origin, which in terms of Nahm data is characterized by the space of smooth solutions to (\ref{5.21}) satisfying the boundary conditions
\beq
\lim_{s\to\infty}\, Y_\tau(s)=0 \ , \qquad \lim_{s\to\infty}\, Y_a(s)=T_a \qquad \mbox{and} \qquad \lim_{s\to0}\, s\, Y_a(s)=I_a \qquad \mbox{for} \quad a=1,2,3 \ ,
\label{Nahmbcs}\eeq
where $T_a$ are fixed elements of the Cartan subalgebra $\tfrak$ of the Lie
algebra $\gfrak$ of $\Gcal$ whose common centralizer is $\T$, and
\beq
[I_a,I_b] = \varepsilon_{ab}^c\, I_c
\eeq
where the generators $I_a$ define an embedding of $\su$ in $\Gcal$. The elements $T_a$ parameterize a hyper-K\"ahler
structure on the regular coadjoint orbit $\Gcal^\C/\T^\C$, while the
elements $I_a$ specify the holonomy of the gauge connection at
infinity which generically reduces the orbit to a smaller space. The
quotient defining the moduli space is taken by the action of the gauge
group $\widehat{\Gcal}$ consisting of gauge transformations $g:\R_{>0}\to \Gcal$ which
are trivial at infinity.

One can solve (\ref{5.21}) by taking $Y_a,Y_\tau$ as constant matrices from $\tfrak$, which produces singular abelian instanton solutions with delta-function sources in the Maxwell equations~\cite{13,14,Ivanova:2013mea}. By considering $s$-dependent solutions with boundary conditions (\ref{Nahmbcs}), the moduli space can be obtained by an infinite-dimensional hyper-K\"ahler quotient construction~\cite[Sect.~5]{12}.
For the subregular representation $I=(I_a)$, the resulting manifold is again a four-dimensional hyper-K\"ahler ALE space
$M_\xi'$ which is a resolution of singularities of the orbifold
$M_0'=\C^2/\Gamma'$~\cite{13}, where the finite subgroup
$\Gamma'\subset\su$ is obtained from the homogeneous Dynkin diagram of
$\Gcal$; if $\Gcal$ is not simply-laced then this graph is understood
as the \emph{associated} homogeneous Dynkin diagram of ADE-type whose
quotient by a finite group of diagram symmetries yields the Dynkin diagram of
$\Gcal$, see~\cite{Slodowy}. The resolution parameters $\xi=(\xi_\ell^a)$ for
$a=1,2,3$ and $\ell=0,1,\dots,r_{\Gamma'}$ are the periods of the
trisymplectic structure determined by $T_a$ under the isomorphism
${\rm H}^2(\Gcal^\C/\T^\C; \R)={\rm H}^2(\Gcal/\T; \R)\cong\tfrak\, $; if
$\Gcal$ is not of ADE-type then the hyper-K\"ahler cohomology classes
are pullbacks
of those associated to $T_a$ by the surjective quotient map
${\tfrak}^\prime\to {\tfrak}$ from the Cartan subalgebra ${\tfrak}^\prime$ of the
associated Lie group with homogeneous Dynkin diagram. For nilpotent
orbits with $T_a=0$, the moduli spaces of dimension four are cones
$C(S^3/\Gamma'\,)$, while in dimensions $\geq8$
they correspond to the minimal nilpotent orbit, i.e. the orbit of the
highest root vector in $\gfrak$, which is the cone over the 3-Sasakian homogeneous manifold associated to $\Gcal$~\cite{15}. For generic non-regular
orbits and smaller $\su$ representations, if the moduli space is
four-dimensional then by~\cite{K2} it is a disjoint union of ALE
spaces and cones.

For a non-trivial orbifold group $\Gamma$, we can repeat this
construction with $\Gcal=\Gcal(R)$. Then the
moduli space of solutions to the Nahm equations (\ref{5.21}) with the
boundary conditions (\ref{Nahmbcs}) is a product of $r_\Gamma+1$
moduli spaces of Nahm data associated with each factor $\urm(N_\ell)$
of the gauge group (\ref{3.12}) for $\ell=0,1,\dots,r_\Gamma$. If each
block of $I$ in $\Gcal(R)$ is the subregular representation of $\su$
in $\urm(N_\ell)$, then the moduli space is a product of ALE spaces
$M'_{\xi_0}\times M'_{\xi_1}\times\cdots \times M'_{\xi_{r_\Gamma}}$
where $M'_{\xi_\ell}$ is a minimal resolution of the orbifold
$\C^2/\Z_{N_\ell}$ for $\ell=0,1,\dots,r_\Gamma$; note that
$M'_{\xi_\ell}$ is a point if $N_\ell=1$. Due to $\Gamma$-invariance
of the frame $(\diff\tau,\hat e^a)$ and hence of the components of $\underline{A}$, irreducible connections
occur only for irreducible actions of $\Gamma$ on the V-bundle $\Rbun$,
i.e. when $N_\ell=N$ for some $\ell\in\{0,1,\dots,r_\Gamma\}$ and
$N_{\ell'}=0$ for $\ell'\neq \ell$, in which case the moduli space is
the four-dimensional hyper-K\"ahler manifold $M_{\xi_\ell}'\to
\C^2/\Z_{N}$; for example, the four-parameter family of 't~Hooft
one-instanton solutions~\cite[Sect.~5]{Ivanova:2013mea}
\beq
Y_\tau=0 \qquad \mbox{and} \qquad Y_a= \frac1{s+\lambda}\, I_a \qquad
\mbox{with} \quad \lambda\in\R_{\geq0} 
\label{tHooftsol}\eeq
live on the base of this deformation for $\lambda=0$. In dimensions $\geq8$ with $T_a=0$ the moduli space is the cone over the 3-Sasakian homogeneous manifold $\suk\big/\Sp\big(\urm(N-2)\times\uo\big)$ of dimension $4(N-1)$ with~$N\geq3$.

\bigskip

\noindent
{\bf Sasakian quiver gauge theory. \ } In Sect.~\ref{sec:dimred} we considered the $\su$-equivariant dimensional
reduction of Yang-Mills theory from $M\times S^3/\Gamma$ to $M$. For
the cones we consider instead the reduction from $\tilde M\times
C(S^3/\Gamma)$ to $\tilde M\times\R_{>0}$, which by the $\Gamma$-action in
(\ref{3.10}) is described by the same quiver (in applications to
holography $\tilde M$ is the boundary of $M$).

The only change in the resulting
quiver gauge theory action is the addition of a term proportional to $\tr_N\,
\Fcal_{\tau a}\,\Fcal^{\tau a}$ in (\ref{SYMreduced}). This additional
term allows for a vacuum state of the quiver gauge theory on $\tilde M$ with not
only flat gauge fields $\Fcal_{ab}=0=\Fcal_{\tau a}$ but also
anti-self-dual gauge fields $\Fcal$ on $C(S^3/\Gamma)$. The reduction of the
anti-self-dual Yang-Mills equations on the cone $C(S^3/\Gamma)$, with
the conformally equivalent metric (\ref{2.20}), for
$\su$-invariant connections
\beq
\Acal= X_\tau\, \diff\tau + X_a\, e^a
\label{AcalSU2}\eeq
on the V-bundle $\Rbun$ is carried out in~\cite[Sect.~5]{Ivanova:2013mea} and shown to be given by Nahm-type equations
\beq
\frac{\diff X_a}{\diff\tau}=2X_a-\frac12\, \ve_{a}^{bc}\, [X_b,X_c]-[X_\tau,X_a] \ .
\label{Nahmtypeeqs}\eeq
As the one-forms $e^a$ are not $\Gamma$-invariant, the matrices $X_a$ now decompose as in
(\ref{Xablocks}), allowing for reducible $\Gamma$-actions on $\Rbun$.

These equations have a solution with
\emph{constant} matrices $X_a=2I_a$ and $X_\tau=0$ which yields
vanishing curvature $\Fcal=0$, where $I_a\in \Rep {\rm
  ep}_{\Qsf_\Gamma,\Rsf_\Gamma}(R) \, \big/\!\!\big/ \ {\rm P}\Gcal(R)$ are $\su$ generators in the
$N$-dimensional irreducible representation on the vector space $V_R$; this is
in contrast to the case of flat space considered earlier where all such matrices are
necessarily diagonal~\cite{10}. For the multiplicity space $R=\widehat{V}$ of the regular representation of $\Gamma$, i.e. when $N_\ell=n_\ell$ for all $\ell=0,1,\dots,r_\Gamma$, the moduli space
of all constant matrices is the orbifold
$M_0=\C^2/\Gamma$, as discussed above. We shall find that the moduli space
of $\tau$-dependent solutions with suitable boundary
conditions at $\tau\to\pm\,\infty$ is also an orbifold $M_0'=\C^2/\Gamma'$ for $\Gamma'=\Z_{|\Gamma|}$,
where $\Gamma'=\Gamma$ only for the A~series; in this instance $I_a$ embed the
group $\su$ irreducibly into ${\rm P}\Gcal= \surm(|\Gamma|)$. This expectation is supported by the
explicit BPST-type instanton solutions $X_a=f(\tau)\, I_a$, $X_\tau=0$ constructed
by~\cite[Sect.~5]{Ivanova:2013mea} in this case (see below) which are
parameterized by a four-dimensional moduli space consisting of one
dilatation parameter valued in $\R_{\geq0}$ and three gauge rotational $\su$
parameters,
analogously to~\cite[Sect.~7]{BFRM}; since the subgroup of
$\surm(|\Gamma|)$ which commutes with $I_a$ in this case is its center
$\Z_{|\Gamma|}$, the gauge rotations are actually valued in
$\su/\Z_{|\Gamma|}=S^3/\Z_{|\Gamma|}$ and so the
moduli space is the cone $C\big(S^3\big/\Z_{|\Gamma|} \big)$.

\bigskip

\noindent
{\bf Sasakian quiver varieties. \ } 
To substantiate and extend these statements, we generalize the
infinite-dimensional hyper-K\"ahler quotient construction of the
moduli space of $\su$-invariant instantons on $C(S^3)$ (see e.g.~\cite[Sect.~5]{12}) to the general
setting of quiver varieties. We put
\beq
C_1=-Z_t+\im\, Z_3:=\mbox{$\frac1{2t}$}\,(-X_\tau+\im\, X_3) \und
C_2= Z_1+\im\, Z_2 :=\mbox{$\frac1{2t}$}\, (X_1+\im\, X_2)
\label{CZX}\eeq
where $t:=\e^{2\tau}=r^2$. For a $\Gamma$-module (\ref{3.5}), we may identify the $t$-dependent
$\Gamma$-equivariant matrices $C_1,C_2:\R_{>0}\to\End_\C(V_R)$ 
with maps $\R_{>0}\to\Rep{\rm
  ep}_{\Qsf_\Gamma^{\rm op}}(R)$ into the representation space of
the \emph{opposite} quiver associated to the Sasakian quiver
$\Qsf_\Gamma$; we denote this infinite-dimensional affine space by $\widehat{\Rep{\rm ep}}_{\Qsf_\Gamma^{\rm op}}(R)$. For this, let
\beq
\C^2_\Gamma=\bigoplus_{\ell=0}^{r_\Gamma}\, d_\ell^\Gamma\ V_\ell
\label{SasakiC2}\eeq
be a two-dimensional representation of the orbifold group $\Gamma$
determined by the $\Gamma$-equivariance conditions on $C_1,C_2$
described in Sects.~\ref{sec:SEgeometry}--\ref{se:Hvb}. Let
$A^\Gamma=(a^\Gamma_{\ell\ell'})$ be the adjacency matrix of the
quiver $\Qsf_\Gamma$, i.e. $a^\Gamma_{\ell\ell'}$ is the number of
arrows joining vertex $\ell$ to vertex $\ell'$; note that $A^\Gamma$
is not a symmetric matrix in general. Then the multiplicities $d_\ell^\Gamma$ can
be determined by using the tensor product multiplicities
$m_{\ell\ell'}{}^{\ell^{\prime\prime}}$ appearing in the
Clebsch-Gordan decomposition
\beq
V_\ell\otimes V_{\ell'}=\bigoplus_{\ell^{\prime\prime}=0}^{r_\Gamma} \,
m_{\ell\ell'}{}^{\ell^{\prime\prime}}\ V_{\ell^{\prime\prime}}
\eeq
via the relations
\beq
\sum_{\ell^{\prime\prime}=0}^{r_\Gamma}\,
d^\Gamma_{\ell^{\prime\prime}} \
m_{\ell^{\prime\prime}\ell}{}^{\ell'} = a_{\ell\ell'}^\Gamma \qquad
\mbox{for} \quad \ell,\ell'=0,1,\dots, r_\Gamma \ .
\label{CGammarels}\eeq
Then by analogous arguments to those which led to
(\ref{eq:GammaeqMcKay}), we find
\beq
\Hom_\Gamma(V_R,\C_\Gamma^2\otimes V_R)=\bigoplus_{\ell,\ell'=0}^{r_\Gamma} \,
a^\Gamma_{\ell\ell'}\ \Hom_\C(R_{\ell'},R_{\ell}) = \Rep {\rm ep}_{\Qsf^{\rm
    op}_{\Gamma}}(R)
\label{repspaceSE}\eeq
from which the identification follows.

The space $\widehat{\Rep{\rm ep}}_{\Qsf_\Gamma^{\rm op}} (R)$ has the natural structure of an
infinite-dimensional quaternionic vector space by identifying the $\Gamma$-module $\C_\Gamma^2$ as a module of rank one over the quaternions $\Hbb$. With suitable boundary conditions that we describe
below, it carries a metric defined by the ${\rm L}^2$-norm
\beq
\big\|(c_1,c_2) \big\|^2:= \int_{\R_{>0}}\, \diff t \ \tr_N\big(c_1^\dag\, c_1+c_2^\dag\, c_2\big)
\label{L2metric}\eeq
and a holomorphic symplectic form
\beq
\omega_\C\big((c_1,c_2)\,,\,(c_1',c_2')\big):= \int_{\R_{>0}}\, \diff t \ \tr_N\big(c_1\, c_2'-c_1'\, c_2\big) \ ,
\eeq
where $c_i:=\delta C_i$ are solutions of the linearised (around $C_i$
given by (\ref{CZX})) equations (\ref{Nahmtypeeqs}).
Let $\widehat{\Gcal}(R)$ be the group of gauge transformations $g:\R_{>0}\to\Gcal(R)$ which are trivial at infinity; it acts on $\widehat{\Rep{\rm ep}}_{\Qsf_\Gamma^{\rm op}} (R)$ as
\beq
C_1 \ \longmapsto \ g\, C_1\, g^{-1} - \frac12\, \frac{\diff g}{\diff t}\, g^{-1} \und C_2 \ \longmapsto \ g\, C_2\, g^{-1} \ .
\label{Cgaugetransf}\eeq
These ingredients endow $\widehat{\Rep{\rm ep}}_{\Qsf_\Gamma^{\rm op}}
(R)$ with the structure of a flat
hyper-K\"ahler Banach manifold which is invariant under the action of
$\widehat{\Gcal}(R)$. The corresponding $(1,1)$-form is
\beq
\omega_\R\big((c_1,c_2)\,,\,(c_1',c_2')\big):= \frac12\,
\int_{\R_{>0}}\, \diff t \ \tr_N\big(\,
c_1\,c_1'\,^\dag-c_1'\,c_1^\dag+c_2\, c_2'\,^\dag-c_2'\,c_2^\dag \, \big) \ .
\eeq

With $\widehat{\urmL}(N)$ denoting the Lie
algebra of infinitesimal gauge transformations $\R_{>0}\to\urmL(N)$
which are trivial at infinity, the corresponding hyper-K\"ahler moment map
$\mu=(\mu_\C,\mu_\R):\widehat{\Rep{\rm ep}}_{\Qsf_\Gamma^{\rm op}} (R)\to \R^3\otimes \widehat{\urmL}(N)$ is
given by
\bea
\mu_\C(C_1,C_2) &=& \frac{\diff C_2}{\diff t}+[C_1,C_2] \ , \label{complexNahm}
\\[4pt]
\mu_\R(C_1,C_2) &=& \frac\im2 \, \Big(\, \frac{\diff C_1}{\diff t}+\frac{\diff C_1^\dag}{\diff t} 
+\big[C_1,C_1^\dag\,\big]+ \big[C_2,C_2^\dag\,\big]\, \Big) \ . \label{realNahm}
\eea
The vanishing locus $\mu^{-1}(0)$ then coincides with the solution
space of the Nahm equations (\ref{Nahmtypeeqs}). Notice how this
moment map formally compares with that of the McKay quiver variety
from Sect.~\ref{sec:instantons} by setting $W_4=\frac\diff{\diff
  t}+Z_t$ and $W_a=Z_a$ for $a=1,2,3$. Its image belongs to the set of gauge
equivalence classes of elements valued in the subspace $\R^3\otimes \widehat{\Rep{\rm ep}}_{\Qsf_\Gamma^{\rm op}} (R)\subset\R^3\otimes \widehat{\urmL}(N) $: Via a gauge transformation (\ref{Cgaugetransf}) one can go to a temporal gauge with $Z_t=0$ in which the components of the moment map are given by
\beq
\mu_a(Z)= \frac{\diff Z_a}{\diff t} +\frac12\, \varepsilon^{bc}_a \, [Z_b,Z_c] 
\label{NahmZa}\eeq
for $a=1,2,3$.

However, in contrast to the moment maps used in the
construction of Nakajima quiver varieties, here $\mu$ is not
$\Gamma$-invariant. Hence our vacuum moduli are generically
parameterized by an orbifold
$\mu^{-1}(0)/\, {\rm P}\widehat{\Gcal}(R)$ which cannot be described
as a hyper-K\"ahler quotient. As long as the action of the Lie group
${\rm P}\widehat{\Gcal}(R)$ is proper, the space of orbits
$\mu^{-1}(0)/\, {\rm P}\widehat{\Gcal}(R)$ has the structure of a
stratified Hausdorff differential space (see e.g.~\cite{Sniatycki}). Here we shall take an explicit and illuminating route that avoids the intricate
technical stacky issues involved in taking such quotients: We first
describe the hyper-K\"ahler quotient corresponding to the full
unbroken gauge group $\Gcal=\urm(N)$, and then implement
$\Gamma$-equivariance; we denote the (singular) quotient space obtained in this
way by
$\widehat{\Rep{\rm ep}}_{\Qsf_\Gamma^{\rm op}} (R)\, \big/\!\!\big/ \,
{\rm P}\widehat{\Gcal}(R)$.

It remains to specify suitable boundary conditions for the equations
$\mu_\C=\mu_\R=0$. Using results of~\cite{Biquard}, one can put solutions of the Nahm equations into a Coulomb gauge such that $C_1,C_2$ converge as $t\to\infty$ to the exact ``model" solution
\beq
\tilde C_1=\frac\im2\, T_3+\frac{\im}{t+\lambda}\, J_3 \qquad \mbox{and} \qquad \tilde C_2=\frac12\, T_++\frac1{t+\lambda}\, J_+ \ ,
\label{modelsol}\eeq
where $\lambda\in\R_{\geq0}$ is a scale parameter, the elements $T_3, T_+=T_1+\im\,
T_2$ are valued in the Cartan subalgebra of the Lie algebra
(\ref{gfrakR}), and $J_3,J_+=J_1+\im\, J_2$ generate a representation
of $\su$ in $\urm(N)$ which resides in the representation space
(\ref{repspaceSE}) and which commutes with $T_a$, i.e. $J_a$ for
$a=1,2,3$ take values in the Lie algebra of the common centralizer of
$T_a$ in $\urm(N)$. Suitable moduli spaces of Nahm data asymptotic to
this model solution yield complex coadjoint orbits of
$\urm(N)$~\cite{13,14,Biquard}. The model solution with $T_a=0$ is
exactly the BPST-type instanton solution on the orbifold $\R^4/\Gamma$
which is discussed in~\cite[Sect.~5]{Ivanova:2013mea}; it generates
nilpotent orbits~\cite{13}. On the other hand, if the joint
centralizer of $T_a$ is the maximal torus $\urm(1)^N$, then $J_a=0$
and the orbits are regular~\cite{14}. Henceforth we set $J_a=0$ but
keep $T_a$ arbitrary corresponding to generic semisimple orbits.

Thus analogously to the boundary conditions in (\ref{Nahmbcs}), we consider solutions with the asymptotics
\beq
\lim_{t\to\infty}\, C_1(t)=\mbox{$\frac\im2$} \, T_3 \qquad \mbox{and} \qquad \lim_{t\to \infty}\, C_2(t) = \mbox{$\frac12$}\, T_+ \ ,
\label{quiverasympt}\eeq
and which acquire simple poles
\beq
\lim_{t\to0}\, t\, C_1(t)= \mbox{$\frac\im2$} \, I_3 \qquad \mbox{and} \qquad \lim_{t\to 0}\, t\, C_2(t)= \mbox{$\frac12$}\, I_+
\label{quiverpoles}\eeq
at $t=0$ with residues defining a representation $I_3,I_+=I_1+\im\,
I_2$ of $\suL$ in (\ref{repspaceSE}). We denote by
$\widehat{\Rep{\rm ep}}_{\Qsf_\Gamma^{\rm op}}(R)_{\xi,I}$ the
subspace of $\widehat{\Rep{\rm ep}}_{\Qsf_\Gamma^{\rm op}}(R)$
consisting of pairs $(C_1,C_2)$ satisfying these boundary conditions
and with suitable analytic behaviour, and by
$\widehat{\Gcal}(R)_I\subset \widehat{\Gcal}(R)$ the subgroup of gauge
transformations preserving the boundary conditions
(\ref{quiverpoles}). Here $\xi=(\xi_\ell^a)$ with $\xi_\ell^a\in\R$
for $a=1,2,3$ and $\ell=0,1,\dots,r_\Gamma$ parameterize the periods
of the trisymplectic structure determined by $T_a$, while fixing the
singular part of the Nahm data to $I=(I_a)$ guarantees that tangent
vectors (infinitesimal deformations) are regular and square-integrable
so that the ${\rm L}^2$-metric (\ref{L2metric}) is well-defined; these
boundary conditions are the main distinguishing feature from the
construction of moduli spaces of (framed) monopoles. In terms of our
original $\su$-invariant instantons which are parameterized by the connections
(\ref{AcalSU2}), the boundary conditions
(\ref{quiverasympt})--(\ref{quiverpoles}) mean that they have
regular values at the origin which determine an $\su$ representation
$I=(I_a)$ in $\urm(N)$,
whereas their behaviour at infinity is governed by a pole at
$r=\infty$ with residue which gives the Fayet-Iliopoulos parameters of
the quiver gauge theory; this
somewhat undesirable asymptotic behaviour will be eliminated by our
construction below.

The corresponding quiver variety is then the infinite quotient
\beq
\Mcal_{\xi,I}(\Qsf_\Gamma,R):=\widehat{\Rep{\rm ep}}_{\Qsf_\Gamma^{\rm
  op}}(R)_{\xi,I} \,
\big/\!\!\big/ \ {\rm P}\widehat{\Gcal}(R)_I \ .
\label{infinitequivervar}\eeq
This quotient space is finite-dimensional: In the temporal gauge the moment map equations $\mu_a=0$ from (\ref{NahmZa}) can be written as
\beq
[Z_a,Z_b]+\varepsilon_{ab}^c\, \frac{\diff Z_c}{\diff t}=0
\label{Zadiffeqs}\eeq
for $a,b=1,2,3$; this leaves the space of solutions to the ordinary
differential equations (\ref{Zadiffeqs}) modulo the action of the
finite-dimensional gauge group (\ref{3.12}). Putting
$Z_a=\frac1{2t}\, X_a$ for constant $X_a$ sets up a one-to-one
correspondence between such solutions of the equations
(\ref{Zadiffeqs}) and solutions of the BPS equations (\ref{flatS3gamma}) describing the vacuum moduli space of quiver gauge theory based on the Sasakian quiver with relations $(\Qsf_\Gamma,\Rsf_\Gamma)$.

\bigskip

\noindent
{\bf Orbits and slices. \ }
Let us now describe the Higgs branch of vacuum states
$\Mcal_{\xi,I}(\Qsf_\Gamma,R)$ more explicitly. By a standard
symplectic quotient argument~\cite{15}, the hyper-K\"ahler quotient by
${\rm P}\widehat{\Gcal}_I$ is equivalent to the holomorphic symplectic quotient $\mu_\C^{-1}(0) \,
\big/ \ {\rm P}\widehat{\Gcal}_I^{\,\C}$ by the action (\ref{Cgaugetransf}) of
the complexification $\widehat{\Gcal}_I^{\,\C}$ of the gauge group
$\widehat{\Gcal}_I$. The complex Nahm equation $\mu_\C(C_1,C_2)=0$
implies that the path $C_2(t)$ lies in the same adjoint orbit in the
complex Lie algebra $\gfrak^\C:=\gfrak\otimes\C$ for all
$t\in\R_{>0}$. It
also implies that the Casimir invariants of $C_2$ are independent of
$t$. Since we quotient only by gauge transformations which are trivial
at infinity, the boundary condition (\ref{quiverasympt})
then implies that the Casimir invariants of $C_2$ coincide with those
of $T_+$.

It follows that the moduli space of solutions to the Nahm
equations with the boundary conditions (\ref{quiverasympt}) is the
\emph{closure} $\overline{\Ocal_{T_+}}$ of the adjoint orbit
$\Ocal_{T_+}$ of $T_+$ in $\gfrak^\C$ obtained by adding the (finitely
many) orbits of elements that have the same Casimir invariants as
$T_+$. The orbit $\Ocal_{T_+}$ is a complex symplectic manifold, with
the standard Kirillov-Kostant-Souriau symplectic form, of
dimension $\dim_\C\Ocal_{T_+}= \dim_\C\Gcal^\C-\dim_\C\Zcal_{T_+}$
where $\Zcal_{T_+}\subset \Gcal^\C$ is the subgroup that commutes with
$T_+$; note that $\T^\C\subseteq\Zcal_{T_+}$. If $T_+$ is regular,
i.e. $ \dim_\C \Zcal_{T_+}= \dim_\C \T^\C$, then $\overline{\Ocal_{T_+}}=\Ocal_{T_+}$ as every element of $\gfrak^\C$ with the same Casimir invariants is conjugate to $T_+$ in this case; in general, the closure $\overline{\Ocal_{T_+}}$ always contains a regular orbit.

Next we have to implement the correct pole structure (\ref{quiverpoles})
at $t=0$ which is determined by a representation $I$ of $\su$ in
$\Gcal$; representations of $\su$ in $\urm(N)$ are in one-to-one
correspondence with ordered partitions
$\vec\lambda=(\lambda_1,\dots,\lambda_s)$ of $N$ with at most $N$
parts, which correspond combinatorially to Young
diagrams with $N$ boxes and at most $N$ rows. By definition they satisfy
\beq
\sum_{i=1}^s\, \lambda_i=N \qquad \mbox{with} \quad \lambda_1\geq \lambda_2\geq \cdots\geq \lambda_s > 0 \ , 
\eeq
where the part $\lambda_i$ is the dimension of the $i$-th irreducible $\su$-module occuring in
the decomposition of the fundamental representation $\C^N$ of
$\urm(N)$ as a representation of $\su$; the integer
$s:=\ell(\vec\lambda\, )$ is called the length of the partition
$\vec\lambda$. Solutions of the complex Nahm equation with this singular behaviour are in a bijective correspondence with points in the Slodowy slice~\cite{13} which is the affine subspace of $\gfrak^\C$ given by
\beq
\Scal_I = I_+ + \zfrak(I_-) \ ,
\label{slice}\eeq
where $\zfrak(I_-)$ is the centralizer of $I_-$ in $\gfrak^\C$; the correspondence associates to points of (\ref{slice}) the solution of the complex Nahm equation given by
\beq
C_1(t)=\frac\im{2t} \, I_3 \qquad \mbox{and} \qquad C_2(t)= \frac1{2t}
\, I_++\sum_{\alpha\in P_-}\, t^{-m_\alpha} \, c_\alpha\,  v_\alpha \ ,
\label{NahmSlodowy}\eeq
where we used a complex gauge transformation to gauge fix $C_1$; here
$c_\alpha\in\C$ and
$P_-$ parameterizes the lowest weight vectors $v_\alpha\in\gfrak^\C$
of weight $m_\alpha\in\frac12\, \Z_{<0}$ for the adjoint action of
$\su$ on $\gfrak^\C$, i.e. $[I_-,v_\alpha]=0$ and $[\, \im\,
I_3,v_\alpha]=m_\alpha\, v_\alpha$, so that $\sum_{\alpha\in P_-}\,
c_\alpha\, v_\alpha\in \zfrak(I_-)$.
The Slodowy slice $\Scal_I$ intersects $\Ocal_{I_+}$ in the single
point $I_+$ transversally, i.e. $\Scal_I\oplus
T_{I_+}\Ocal_{I_+}=\gfrak^\C$, and it meets only those orbits whose
closures contain $\Ocal_{I_+}$ where it has transverse intersections
which are thereby submanifolds of $\gfrak^\C$.

It follows that the moduli space of
solutions to the Nahm equations with boundary conditions
(\ref{quiverasympt})--(\ref{quiverpoles}) is the intersection
$\overline{\Ocal_{T_+}}\cap \Scal_I$ of dimension
$\dim_\C\Zcal_{I_-}-\dim_\C\T^\C$; it is a complex symplectic manifold
with the restriction of the Kirillov-Kostant-Souriau symplectic form. Here the dimension of the
centralizer $\Zcal_{I_-}$ of $I_-$ coincides with the number of 
summands in the decomposition of $\gfrak^\C$ into irreducible
representations of $\su$, as each irreducible representation has a
one-dimensional subspace of lowest weight vectors; as 
$\su$-modules are self-dual, we can explicitly
decompose $\gfrak^\C\cong\C^N\otimes\C^N$ under the $\su$ embedding
$I$ by using the fact that for each positive integer $n=2j+1$ with
$j\in\frac12\, \Z_{\geq0}$ the Lie
group $\su$ has a unique irreducible spin-$j$ representation on $\C^n$ which
obey the Clebsch-Gordan rules
\beq
\C^n\otimes\C^{n'}\cong \bigoplus_{j^{\prime\prime}=|j-j'\,|}^{j+j'}\,
\C^{n^{\prime\prime}} \ ,
\label{SU2CG}\eeq
with $n'=2j'+1$ and $n^{\prime\prime}=2j^{\prime\prime}+1$.
It is shown
by~\cite{15} that this manifold can be naturally identified with the
hyper-K\"ahler quotient $\overline{\Ocal_{T_+}}\cap \Scal_I\cong
\big(\, \overline{\Ocal_{T_+}}\times {\rm P}\Gcal^\C \times \Scal_I \big)\,
\big/\!\!\big/\!\!\big/\!^{~}_0 \ {\rm P}\Gcal$ which is interpreted as
matching the two solutions to the Nahm equations coming from
$\overline{\Ocal_{T_+}}$ on $\R_{>0}$ and from ${\rm P}\Gcal^\C \times
\Scal_I$ on $(0,1]$; the latter moduli space consists of pairs
$(g(1),C_2)$, where $g:(0,1]\to{\rm P}\Gcal^\C$ is the unique complex gauge
transformation which gauge fixes $C_1$ in (\ref{NahmSlodowy}).

It remains to implement $\Gamma$-equivariance. This requires that the
pair $(C_1,C_2)$ belong to the representation space
(\ref{repspaceSE}), and hence our moduli space (\ref{infinitequivervar}) can be described easily
as the intersection
\beq
\Mcal_{\xi,I}(\Qsf_\Gamma,R)\cong \overline{\Ocal_{T_+}}\cap
\Scal_I\cap {\Rep{\rm ep}}_{\Qsf_\Gamma^{\rm op}} (R)
\label{moduliint}\eeq
which is naturally a hyper-K\"ahler variety with the restricted
hyper-K\"ahler structure of $\overline{\Ocal_{T_+}}\cap
\Scal_I$.
Since the complex gauge transformation which fixes $C_1$ in
(\ref{NahmSlodowy}) resides in ${\rm P}\Gcal(R)^\C$, and with the understanding that the $\su$ representation $I$ already
resides in (\ref{repspaceSE}) (or else (\ref{moduliint}) is empty), by
restricting the gauge group $\Gcal$ to the subgroup
$\Gcal(R)\subset\urm(N)$ we can compute the dimension of the moduli space of
vacua using the hyper-K\"ahler quotient construction above to get
\bea
\dim_\R \Mcal_{\xi,I}(\Qsf_\Gamma,R)&=& \dim_\R \big(\,
(\, \overline{\Ocal_{T_+}}\cap {\Rep{\rm ep}}_{\Qsf_\Gamma^{\rm
    op}} (R) ) \times \ {\rm P}\Gcal(R)^\C \times
(\Scal_I \cap {\Rep{\rm ep}}_{\Qsf_\Gamma^{\rm
    op}} (R) ) \, \big)\nonumber \\ &&  -\, 4\dim_\R {\rm P}\Gcal(R) \nonumber\\[4pt]
&=&
\dim_\R{\Rep{\rm ep}}_{\Qsf_\Gamma^{\rm
    op}} (R)-2\dim_\R\T(R) +\dim_\R \big(\zfrak(I_-) \cap {\Rep{\rm ep}}_{\Qsf_\Gamma^{\rm
    op}} (R) \big)\nonumber \\ && -\, 2\dim_\R\Gcal(R) \label{equivdim}
\\[4pt]
&=&\dim_\R \big(\zfrak(I_-) \cap {\Rep{\rm ep}}_{\Qsf_\Gamma^{\rm
    op}} (R) \big) +2\vec N\cdot \big(A^\Gamma\,\big)^\top\vec N
-2\, \sum_{\ell=0}^{r_\Gamma}\, N_\ell\, (N_\ell+1) \ . \nonumber
\eea
For Sasakian quivers this dimension is always a multiple of four.

\bigskip

\noindent
{\bf Nilpotent cones. \ }
On imposing $\Gamma$-equivariance, one generically encounters further
phenomena. The $\Gamma$-equivariance of $C_2$ generally
requires $T_+=0$ and therefore the Casimir invariants of $T_+$ all
vanish; the corresponding spherically symmetric instanton solutions
are then regular at $r=\infty$. The only elements $C_2$ of $\gfrak^\C$ which have vanishing
Casimir elements are nilpotent elements. For $\gfrak=\urmL(N)$, one
can conjugate any $N\times N$ complex matrix to its Jordan normal form
which for a traceless nilpotent matrix takes a block diagonal form
$\mbf J_+= \diag(\mbf J_{d_1},\dots,\mbf J_{d_m})$ where $\mbf J_{d_p}$ for $p=1,\dots,m$ is a regular nilpotent $d_p\times d_p$ matrix of the form
\beq 
\mbf J_{d_p} = \begin{pmatrix}
   0&1&0&\cdots&0\\0&0&1&\ddots&\vdots\\
   \vdots&\vdots&\ddots&\ddots&0\\0&0& \cdots &0&1 \\
   0&0& \cdots &0&0 \end{pmatrix} 
\label{Xpnilpotent}\eeq
with $(\mbf J_{d_p})^{d_p}=0$, whose centralizer is generated by
$\mbf J_{d_p},(\mbf J_{d_p})^2,\dots,(\mbf J_{d_p})^{d_p-1}$; here we have decomposed
$N=\sum_{p=1}^m\, d_p$ into integers satisfying $d_1\geq d_2\geq
\cdots \geq d_m>0$. Note that $\vec d=(d_1,\dots,d_m)$ defines an
ordered partition of $N$ and hence corresponds to an embedding of
$\su$ in $\urm(N)$ with nilpotent generator $\mbf J_+$; that every nilpotent element arises in this way is
a consequence of the Jacobson-Morozov theorem. 

In particular, there is
a unique regular nilpotent element $\mbf J_N$, corresponding to the
irreducible representation of $\su$ on $\C^N$ with partition $\vec
d=(N)$, which has vanishing Casimir
invariants and which generates the regular nilpotent orbit $\Ocal_{\mbf J_N}$ of maximal dimension. The closure of this orbit is the nilpotent cone
$\Ncal=\overline{\Ocal_{\mbf J_N}}$, of dimension $\dim_\C\Ncal=
\dim_\C\Gcal^\C-\dim_\C\T^\C$, consisting of all nilpotent elements of
$\gfrak^\C$, each generating finitely many orbits; the irreducible
subvariety $\Ncal$ has singularities corresponding to non-regular
nilpotent orbits. Among these orbits there is the unique subregular nilpotent
orbit whose closure contains all non-regular nilpotent orbits and has
complex codimension two in $\Ncal$; it corresponds to the subregular
representation of $\su$ in $\urm(N)$ with $\vec d=(N-1,1)$ and appears
as the locus of Kleinian quotient singularities $\C^2/\Z_{N}$ in
$\Ncal$~\cite{Brieskorn} (the cone over $S^3/\Z_{N}$). On
the other hand, the element $T_+=0$ generates the unique nilpotent
orbit consisting of a single singular point in $\Ncal$. The minimal nilpotent orbit is the
unique nilpotent orbit of smallest non-zero dimension, which is
generated by the highest root vector of $\gfrak^\C$ and thereby
consists of $N\times N$ matrices $H$ of rank one with $H^2=0$; it corresponds
to the $\su$ embedding with $\vec d=(2,1,1,\dots,1)$ and is the cone
over $\suk\big/\Sp\big(\urm(N-2)\times\uo\big)$ of complex dimension~$2(N-1)$.

It follows that our moduli space (\ref{moduliint}) in this case
is the \emph{singular} variety
\beq
\Mcal_{0,I}(\Qsf_\Gamma,R)\cong \Ncal \cap
\Scal_I\cap {\Rep{\rm ep}}_{\Qsf_\Gamma^{\rm op}} (R)
\label{modulinilpotent}\eeq
of dimension given by the formula (\ref{equivdim}); the appearence of orbifold
singularities here is not surprising given our earlier observation
concerning the stacky nature of the quotient parameterizing the
vacuum moduli. The structure of
the singular locus and the dimension of the moduli space now depend
on the embedding $I$ of $\suL$ in ${\Rep{\rm ep}}_{\Qsf_\Gamma^{\rm op}}
(R)$ that determines the transverse slice $\Scal_I$ to the orbit of the
nilpotent element $I_+$, and on the particular Sasakian quiver
$\Qsf_\Gamma$. In general, our moduli
spaces generically have higher dimension than the naive (non-equivariant)
prediction because of two non-standard features: The
representation varieties $ {\Rep{\rm ep}}_{\Qsf_\Gamma^{\rm op}} (R) \subset\urmL(N)$
contain the Lie algebras (\ref{gfrakR}) as proper subalgebras, while the
quotient defining the moduli space is taken with respect to the broken
gauge group (\ref{3.12}) which is a proper subgroup of $\urm(N)$. For example, if $I=0$ is the trivial representation corresponding to the partition $\vec\lambda=(1,1,\dots,1)$, then $\Scal_0=\gfrak^\C$ and
hence the moduli space is simply the cone
\beq
\Mcal_{0,0}(\Qsf_\Gamma,R)\cong \Ncal \cap
{\Rep{\rm ep}}_{\Qsf_\Gamma^{\rm op}} (R)
\eeq
consisting of all $\Gamma$-equivariant nilpotent endomorphisms of
$V_R$; it has dimension
\beq
\dim_\R \Mcal_{0,0}(\Qsf_\Gamma,R) = 4\vec N\cdot \big(A^\Gamma\,\big)^\top\vec N
-2\, \sum_{\ell=0}^{r_\Gamma}\, N_\ell\,(N_\ell+1) \ .
\eeq

\bigskip

\noindent
{\bf $\mbf{A_k}$. \ }
For the cyclic group
$\Gamma=\Z_{k+1}$ with $k=2q$, the adjacency matrix of the quiver
(\ref{3.19}) is given by $a_{\ell\ell'}^{\Gamma}=
\delta_{\ell'\ell}+\delta_{\ell',\ell+1}$ and the tensor product
multiplicities are
$m_{\ell\ell'}{}^{\ell^{\prime\prime}}=\delta_{\ell^{\prime\prime},\ell+\ell'}$. By
(\ref{CGammarels}) the two-dimensional $\Gamma$-module (\ref{SasakiC2}) decomposes into irreducible representations as $\C_{\Gamma}^2=V_0\oplus V_1$ so that
\beq
\C_{\Gamma}^2\otimes V_\ell= V_\ell\oplus V_{\ell+1}
\label{CGammaVell}\eeq
for $\ell=0,1,\dots,k$; note that $\C_\Gamma^2$ is not a self-dual
representation of $\Gamma$. With $C_{\Qsf_\Gamma}=2\, \one_{\C^{k+1}}-A^\Gamma-(A^\Gamma\,)^\top$ the generalized Cartan matrix of the Sasakian quiver $\Qsf_\Gamma$, by
(\ref{dimquiver}) the dimension of the corresponding Nakajima quiver
variety is given by
\beq
\dim_\R \Xcal_\xi(\Qsf_\Gamma ,R)= 4+4 \, \sum_{\ell=0}^k\, N_\ell\, N_{\ell+1} \ .
\eeq
It follows that all representations $R$ of the Sasakian quiver in this
case are indecomposable and correspond to imaginary roots; the
Nakajima quiver variety is four-dimensional for irreducible quiver bundles (\ref{4.3}) corresponding to
simple representations $R$ with $N_\ell=N$ for some
$\ell\in\{0,1,\dots,r_\Gamma\}$ and $N_{\ell'}=0$ for $\ell'\neq\ell$.

The respective summands in (\ref{CGammaVell}) correspond to the matrix pairs $(C_1,C_2)$ which can be decomposed analogously to (\ref{4.7})--(\ref{4.8}) as
\beq \label{C1C2Sasaki}
C_1=\diag (\rho_0, \rho_1,\ldots,\rho_k) \und C_2=\begin{pmatrix}
   0&\phi_1&0&\cdots&0\\0&0&\phi_2&\ddots&\vdots\\
   \vdots&\vdots&\ddots&\ddots&0\\0&0& \cdots &0&\phi_k \\
   \phi_{k+1}&0& \cdots &0&0 \end{pmatrix} \ ,
\eeq
with $\rho_\ell:\R_{>0}\to \End_\C(R_\ell)$ and $\phi_{\ell+1}:\R_{>0}\to \Hom_\C(R_{\ell+1},R_\ell)$ for $\ell=0,1,\dots,k$. The complex Nahm equations determined by the complex moment map $\mu_\C$ from (\ref{complexNahm}) read as
\beq
\frac{\diff \phi_{\ell+1}}{\diff t}= \phi_{\ell+1} \,\rho_{\ell+1}- \rho_{\ell}\, \phi_{\ell+1}
\label{complexNahmeq}\eeq
in $\Hom_\C(R_{\ell+1},R_\ell)$ for $t\in\R_{>0}$ and $\ell=0,1,\dots,k$, while the real Nahm equations from $\mu_\R$ in (\ref{realNahm}) are given by
\beq
\frac{\diff\rho_\ell}{\diff t}+\frac{\diff\rho_\ell^\dag}{\diff t}= \phi_{\ell+1}\, \phi_{\ell+1}^\dag - \phi_\ell^\dag\, \phi_\ell - \big[\rho_\ell,\rho_\ell^\dag\,\big]
\label{realNahmeq}\eeq
in $\End_\C(R_\ell)$ for $t\in\R_{>0}$ and $\ell=0,1,\dots,k$; here the boundary conditions (\ref{quiverasympt}) require setting $T_+=0$. Note the formal similarily between these equations and the constant F-term and D-term relations (\ref{Ftermrels}) and (\ref{Dtermrels}): In the temporal gauge $Z_t=0$ the path $\rho_\ell$ is Hermitian, and the right-hand sides of (\ref{Ftermrels}) and (\ref{Dtermrels}) are replaced with the corresponding radial variations in (\ref{complexNahmeq}) and~(\ref{realNahmeq}).

By (\ref{equivdim}) the dimension of the corresponding Sasakian quiver
variety is given by
\beq
\dim_\R \Mcal_{0,I}(\Qsf_\Gamma,R) = \dim_\R \big(\zfrak(I_-) \cap {\Rep{\rm ep}}_{\Qsf_\Gamma^{\rm
    op}} (R) \big)+2\,
\sum_{\ell=0}^k\, N_\ell\, (N_{\ell+1}-1) \ .
\eeq
In particular, when $N_\ell=1$ for all $\ell=0,1,\dots,k$ (so that
$N=k+1$) this
dimension formula becomes
\beq
\dim_\R \Mcal_{0,I}\big(\, \Qsf_\Gamma\,,\,\widehat{V}\, \big) =
\dim_\R \big(\zfrak(I_-) \cap {\Rep{\rm ep}}_{\Qsf_\Gamma^{\rm
    op}} (\,\widehat{V}\,) \big) \ ,
\label{dimNell1}\eeq
whereas the real dimension of the nilpotent cone $\Ncal$ in
$\slrmL(k+1,\C)$ is $2k\,(k+1)$.
Let us examine some particular cases for illustration.

For the trivial representation $I=0$, the moduli space has dimension
$4(k+1)\geq8$ and it contains the BPST-type instanton solutions
from~\cite[Sect.~5]{Ivanova:2013mea}, i.e. the model solution
(\ref{modelsol}) with $T_a=0$, $\lambda>0$ and $J_a$ the generators of the regular
embedding of the group $\su$ into $\surm(k+1)$; this four-parameter family lies in the
subcone $C(S^3/\Z_{k+1})$ along the subregular
orbit of the equivariant nilpotent cone $\Mcal_{0,0}\big(\,
\Qsf_\Gamma\,,\,\widehat{V}\, \big)  =\Ncal \cap
{\Rep{\rm ep}}_{\Qsf_\Gamma^{\rm op}} \big(\,\widehat{V}\,\big)$ in
the complex Lie algebra $\slrmL(k+1,\C)$. 

Now let us consider 
the regular representation $I=I^{\rm reg}$ of $\su$ in
$\surm(k+1)$. To determine the centralizer
$\zfrak(I_-)$ in ${\Rep{\rm ep}}_{\Qsf_\Gamma^{\rm
    op}} (\,\widehat{V}\,) $ in this case,
we need to determine the space of matrices
(\ref{C1C2Sasaki}) which commute with the nilpotent matrix $\mbf J_{k+1}$
from (\ref{Xpnilpotent}). It is easy to see that the general form of
such matrices is given by $C_1=\rho\, \one_{\C^{k+1}}$ and $C_2=\phi\,
\mbf J_{k+1}$ for arbitrary $\rho,\phi\in\C$; hence the
moduli space in this case is four-dimensional. Upon intersecting with the
nilpotent cone $\Ncal$ in $\slrmL(k+1,\C)$, we expect to see the singular
locus $\C^2/\Z_{k+1}$ by Brieskorn's theorem~\cite{Brieskorn}. This
can be checked directly: It remains to quotient the space
$(\rho,\phi)\in\C^2$ by the commutant subgroup of $\su$ in $\surm(k+1)$, which
in this case is simply the center $\Z_{k+1}$ of $\surm(k+1)$ and hence the moduli space is biholomorphic to the orbifold
singularity
\beq
\Mcal_{0,I^{\rm reg}}\big(\,
\Qsf_\Gamma\,,\,\widehat{V}\, \big) \cong C\big(S^3\big/\Z_{k+1}\big) \ .
\eeq
This is in marked contrast to
the non-equivariant case where the moduli space would consist of just the
single element $I_+$. The model solution
(\ref{modelsol}), with $T_a=0$, $\lambda=0$ and $J_a$ the generators of the regular
embedding of $\su$, lives in this moduli space. Recall that this was
precisely the situation for the solution (\ref{tHooftsol}); in this
sense the 't~Hooft and BPST instantons are ``equivalent'' for fixed
${\rm P}\Gcal(R)=\surm(N)$. Moreover, for the minimal charge
instantons and suitable boundary conditions, by taking a trivial
$\Gamma$-action one can extend instantons from $\R^4$ to $\R^4/\Gamma$
and its deformation $M_\xi$, giving the same four-dimensional moduli
space. On the other hand, by an appropriate choice of non-trivial $\Gamma$-action on
the rank $N$ vector bundle $E\to M_\xi$ and suitable embedding of
$\su$ in $\surm(N)$ at infinity, one obtains BPST-type instantons on
$\C^2/\Gamma$ in a four-dimensional moduli space via both ADHM and
Nahm equations.

A completely analogous calculation shows that for the subregular representation
$I=I^{\rm subreg}$
the dimension of the moduli space is equal to $8$, and upon dividing by the
commutant subgroup the moduli space can be described topologically as the cone
\beq
\Mcal_{0,I^{\rm subreg}}\big(\,
\Qsf_\Gamma\,,\,\widehat{V}\, \big) \cong C\big(\sut\,\big/\, \Sp(\uo\times
\uo)\big) \ .
\eeq
One can carry on with smaller representations of $\su$ whose
centralizers have complex dimension two or more; in general the
commutant subgroup of $\surm(k+1)$ is a Lie group of rank
$\ell(\vec\lambda\, )-1$, where
$\vec\lambda$ is the partition corresponding to the $\su$
embedding $I$, and it is abelian if and
only if the parts of $\vec\lambda$ are all distinct integers. The construction continues
until we reach the trivial representation $I=0$ with maximal moduli space
dimension $4(k+1)$. When some $N_\ell>1$ one encounters moduli spaces
of even higher dimensions.

\bigskip

\noindent
{\bf Non-abelian affine Toda field theory. \ }
We have found that imposing  
$\Z_{k+1}$-equivariance on the matrix pairs $(C_1,C_2)$ yields
matrices (\ref{C1C2Sasaki}) and reduces the anti-self-duality equations on $\C^2/\Z_{k+1}$  
to the equations (\ref{complexNahmeq})--(\ref{realNahmeq}). With
$N_\ell=1$ for all $\ell=0,1,\dots,k$, the equation (\ref{complexNahmeq}) can be
written as
\beq
\frac{\diff\log\phi_{\ell+1}}{\diff t} = \rho_{\ell+1} - \rho_\ell \ .
\label{complexNahmeqN1}\eeq
By taking $\rho_\ell, \phi_\ell\in\R$ (where Hermiticity of $C_1$ is
automatic in the temporal gauge $Z_t=0$), differentiating
(\ref{complexNahmeqN1}) with respect to $t$  
and using (\ref{realNahmeq}) we get the equations
\beq
  2\, \frac{\diff^2 \log\phi_{\ell+1}}{\diff t^2} =(\phi_{\ell+2})^2 -  
2\, (\phi_{\ell+1})^2 + (\phi_{\ell})^2
\eeq
which are the equations of the affine Toda lattice associated with the
$\widehat{A}_k$-type Lie algebra for rotationally symmetric fields in two-dimensions, see
e.g.~\cite[App.~A]{Dunne}. These equations are explicitly
integrable and their solutions are parameterized in terms of $2(k+1)$ arbitrary constants, where the
integrability is based on the underlying group theory structure and
the solutions can be expressed as particular matrix elements in the
fundamental representation of $\surm(k+1)$. Imposing the requirements that these
solutions vanish as $t\to\infty$ and that they admit the appropriate
pole structure (\ref{quiverpoles}) at $t=0$ reduces the number of free
parameters accordingly. For example, the $2(k+1)$-parameter family of
solutions presented in~\cite[eqs.~(A31)--(A33)]{Dunne} vanish at
$t=\infty$ and are regular at the origin $t=0$; for generic values of
these parameters the solutions correspond to the trivial
representation $I=0$ of $\su$. Solutions with residues at $t=0$
defining non-trivial representations of $\su$ in $\surm(k+1)$ require
fixing some of these parameters appropriately. All of this agrees with
our analysis of the moduli space of solutions above for the
$\widehat{A}_{k}$-type quiver gauge theory.

For $N_\ell> 1$, our equations (\ref{complexNahmeq})--(\ref{realNahmeq}) are a variant of non-abelian  
affine Toda lattice equations; however, they do not coincide exactly
with the existing
non-abelian generalizations considered previously in the literature,
see e.g.~\cite{Gekhtman}.
Nevertheless, our equations have a Lax pair and zero curvature presentation  
from their origin as anti-self-duality equations in four  
dimensions, and in the abelian limit they coincide with the affine Toda  
lattice equations; hence we may refer to
(\ref{complexNahmeq})--(\ref{realNahmeq}) as non-abelian  
affine Toda lattice equations. We also have an explicit
description of their moduli spaces of solutions as
real slices of the (singular) hyper-K\"ahler moduli spaces of our Sasakian quiver
gauge theories. This description may have important uses as
generalizations of the standard conformal two-dimensional $\widehat{A}_k$ Toda field theories which are well
studied in the literature. In particular, our approach is reminescent
of the recent AGT
duality~\cite{Alday,Wyllard} which relates them to four-dimensional
$\Ncal=2$ superconformal quiver gauge theories of
$\widehat{A}_k$-type.

\bigskip

\noindent
{\bf $\mbf{A_n}$ quiver gauge theory. \ }
In the non-equivariant case $\Gamma=\{1\}$, Kronheimer's moduli spaces
of $\su$-invariant $\urm(N)$-instantons on $C(S^3)$ with $T_a=0$ can
be regarded in certain cases as
particular classes of Nakajima quiver varieties associated to a linear
$A_n$ quiver (\ref{Akquiver}) for some $n\leq
N$ determined by the $\su$ embedding $I$~\cite[Sect.~8]{Nakajima}; the identification is based on the ADHM
transform of instantons on $\R^4$. The ADHM moduli space
is itself a quiver variety based on the Jordan quiver
\beq
\xymatrix{
{\bullet} \ar@(dr,ur)[]
} 
\eeq
which is the oriented graph of the
$\widehat{A}_0$ Dynkin diagram corresponding to the $k=0$ limit of the
cyclic group $\Z_{k+1}$; the representation space of the corresponding
double
quiver is $\Hom_\C(W_Q,\C^2\otimes W_Q)$ where $W_Q\cong\C^N$. The
Kronheimer moduli space is constructed from the $\su$-invariant part of
$\Hom_\C(W_Q,\C^2\otimes W_Q)$, in much the same way that the
representation space (\ref{repspaceMcKay}) parameterizes
$\Gamma$-equivariant instantons on $\R^4$. Now we
decompose the vector space $W_Q$ into irreducible representations of
$\su$ on $\C^\ell$ as
\beq
W_Q=\bigoplus_{\ell=1}^n\, Q_\ell\otimes\C^\ell \qquad \mbox{with}
\quad Q_\ell\cong\C^{v_\ell} \ ,
\eeq
where the dimension vector $\vec v=(v_1,\dots,v_n)$
represents the $\su$-module structure of the instantons at the
origin. From the Clebsch-Gordan decomposition (\ref{SU2CG}) we have
$\C^2\otimes\C^\ell= \C^{\ell-1}\oplus\C^{\ell+1}$, and hence by
Schur's lemma $\Hom_\su(W_Q,\C^2\otimes W_Q)$ coincides with the
representation variety of the double of the $A_n$ quiver
(\ref{Akquiver}). To accomodate a non-trivial holonomy of the
anti-self-dual connection at infinity which is specified by an $\su$
embedding $I$ corresponding to an ordered partition
$\vec\lambda$ of $N$, we consider the
corresponding framed quiver with framing nodes specifying a
representation $W_S$ of $\su$ in $\urm(N)$ with $S_\ell\cong\C^{w_\ell}$, where the dimension
vector $\vec w=(w_1,\dots,w_n)$ labels the number $0\leq w_\ell\leq N$ of parts of
$\vec\lambda$ with
$\lambda_i=\ell$ for $\ell=1,\dots,n$; then 
\beq
N=\sum_{\ell=1}^n\, \ell\, w_\ell \qquad \mbox{and} \qquad 
\ell(\vec\lambda\,) = \sum_{\ell=1}^n\, w_\ell \ .
\eeq
We denote this
framed Nakajima quiver variety by
$\Xcal_0(\Qsf_{A_n};Q,S)\cong\Ncal\cap\Scal_I$; the dimension vectors
$\vec v$ and $\vec w$ obey certain consistency relations with the
nilpotent orbits in $\Ncal$ which are described
in~\cite[Sect.~8]{Nakajima}.

In the $\Gamma$-equivariant case, we have to take the intersection
(\ref{modulinilpotent}). Moreover the roles of the vector spaces $W_Q$ and
$W_S$ are interchanged: The $\su$-module $W_Q$ now describes the behaviour
of the instanton connection at infinity, while $W_S$ describes the
$\su$-module structure of its regular value at the origin $r=0$. Demanding as
usual that the $\su$ representation fit into the $\Gamma$-equivariant
structure, it follows that our moduli space admits a presentation as
the subvariety
\beq
\Mcal_{0,I}(\Qsf_\Gamma,R)\cong \Xcal_0(\Qsf_{A_n};Q,S) \cap {\Rep{\rm
    ep}}_{\Qsf_\Gamma^{\rm op}} (R)
\eeq
of $\Gamma$-equivariant maps in a framed Nakajima quiver variety
associated to a linear $A_n$ quiver. This gives a realization of the
vacuum states of the Sasakian quiver gauge theory in the Higgs
moduli spaces of certain $A_n$ quiver gauge theories.

Because of the role reversal of
boundary conditions, we cannot interpret this presentation as an ADHM
matrix model of the type which arises from
systems of D$p$--D$(p+4)$
branes. Instead, as the scalar fields $Z_a$ for $a=1,2,3$ have
a Nahm pole boundary condition parameterized by the partition
$\vec\lambda$, we may give an (indirect) interpretation of our moduli of Nahm data in terms of
configurations of D$p$--D$(p+2)$--D$(p+4)$ branes following~\cite{Gaiotto}. For
this, we
consider a system of $N$ parallel D$(p+2)$-branes wrapping
$C(S^3/\Gamma)$ which transversally intersect $\ell(\vec\lambda\,)$
D$(p+4)$-branes with $\lambda_i$ D$(p+2)$-branes ending on the $i$-th
D$(p+4)$-brane at the apex $r=0$ of the cone
$C(S^3/\Gamma)$ for $i=1,\dots,\ell(\vec\lambda\,)$. The D$(p+2)$-branes
support a four-dimensional $\Ncal=2$ quiver gauge theory based on the
$A_n$ quiver, with scalar fields $Z_a$ for $a=1,2,3$ along the
D$(p+4)$-branes such that $(C_1,C_2)\in \Hom_\su(W_Q,\C^2\otimes
W_Q)$; the gauge group at the $\ell$-th node of the quiver
(\ref{Akquiver}) is $\urm(w_\ell)$ for $\ell=1,\dots, n$. Due to the
orbifold singularity at $r=0$, there are an additional $N_\ell$
constituent fractional D$p$-branes probing the D$(p+4)$-branes. It would be interesting to give a more
direct picture in terms of the original scalar fields $X_a$ from the
$\su$-equivariant 
dimensional reduction over the cone $C(S^3/\Gamma)$.

\bigskip

\section*{Acknowledgments}

\noindent
The work of OL and ADP was partially supported by the Deutsche  
Forschungsgemeinschaft under Grant LE 838/13. The work of RJS was partially supported by the Consolidated Grant
ST/L000334/1 from the
UK Science and Technology Facilities Council. This work was completed
while RJS was visiting the Hausdorff Research Institute for
Mathematics in Bonn during the 2014 Trimester Program ``Noncommutative
Geometry and its Applications''; he would like to thank Alan Carey,
Victor Gayral, Matthias Lesch, Walter van Suijlekom and Raimar
Wulkenhaar for the invitation, and all the staff at HIM for the warm hospitality.


\bigskip

\begin{thebibliography}{99}
\addtolength{\itemsep}{-6pt}

\bibitem{AlGar12}
L.~Alvarez-C\'onsul and O.~Garc\'ia-Prada, ``{Dimensional reduction and quiver
  bundles},'' {J. Reine Angew. Math.} {\bf 556} (2003) 1 [arXiv:math-dg/0112160].

\bibitem{Lechtenfeld:2007st}
O.~Lechtenfeld, A.D. Popov and R.J. Szabo, ``{Quiver gauge theory and
  noncommutative vortices},''
  Prog. Theor. Phys. Suppl.
  {\bf 171} (2007) 258 [arXiv:0706.0979 [hep-th]].

\bibitem{Dolan:2010ur}
B.P. Dolan and R.J. Szabo, ``{Equivariant dimensional reduction and quiver
  gauge theories},'' 
  Gen. Rel. Grav. {\bf 43} (2010) 2453 [arXiv:1001.2429 [hep-th]].

\bibitem{DM}
  M.R.~Douglas and G.W.~Moore,
  ``D-branes, quivers, and ALE instantons,''
  arXiv:hep-th/9603167.
  
\bibitem{JM}
  C.V.~Johnson and R.C.~Myers,
  ``Aspects of type IIB theory on ALE spaces,''
  Phys.\ Rev.\ D {\bf 55} (1997) 6382
  [arXiv:hep-th/9610140].
  
\bibitem{DGM}
  M.R.~Douglas, B.R.~Greene and D.R.~Morrison,
  ``Orbifold resolution by D-branes,''
  Nucl.\ Phys.\ B {\bf 506} (1997) 84
  [hep-th/9704151].
  
\bibitem{Gubser}
  S.S.~Gubser, N.A.~Nekrasov and S.L.~Shatashvili,
  ``Generalized conifolds and 4-dimensional $\Ncal=1$ superconformal field theory,''
  JHEP {\bf 9905} (1999) 003
  [arXiv:hep-th/9811230].
  
\bibitem{Douglas:2000qw}
  M.R.~Douglas, B.~Fiol and C.~Romelsberger,
  ``The spectrum of BPS branes on a noncompact Calabi-Yau,''
  JHEP {\bf 0509} (2005) 057
  [arXiv:hep-th/0003263].

\bibitem{PS1}
  A.D.~Popov and R.J.~Szabo,
  ``Quiver gauge theory of non-abelian vortices and noncommutative instantons in higher dimensions,''
  J.\ Math.\ Phys.\  {\bf 47} (2006) 012306
  [arXiv:hep-th/0504025].

\bibitem{Dolan:2009ie}
  B.P.~Dolan and R.J.~Szabo,
  ``Dimensional reduction, monopoles and dynamical symmetry breaking,''
  JHEP {\bf 0903} (2009) 059
  [arXiv:0901.2491 [hep-th]].

\bibitem{Szabo:2014zua}
  R.J.~Szabo and O.~Valdivia,
  ``Covariant quiver gauge theories,''
  JHEP {\bf 1406} (2014) 144
  [arXiv:1404.4319 [hep-th]].
  
\bibitem{10}
P.B. Kronheimer, ``The construction of ALE spaces as hyper-K\"ahler quotients,''
J. Diff. Geom. {\bf 29} (1989) 665.

\bibitem{K2} 
P.B. Kronheimer, ``A Torelli-type theorem for gravitational instantons,'' J. Diff. Geom. 
{\bf 29} (1989) 685.

\bibitem{KN}  
P.B. Kronheimer and H. Nakajima, ``Yang-Mills instantons on ALE
gravitational instantons,'' Math. Ann. {\bf 288} (1990) 263.

\bibitem{13}
P.B. Kronheimer, ``Instantons and the geometry of the nilpotent variety,'' 
J. Diff. Geom.
{\bf 32} (1990) 473.

\bibitem{14}
P.B. Kronheimer, ``A hyper-K\"ahlerian structure on coadjoint orbits of a semisimple
complex group,'' J. London Math. Soc. {\bf 42} (1990) 193.

\bibitem{15}
 R. Bielawski, ``Hyper-K\"ahler structures and group actions,'' 
 J. London Math. Soc. {\bf 55} (1997) 400;
 ``On the hyper-K\"ahler metrics associated to singularities of nilpotent varieties,''
 Ann. Global Anal. Geom. {\bf 14} (1996) 177.
 
 \bibitem{Ivanova:2013mea}
  T.A.~Ivanova, O.~Lechtenfeld, A.D.~Popov and R.J.~Szabo,
  ``Orbifold instantons, moment maps and Yang-Mills theory with sources,''
  Phys.\ Rev.\ D {\bf 88} (2013) 105026
  [arXiv:1310.3028 [hep-th]].

\bibitem{Gaiotto}
  D.~Gaiotto and E.~Witten,
  ``Supersymmetric boundary conditions in $\Ncal=4$ super Yang-Mills theory,''
  J.\ Statist.\ Phys.\  {\bf 135} (2009) 789
  [arXiv:0804.2902 [hep-th]]; ``S-duality of boundary conditions in $\Ncal=4$ super Yang-Mills theory,''
  Adv.\ Theor.\ Math.\ Phys.\  {\bf 13} (2009) 721
  [arXiv:0807.3720 [hep-th]].
  
\bibitem{Boyer}
C.P.~Boyer and K.~Galicki,
{\sl Sasakian Geometry} (Oxford University Press, 2008).

\bibitem{PS2}
  A.D.~Popov and R.J.~Szabo,
  ``Double quiver gauge theory and nearly K\"ahler flux compactifications,''
  JHEP {\bf 1202} (2012) 033
  [arXiv:1009.3208 [hep-th]].

\bibitem{GP}
 L. Alvarez-C\'onsul and O. Garc\'ia-Prada,
``Dimensional reduction, ${\rm SL}(2,\C)$-equivariant bundles
and stable holomorphic chains,'' Int. J. Math. {\bf 12}
(2001) 159.
  
\bibitem{LM}
  H.~Lin and J.M.~Maldacena,
  ``Fivebranes from gauge theory,''
  Phys.\ Rev.\ D {\bf 74} (2006) 084014
  [arXiv:hep-th/0509235].
  
\bibitem{Reid} M. Reid,  
``McKay correspondence,'' Proc. Alg. Geom. Symp. (Kinosaki) (1996) 14 [arXiv:alg-geom/9702016].
  
\bibitem{Cirafici:2012qc}
  M.~Cirafici and R.J.~Szabo,
  ``Curve counting, instantons and McKay correspondences,''
  J.\ Geom.\ Phys.\  {\bf 72} (2013) 54
  [arXiv:1209.1486 [hep-th]].
  
\bibitem{D}
S.K. Donaldson, ``Anti-self-dual Yang-Mills connections over complex algebraic surfaces and
stable vector bundles,'' Proc. London Math. Soc. {\bf 50} (1985) 1.

\bibitem{UY}
 K. Uhlenbeck and S.-T. Yau, ``On
the existence of Hermitian Yang-Mills connections in stable vector bundles,'' Commun. Pure
Appl. Math. {\bf 39} (1986) 257.
  
\bibitem{Cassens}
H.~Cassens and P.~Slodowy,
``On Kleinian singularities and quivers,''
Prog. Math. {\bf 162} (1996) 263.

\bibitem{Nakajima}
H.~Nakajima,
``Instantons on ALE spaces, quiver varieties and Kac-Moody algebras,''
Duke Math. J. {\bf 76} (1994) 365.

\bibitem{12}
N.J.~Hitchin,
``Hyper-K\"ahler manifolds,'' Ast\'erisque {\bf 206} (1992) 137.

\bibitem{Kac}
V.G.~Kac,
``Infinite root systems, representations of graphs and invariant theory,''
Invent. Math. {\bf 56} (1980) 57.

\bibitem{BFRM}
  M.~Bianchi, F.~Fucito, G.~Rossi and M.~Martellini,
  ``Explicit construction of Yang-Mills instantons on ALE spaces,''
  Nucl.\ Phys.\ B {\bf 473} (1996) 367
  [arXiv:hep-th/9601162].
  
\bibitem{Nak90}
H.~Nakajima,
``Moduli spaces of anti-self-dual connections on ALE gravitational
instantons,''
Invent. Math. {\bf 102} (1990) 267.

\bibitem{Slodowy}
P.~Slodowy,
{\sl Simple Singularities and Simple Algebraic Groups} (Springer,
1980).

\bibitem{Sniatycki}
J.~\'Sniatycki,
{\sl Differential Geometry of Singular Spaces and Reduction of
  Symmetries}
(Cambridge University Press, 2013).

\bibitem{Biquard}
O.~Biquard, ``Sur les \'equations de Nahm et la structure de Poisson
des alg\`ebres de Lie semi-simples complexes,''
Math. Ann. {\bf 304} (1996) 253.

\bibitem{Brieskorn}
E.~Brieskorn,
``Singular elements of semisimple algebraic groups,''
Actes Congr\'es Intern. Math. {\bf 2} (1970) 279.

\bibitem{Dunne}
  G.V.~Dunne, R.~Jackiw, S.-Y.~Pi and C.A.~Trugenberger,
  ``Self-dual Chern-Simons solitons and two-dimensional nonlinear equations,''
  Phys.\ Rev.\ D {\bf 43} (1991) 1332.

\bibitem{Gekhtman}
M.~Gekhtman,
``Hamiltonian structure of non-abelian Toda lattice,''
Lett. Math. Phys. {\bf 46} (1998) 189.

\bibitem{Alday}
  L.F.~Alday, D.~Gaiotto and Y.~Tachikawa,
  ``Liouville correlation functions from four-dimensional gauge theories,''
  Lett.\ Math.\ Phys.\  {\bf 91} (2010) 167
  [arXiv:0906.3219 [hep-th]].

\bibitem{Wyllard}
  N.~Wyllard,
  ``$A_{N-1}$ conformal Toda field theory correlation functions from conformal $\Ncal = 2$ $\surm(N)$ quiver gauge theories,''
  JHEP {\bf 0911} (2009) 002
  [arXiv:0907.2189 [hep-th]].

\end{thebibliography}
\end{document}